\documentclass[aps,prd,superscriptaddress,groupedaddress,nofootinbib]{article}
\usepackage{dcolumn}
\usepackage{bm}
\usepackage{lscape}
\usepackage{amsmath,amssymb,graphicx}
\usepackage{epsfig}
\usepackage{pstricks}
\usepackage{multirow}
\usepackage[bookmarks,linktocpage]{hyperref}
\usepackage{booktabs}
\usepackage{graphicx}
\usepackage{subfigure}
\usepackage{cite}
\usepackage{rotating}
\usepackage{float}
\usepackage{soul}
\usepackage[a4paper, left=2.5cm, right=2.5cm, top=2.5cm, bottom=2.5cm]{geometry}
\usepackage{comment}
\usepackage{textcomp}
\usepackage{afterpage}

\usepackage{array}
\newcolumntype{L}[1]{>{\raggedright\let\newline\\\arraybackslash\hspace{0pt}}m{#1}}
\newcolumntype{C}[1]{>{\centering\let\newline\\\arraybackslash\hspace{0pt}}m{#1}}
\newcolumntype{R}[1]{>{\raggedleft\let\newline\\\arraybackslash\hspace{0pt}}m{#1}}

\RequirePackage{lineno}

\newcommand{\Lag}{\mathcal L}

\newcommand{\MET}{\textbf{$E_{\rm T}^{\rm miss}$}}

\begin{document}

\setlength{\arraycolsep}{1.5pt}

\begin{center}
{\Large \textbf{Production of extra quarks decaying to Dark Matter \\ 
beyond the Narrow Width Approximation at the LHC }}

\vskip.1cm
\end{center}

\vskip0.2cm

\begin{center}
\textbf{{Stefano Moretti$^{1,2,3}$, Dermot O'Brien$^{1,2}$, Luca Panizzi$^{4,1,2}$ and Hugo Prager$^{1,2}$} \vskip 8pt }

{\small
$^1$\textit{School of Physics and Astronomy, University of Southampton, Highfield, Southampton SO17 1BJ, UK}\\[0pt]
\vspace*{0.1cm} $^2$\textit{Particle Physics Department, Rutherford Appleton Laboratory, Chilton, Didcot, Oxon OX11 0QX, UK}\\[0pt]
\vspace*{0.1cm} $^3$\textit{Physics Department, CERN, CH-1211, Geneva 23, Switzerland}\\[0pt]
\vspace*{0.1cm} $^4$\textit{Dipartimento di Fisica, Universit\`a di Genova and INFN, Sezione di Genova,\\via Dodecaneso 33, 16146 Genova, Italy}\\[0pt]
}
\end{center}

\begin{abstract} 
\noindent
This paper explores the effects of finite width in processes of pair production of an extra heavy quark with charge 2/3 (top partner) and its subsequent decay into a bosonic Dark Matter (DM) candidate -- either scalar or vector -- and SM up-type quarks at the Large Hadron Collider (LHC). This dynamics has been ignored so far in standard experimental searches of heavy quarks decaying to DM and we assess herein the regions of validity of current approaches, based on the assumption that the extra quarks have a narrow width. Further, we discuss the configurations of masses, widths and couplings where the latter breaks down.
\end{abstract}


\tableofcontents

%%%%%%%%%%%%%%%%%%%%%%%%%%%%%%%%%%%%%%%%%%%%%%%%%%%%%%%%%%%%%%%%%%%%%%%%%
%%%%%%%%%%%%%%%%%%%%%%%%%%%%%%%%%%%%%%%%%%%%%%%%%%%%%%%%%%%%%%%%%%%%%%%%%
\section{Introduction}

The existence of new extra-quarks (XQs) besides the Standard Model (SM) ones is among the open problems of particle physics to which the LHC may soon provide an answer. Searches for new quarks are actively undertaken by both ATLAS and CMS experiments, though no signals have been found so far. Limits on extra-quark masses are then provided, depending on their interaction mechanisms and for each given search channel. One of the most intriguing hypotheses about their role in new physics beyond the SM (BSM) is that XQs may be the mediators between the SM sector and the dark sector, which contains the Dark Matter (DM) candidate(s). Indeed, various theoretical scenarios predict the existence of DM candidates which interact with SM states through new fermions with color charge, such as Universal Extra Dimensions (UED)~\cite{Antoniadis:1990ew,Appelquist:2000nn,Servant:2002aq,Csaki:2003sh,Cacciapaglia:2009pa} or Little Higgs models with T-parity~\cite{Cheng:2003ju,Cheng:2004yc,Low:2004xc,Hubisz:2004ft,Hubisz:2005tx,Cheng:2005as}. The general feature of models in which the XQs are mediators between the SM and DM sectors is the presence of a $\mathcal Z_2$ parity, under which the SM particles are even and the XQ and DM particles are odd. 

Scenarios where the XQs decay into DM feature the presence of missing transverse energy ($\MET$) in the final state and, in this respect, their signatures are identical to the corresponding ones from, e.g., Supersymmetry (SUSY). Indeed, a $\mathcal Z_2$ parity is also the origin of the stability of the DM candidate in minimal SUSY scenarios, wherein the scalar partners of SM quarks typically decay into a fermionic DM candidate (the neutralino) and SM quarks. The difference between SUSY and XQ scenarios is only related to the spin of the particles in the interaction: in the XQ case a fermionic mediator decays into bosonic DM, in the SUSY case a scalar mediator decays into fermionic DM. For this reason, it is then possible to interpret the results of SUSY searches in terms of limits in scenarios with XQs and DM, so that studies exist in literature about such reinterpretations with different assumptions on the coupling properties of the XQs~\cite{Cacciapaglia:2013wha,Edelhauser:2015ksa,Arina:2015uea,Kraml:2016eti,Chala:2017xgc}. 

Usually these studies focus on scenarios where the widths of the XQs are small (with respect to their masses), such that it is possible to factorize the production and decay parts of the scattering amplitudes and write the cross section of any process as $\sigma = \sigma(Q) \times {\rm BR}(Q)$, in terms of  the production cross section $\sigma$ and the decay 
Branching Ratio (BR) of a generic XQ, thus neglecting terms of $\mathcal O(\Gamma_Q/M_Q)^n$ (the power $n$ depending on the observable), with $M_Q$ and $\Gamma_Q$ being, respectively, the mass and total width of the XQ. This approximation is particularly useful in processes where the XQs are produced in pairs via Quantum Chromo-Dynamics (QCD) interactions, such that the production cross section depends only on the XQ mass and the assumptions about the XQ interactions with the SM quarks are encoded in their BRs. However, the width of the XQs may not always be small enough for the above approximation to hold: if the XQ couplings are numerically large or if the XQ has a large number of decay channels, the total width may increase to sizable values, so that it is not possible to factorize production from decay. In this case, only the analysis of the full process, from the initial state to the XQ decay products, can provide a good description of the kinematics of the final states and thus of the determination of the limits on the XQ and DM masses from experimental searches.

In this paper we will focus on a simplified scenario where a top-like XQ ($T$) interacts with SM quarks and DM candidates and its width is large relatively to its mass (up to 40\% of it). We will consider final states compatible with processes of pair production of the $T$ and subsequent decay into a bosonic DM candidate and a SM quark. Then, we will evaluate the effects of large width in the determination of the cross section and in the reinterpretation of bounds from experimental searches. We will distinguish scenarios with a scalar DM from scenarios with a vector DM and we will analyse in detail scenarios where the $T$ state interacts either with the SM up or top quark, such that the final states we will consider are either $2j+\MET$ or $t\bar t + \MET$, respectively. For scenarios where $T$ interacts with the charm quark, leading to a final state analogous to the case of the up-quark in terms of reconstructed objects if charm-tagging is not considered, only the main results will be provided. It is important to notice that, unlike in the case of scenarios where the XQs decay only into SM states~\cite{Moretti:2016gkr}, interference terms with the SM background are absent if the XQs decay to DM candidates, as the only (irreducible) source of \MET~in the SM is given by final states containing neutrinos.

The structure of the paper is the following: in Section~\ref{sec:Conventions} we describe the simplified scenarios we consider, providing the relevant Lagrangian terms and relations between couplings, define our naming conventions and identify the final states of interest; in Section~\ref{sec:Monte Carlo} we describe our setup for the numerical computation and we list the experimental searches we consider in our analysis; in Sections~\ref{sec:Tt} and \ref{sec:Tu} we analyse in detail scenarios where the $T$ quarks interacts with third generation and first generation SM quarks, respectively; in Section~\ref{sec:BellPlots} we present the 95\% CL exclusion limits in the $M_T$-$M_{DM}$ plane, including the case of interaction with the charm quark; finally, in Section~\ref{sec:conclusions} we state our conclusions.

%%%%%%%%%%%%%%%%%%%%%%%%%%%%%%%%%%%%%%%%%%%%%%%%%%%%%%%%%%%%%%%%%%%%%%%%%
%%%%%%%%%%%%%%%%%%%%%%%%%%%%%%%%%%%%%%%%%%%%%%%%%%%%%%%%%%%%%%%%%%%%%%%%%
\section{Model and conventions}
\label{sec:Conventions}

%-----------------------------------------------------------------------------------------------------------------------------------------
\subsection{Lagrangian terms}
%-----------------------------------------------------------------------------------------------------------------------------------------

We concentrate here, for illustrative purposes, on a minimal extension of the SM with just one XQ state and one DM state\footnote{More realistic constructs including families of XQs have been dealt with from a phenomenological point of view in previous publications of some of us\cite{Barducci:2013zaa,Barducci:2014ila,Barducci:2014gna}.}. 
Our notation follows the same conventions of Ref.\cite{Kraml:2016eti}, to which we refer for the most general lagrangian terms involving different representations of XQs. In the present analysis we limit our investigation to a top-partner XQ and a DM real scalar or real vector singlet: therefore we label the XQ as $T$ if it is a singlet or $\Psi_{1/6}=(T \ B)^T$ if it belongs to a doublet (where 1/6 indicates the weak isospin of the doublet), and the DM states as $S^0_{\rm DM}$ if scalar or $V^{0\mu}_{\rm DM}$ if vector.
The couplings between the XQ, the DM and SM quarks are introduced using the notation $\lambda_{ij}^q$ if the DM is scalar and $g_{ij}^q$ if the DM is vector, with the labels $i,j \in \{1,2\}$ indicating the representations of the XQ and DM, respectively (1 for singlet, 2 for doublet), while $q \in \{u,d,c,s,t,b\}$ identifies which SM quark the new states couples to. The relevant Lagrangian terms are therefore:
\begin{eqnarray}
\Lag^S_1 &=& 
\left[
\lambda_{11}^{u_f} \bar{T} P_R \; u_f + 
\lambda_{11}^{d_f} \bar{B} P_R \; d_f +
\lambda_{21}^f \; \overline\Psi_{1/6} P_L {u_f \choose d_f} 
\right] 
S^0_{\rm DM} + {\rm h.c.} 
\label{eq:LagSingletDMS}
\\
\Lag^V_1 &=& 
\left[
g_{11}^{u_f} \bar{T} \gamma_\mu P_R \; u_f + 
g_{11}^{d_f} \bar{B} \gamma_\mu P_R \; d_f + 
g_{21}^f  \; \overline\Psi_{1/6} \gamma_\mu P_L {u_f \choose d_f} 
\right] 
V^{0\mu}_{\rm DM} + {\rm h.c.},
\label{eq:LagSingletDMV}
\end{eqnarray}
where $f$ is a flavour index and where we included terms corresponding to a $B$ singlet for completeness.

The XQs can be either vector-like (VLQ) or chiral (ChQ), and the gauge-invariant mass terms depend on the scenario:
\begin{align}
\Lag_{\rm VLQ} = & - M_{T_{\rm VLQ}} \bar T T - M_{B_{\rm VLQ}} \bar B B, \label{eq:VLQmass} \\
\Lag_{\rm ChQ} = & - y_{\rm XQ}^B \bar \Psi_{1/6} H B - y_{\rm XQ}^T \bar \Psi_{1/6} H^c T + {\rm h.c.} \underset{\rm{vev}}{\longrightarrow}\; - M_{T_{\rm ChQ}} \bar T T - M_{B_{\rm ChQ}} \bar B B. \label{eq:ChQmass}
\end{align}
In the following, for simplicity, we will consider the mass of the XQ as a free parameter, though one should bear in mind that in the chiral case the contribution of the new ChQ to Higgs production and decay processes, even if different from scenarios where a fourth chiral generation mixes with the SM quarks, can be used to impose constraints on the coupling between the XQ and  Higgs boson and, consequently, on the maximum mass the ChQ can acquire through the Higgs mechanism. Of course, ChQs can still acquire mass by some different new physics mechanism (e.g., by interacting with a heavier scalar which develops a VEV).

An important consequence of the assumption about the VLQ or ChQ nature of the XQ top-partner is the structure of its couplings. If the XQ is vector-like, either the left-handed or right-handed components of the interactions are allowed, depending if the XQ is a doublet or singlet respectively. If the XQ is chiral, both projections must be present (even if one of the two can be dominant). The difference between a dominant left-handed or right-handed coupling, in the NWA and for couplings with third generation SM quarks, has been explored in Ref.\cite{Kraml:2016eti}, where it was found that the bounds in the XQ-DM mass plane can be slightly different in the two cases, though they have the same qualitative behaviour. In the following analysis we will expore in detail scenarios where the $T$ has a purely left-handed coupling ({\it i.e.} it belongs to a VLQ doublet), but we will show (for specific benchmarks) how the experimental limits change in the large width regime when considering alternative hypotheses, such as pure right-handed couplings (VLQ singlet) or couplings where the left- and right-handed components are equal in size with same or opposite sign (ChQ scenarios).

A further important aspect we ought to consider is the physical origin of the large width. This can be achieved in two ways: both by increasing the couplings of the XQs to the states they can decay to, and by increasing the number of decay channels. Of course, if the coupling is increased to achieve a large enough width, it can reach values for which the perturbative regime is not valid anymore, and therefore it would not be possible to perform our analysis with the same techniques. Increasing the number of decay channels has the advantage of limiting the size of couplings, and the disadvantage of introducing further new particles, such as new $Z_2$-odd neutral or charged bosons or further XQs which then chain-decay to the DM candidate and generate further and complementary final states of phenomenological relevance. To avoid running into the problem of describing {\it how} the XQ gets a large width, and in order to be as model-independent as possible, we will be agnostic about the mechanism at the origin of the XQ width, and we will consider the $\Gamma_{XQ}$ as an independent free parameter.

%-----------------------------------------------------------------------------------------------------------------------------------------
\subsection{Observables and conventions}
%-----------------------------------------------------------------------------------------------------------------------------------------

To understand the effects of large widths on the signal, we will consider two different processes, both leading to the same four-particle final state ${\rm DM} \; q \ {\rm DM} \; \bar{q} \equiv q\bar q+\MET$, where $q(\bar q)$ is an ordinary SM (anti)quark.

\begin{itemize}

\item\textit{QCD pair production and decay of on-shell XQs}

This process is the one usually considered in experimental searches for  XQs. In the so-called Narrow Width Approximation (NWA), wherein the Breit-Wigner propagators of the two $T$ states are suitably replaced by Dirac $\delta$ distribution functions,
 it is possible to separate  production and decay of the heavy quarks, thus allowing for a model-independent analysis of the results. The cross section for this process is given by (hereafter, in our formulae, $Q$ denotes an XQ):
\begin{equation}
 \sigma_X \equiv \sigma_{2 \to 2}~{\rm BR}(Q)~{\rm BR}(\bar Q)
\end{equation}
where, for simplicity, $\sigma_{2 \to 2}$ only takes into account the dominant (pure) QCD topologies. This factorization of production and decay only makes sense in  NWA so this process is {\sl dynamically}
 independent of the width, i.e., $\sigma_X\equiv \sigma_X (M_Q)$, though $\Gamma_Q$ obviously enter in the definition of the BRs of $Q$ and $\bar Q$. 

\item\textit{Full signal}

In this process all the topologies which lead to the same four-particle final state and contain \textit{at least one} XQ propagator are taken into account. The only assumption we make, to allow a consistent comparison with the NWA results, is that the order of the QCD $\alpha_s$ in the full signal topologies is the same as in the NWA case. 
The pair production and decay topologies are included, but for the full signal the XQs are not strictly required to be on-shell. Furthermore, diagrams with only one XQ propagator are also included. We stress that the NWA limit is indeed recovered when the XQ width becomes small with respect to its mass: in this limit, factorisation of production and decay can still be done, as the contribution of all the subleading topologies considered in the full signal becomes negligible and the dominant contribution is given only by pair-production topologies where the XQ is on-shell. If the XQ width is large with respect to its mass, the contribution of other topologies becomes relevant and the factorisation is not possible anymore. 
Hence, this approach, on the one hand, describes  accurately scenarios where the widths of the XQs are large and, on the other hand, is fully gauge invariant (like the NWA approach). Furthermore, it takes into account the spin correlations between the $Q$ quark and antiquark decay branches, which are lost in the NWA. 
The cross section of this process will be labelled as $\sigma_S$ and depends  upon both the mass and width of the XQ: $\sigma_S\equiv\sigma_S (M_Q, \Gamma_Q)$. Some example topologies for this process, which are not included in the previous one, are given in Fig.~\ref{fig:fullsignaltopologies}.

\begin{figure}[ht!]
% \begin{center}
% \begin{picture}(140,10)(0,0)
% \SetWidth{1}
% \Gluon(10,0)(50,0){3}{6}
% \Text(8,0)[rc]{\large $g$}
% \Gluon(10,90)(50,90){3}{6}
% \Text(8,90)[rc]{\large $g$}
% \Line[arrow](90,0)(50,0)
% \Text(92,0)[lc]{\large $\bar u, \bar t$}
% \Line[arrow](50,0)(50,30)
% \Text(46,15)[rc]{\large $u,t$}
% \SetColor{Red}\SetWidth{1.5}
% \Line[arrow](50,30)(50,60)
% \Text(46,45)[rc]{\large\Red{$T$}}
% \SetColor{Black}\SetWidth{1}
% \Line[arrow](50,60)(50,90)
% \Text(46,75)[rc]{\large $u,t$}
% \Line[arrow](50,90)(90,90)
% \Text(92,90)[lc]{\large $u,t$}
% \Line[dash](50,30)(90,30)
% \Photon(50,30)(90,30){3}{5}
% \Text(92,30)[lc]{\large $S^0_{DM},V^0_{DM}$}
% \Line[dash](50,60)(90,60)
% \Photon(50,60)(90,60){3}{5}
% \Text(92,60)[lc]{\large $S^0_{DM},V^0_{DM}$}
% \end{picture}
% \hskip 20pt
% \begin{picture}(140,100)(0,0)
% \SetWidth{1}
% \Line[arrow](10,0)(40,30)
% \Text(8,0)[rc]{\large $q$}
% \Line[arrow](40,30)(10,60)
% \Text(8,60)[rc]{\large $\bar q$}
% \Gluon(40,30)(80,30){3}{5}
% \Text(60,38)[cb]{\large $g$}
% \Line[arrow](80,30)(100,50)
% \Text(83,47)[cc]{\large $u,t$}
% \Line[arrow](110,0)(80,30)
% \Text(112,0)[lc]{\large $\bar u, \bar t$}
% \SetColor{Red}\SetWidth{1.5}
% \Line[arrow](100,50)(120,70)
% \Text(103,67)[cc]{\large\Red{$T$}}
% \SetColor{Black}\SetWidth{1}
% \Line[dash](100,50)(120,30)
% \Photon(100,50)(120,30){3}{5}
% \Text(122,30)[lc]{\large $S^0_{DM},V^0_{DM}$}
% \Line[arrow](120,70)(140,90)
% \Text(142,90)[lc]{\large $u,t$}
% \Line[dash](120,70)(140,50)
% \Photon(120,70)(140,50){3}{5}
% \Text(142,50)[lc]{\large $S^0_{DM},V^0_{DM}$}
% \end{picture}
% \end{center}
\centering
\includegraphics[width=.9\textwidth]{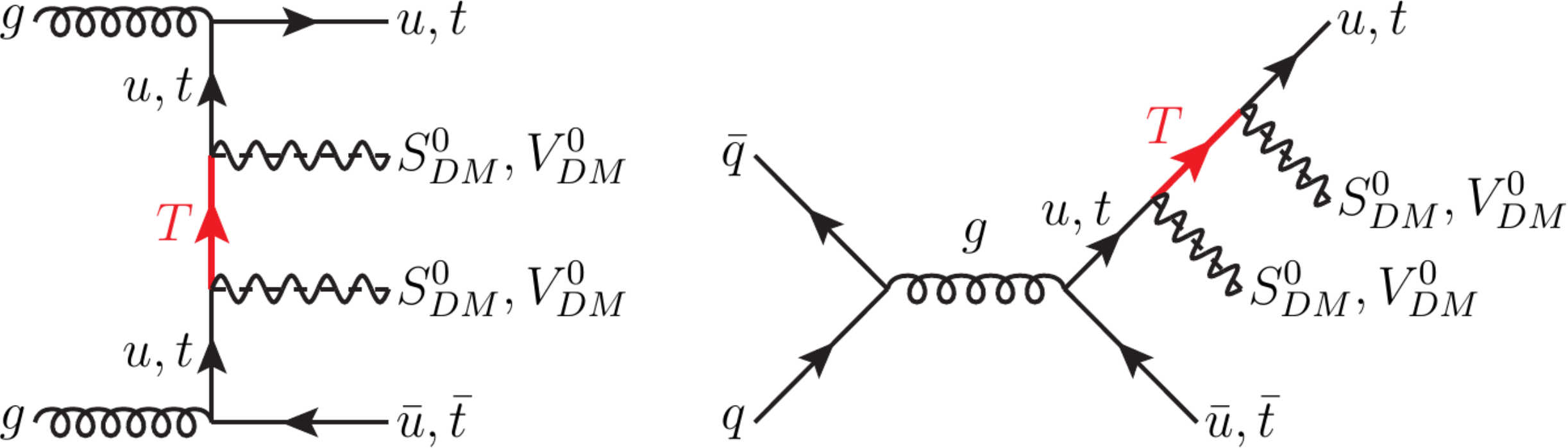}
\caption{Examples of topologies containing only one XQ propagator for final states compatible with XQ pair production and decay into scalar or vector DM and SM quarks of first or third generation.}
\label{fig:fullsignaltopologies}
\end{figure}

\end{itemize}

In order to determine the difference between the two approaches above, we will consider the variable $(\sigma_S - \sigma_X)/\sigma_X$. This ratio takes into account effects of both the off-shellness of $T$ and $\bar T$ in their pair production as well as contributions given by topologies which contain at least one XQ propagator (including interference between the two). It measures in practice how much the full signal differs from the approximate pair-production-plus-decay one computed in the NWA.

%%%%%%%%%%%%%%%%%%%%%%%%%%%%%%%%%%%%%%%%%%%%%%%%%%%%%%%%%%%%%%%%%%%%%%%%%
\subsection{Channels \label{sec:Channels}}

In the present analysis we consider the processes of production of a heavy top-like quark $T$. In principle, from a model-independent point of view, the $T$ quark is allowed to interact with all SM quark generations, but to evaluate the effects of large widths in different scenarios, only specific interactions will be switched on in the different scenarios we will consider. 

Since the purpose of this analysis is to evaluate the effects of large widths on channels commonly explored by experimental analysis, we will consider only final states allowed by $T$ pair production and decay. The full set of channels in which a pair-produced $T$ quark can decay is given by the following matrix:
\begin{eqnarray}
T\bar T\to\left(
\begin{array}{ccc|ccc}
S^0_{DM}u\; S^0_{DM}\bar u & \ \; S^0_{DM}u\; S^0_{DM}\bar c & \ \; S^0_{DM}u\; S^0_{DM}\bar t & \ \; S^0_{DM}u\; V^0_{DM}\bar u & \ \; S^0_{DM}u\; V^0_{DM}\bar c & \ \; S^0_{DM}u\; V^0_{DM}\bar t \\
S^0_{DM}c\; S^0_{DM}\bar u & \ \; S^0_{DM}c\; S^0_{DM}\bar c & \ \; S^0_{DM}c\; S^0_{DM}\bar t & \ \; S^0_{DM}c\; V^0_{DM}\bar u & \ \; S^0_{DM}c\; V^0_{DM}\bar c & \ \; S^0_{DM}c\; V^0_{DM}\bar t \\ 
S^0_{DM}t\; S^0_{DM}\bar u & \ \; S^0_{DM}t\; S^0_{DM}\bar c & \ \; S^0_{DM}t\; S^0_{DM}\bar t & \ \; S^0_{DM}t\; V^0_{DM}\bar u & \ \; S^0_{DM}t\; V^0_{DM}\bar c & \ \; S^0_{DM}t\; V^0_{DM}\bar t \\ 
\hline
V^0_{DM}u\; S^0_{DM}\bar u & \ \; V^0_{DM}u\; S^0_{DM}\bar c & \ \; V^0_{DM}u\; S^0_{DM}\bar t & \ \; V^0_{DM}u\; V^0_{DM}\bar u & \ \; V^0_{DM}u\; V^0_{DM}\bar c & \ \; V^0_{DM}u\; V^0_{DM}\bar t \\
V^0_{DM}c\; S^0_{DM}\bar u & \ \; V^0_{DM}c\; S^0_{DM}\bar c & \ \; V^0_{DM}c\; S^0_{DM}\bar t & \ \; V^0_{DM}c\; V^0_{DM}\bar u & \ \; V^0_{DM}c\; V^0_{DM}\bar c & \ \; V^0_{DM}c\; V^0_{DM}\bar t \\ 
V^0_{DM}t\; S^0_{DM}\bar u & \ \; V^0_{DM}t\; S^0_{DM}\bar c & \ \; V^0_{DM}t\; S^0_{DM}\bar t & \ \; V^0_{DM}t\; V^0_{DM}\bar u & \ \; V^0_{DM}t\; V^0_{DM}\bar c & \ \; V^0_{DM}t\; V^0_{DM}\bar t  
\end{array}
\right)
\label{eq:finalstates}
\end{eqnarray}

To limit ourselves to representative and simple scenarios, we will focus on the diagonal terms of this matrix and analyse in detail XQs coupling either to first or third generation quarks (though the main results for couplings with second generation will also be provided). 
Effects of large width are different depending on the kinematics of the process and by selecting representative scenarios it is always possible to reconstruct intermediate configurations (XQs interacting partly with heavy and partly with light SM generations).

\begin{figure}[ht!]
\centering
\includegraphics[width=.5\textwidth]{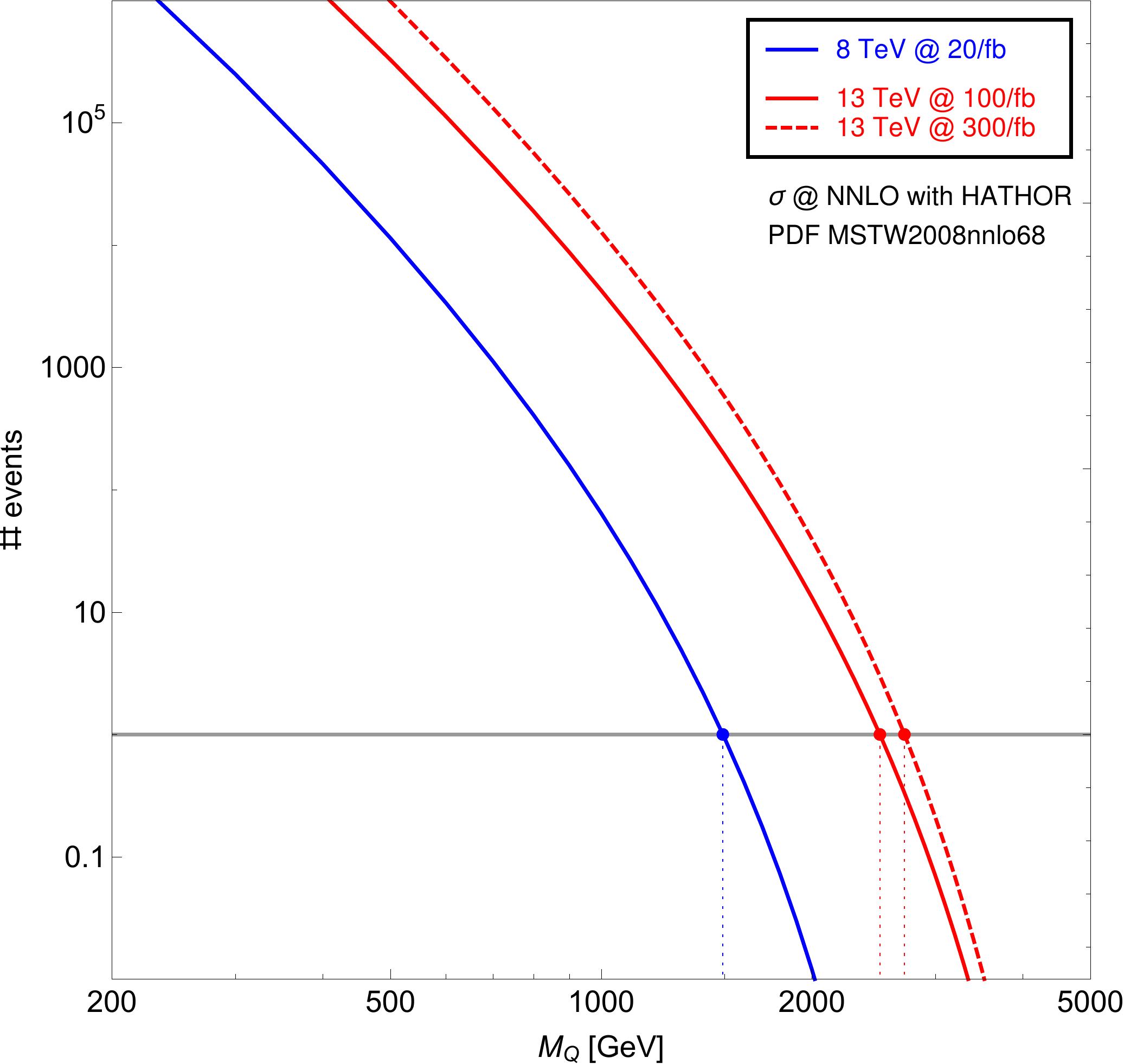} 
\caption{\label{fig:Xsigma}Number of events at partonic level for $Q \bar Q$ pair production and for different LHC energies and luminosities. The corresponding cross sections have been computed using HATHOR\cite{Aliev:2010zk} with MSTW2008nnlo68 Parton Distribution Functions (PDFs) \cite{Martin:2009bu}.}
\end{figure}

This analysis is of phenomenological interest only for mass values for which the number of final events is (ideally) larger than 1. In Fig.~\ref{fig:Xsigma} we show the number of events for different LHC luminosities for XQ pair production. The number of events in Fig.~\ref{fig:Xsigma} has been computed considering a Next-to-Next-to-Leading Order (NNLO) cross section, as accurate guidance for observability. 
For our analysis, however, we will consider only LO cross sections, and defer the evaluation of higher orders effects to future studies.
From Fig.~\ref{fig:Xsigma}, it is possible to see that the ideal practical validity of our results is limited to mass values of around 1500 GeV for LHC@8TeV, 2500 GeV (2700 GeV) for LHC@13TeV with 100/fb (300/fb) integrated luminosity. Of course, we are not considering here effects due to experimental acceptances and efficiencies: this study is only meant to assess the role of the complete signal with respect to the common approximations made in theoretical and experimental analyses. 

%%%%%%%%%%%%%%%%%%%%%%%%%%%%%%%%%%%%%%%%%%%%%%%%%%%%%%%%%%%%%%%%%%%%%%%%%
%%%%%%%%%%%%%%%%%%%%%%%%%%%%%%%%%%%%%%%%%%%%%%%%%%%%%%%%%%%%%%%%%%%%%%%%%
\section{Analysis tools and experimental searches\label{sec:Monte Carlo}}

As intimated, herein, we want to study the ratio of cross sections $(\sigma_S - \sigma_X)/\sigma_X$ (where we recall that $\sigma_S$ corresponds to the full signal and $\sigma_X$ to the NWA) as well as understand which influence the width of the XQ, in turn triggering the contribution of the forementioned new topologies not present in pair production, can have on its mass bounds. To do so we consider an XQ top-partner belonging to the doublet representation $\Psi_{1/6}=(T \ B)^T$ (corresponding to pure left-handed couplings in Eqs.\ref{eq:LagSingletDMS} and \ref{eq:LagSingletDMV}) and scan over the parameters $M_T$, $M_{\rm DM}$ and $\Gamma_T$.

For our simulation we analyse in detail scenarios where the DM state has masses $M_{\rm DM}$ = 10 GeV, 500 GeV and 1000 GeV and with an XQ of mass $M_T > M_{\rm DM} + m_q$, with $q \in \{u,c,t\}$ (such that its on-shell decay is kinematically allowed) up to $M_T^{\rm max}$ = 2500 GeV, which is the maximal value of a $T$ mass so that it can be produced for LHC@13TeV with 100/fb integrated luminosity as shown in Fig. \ref{fig:Xsigma}. We also consider values of the $T$ width from $\Gamma_T / M_T$ $\simeq$ 0\% (NWA) to 40\% of the $T$ mass.

Our numerical results at partonic level are obtained using {\sc MadGraph5} \cite{Alwall:2011uj,Alwall:2014hca} and a model we implemented in {\sc Feynrules} \cite{Alloul:2013bka} to obtain the UFO interface format. The model we used is the same as the one in the analysis of Ref.\cite{Kraml:2016eti}. For the Monte Carlo simulation we use the PDF set {\sc cteq6l1}~\cite{Pumplin:2002vw}. Events are then passed to {\sc Pythia}\,8~\cite{Sjostrand:2007gs,Sjostrand:2006za}, which takes care of the hadronisation and parton showering.

To analyse and compare the effects of a set of 13 TeV analyses considering final states compatible with our scenarios, we employ {\sc CheckMATE 2}~\cite{Dercks:2016npn}, which uses the {\sc Delphes\,3}~\cite{deFavereau:2013fsa} framework for the emulation of detector effects. In our simulations we include all the ATLAS and CMS (carried out at 13 TeV) analyses available within the CheckMATE database but we will only list here the most relevant ones for our study. These analysis are the following ATLAS searches: 
\begin{itemize}
\item ATLAS 1604.07773 \cite{Aaboud:2016tnv}, a search for new phenomena in final states with an energetic jet and large missing transverse momentum,
\item ATLAS 1605.03814 \cite{Aaboud:2016zdn}, a search for squarks and gluinos in final states containing hadronic jets, missing transverse momentum but no electrons or muons, 
\item ATLAS 1605.04285 \cite{Aad:2016qqk}, a search for gluinos in final states with one isolated lepton, jets and missing transverse momentum. This search is however sensitive only to a specific region of the parameter space we explored, {\it i.e.} very light $T$ and DM masses and small $T$ width.
\item ATLAS-CONF-2016-050 \cite{ATLAS-CONF-2016-050}, a search for the stop in final states with one isolated electron or muon, jets and missing transverse momentum.
\end{itemize}

%%%%%%%%%%%%%%%%%%%%%%%%%%%%%%%%%%%%%%%%%%%%%%%%%%%%%%%%%%%%%%%%%%%%%%%%%
%%%%%%%%%%%%%%%%%%%%%%%%%%%%%%%%%%%%%%%%%%%%%%%%%%%%%%%%%%%%%%%%%%%%%%%%%
\section{Extra $T$ quark interacting with Dark Matter and the SM top quark}
\label{sec:Tt}

In this section we will study the case of XQs coupling to third generation SM quarks only. The possible decay channels are therefore $t \bar t + \{S^0_{DM} S^0_{DM}, V^0_{DM} V^0_{DM}\}$, {i.e.}  $t \bar t + \MET$. We start from this channel because, from a theoretical point of view, the top quark is considered the most likely to be affected by new physics phenomena.

%%%%%%%%%%%%%%%%%%%%%%%%%%%%%%%%%%%%%%%%%%%%%%%%%%%%%%%%%%%%%%%%%%%%%%%%%

\subsection{Large width effects at parton level}
\label{sec:Parton3}

\begin{figure}[ht!]
\centering
\includegraphics[width=.32\textwidth]{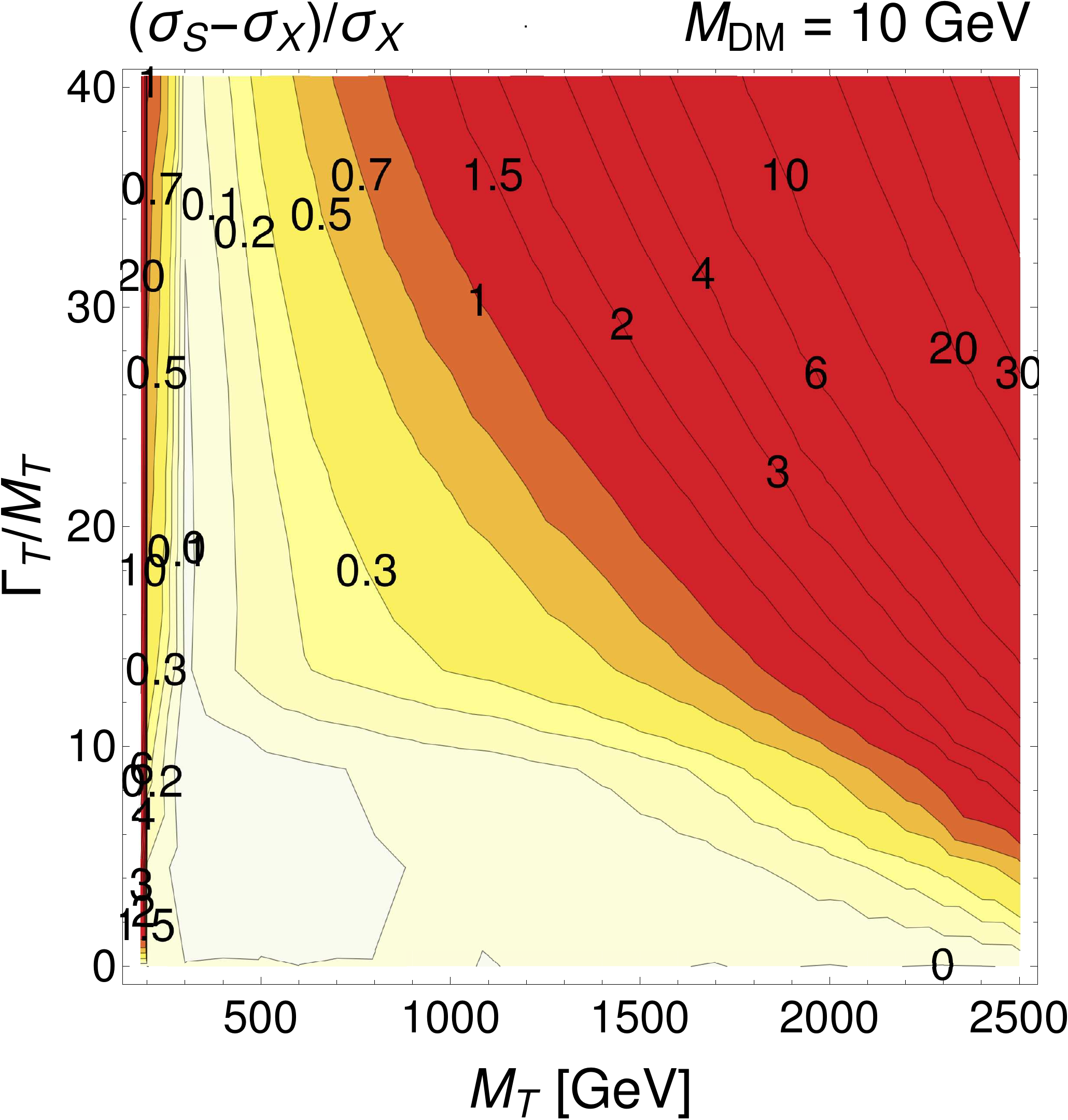} 
\includegraphics[width=.32\textwidth]{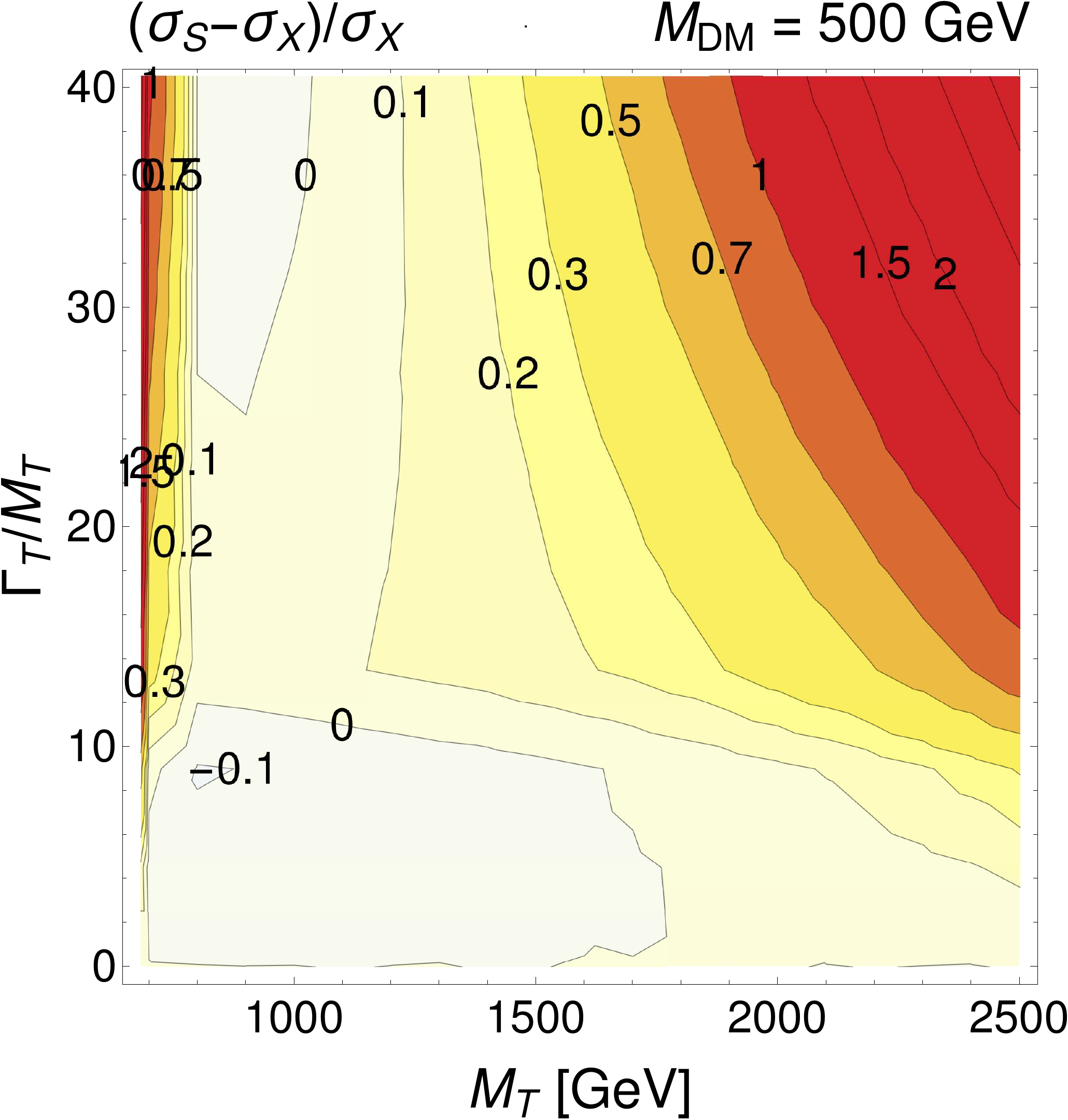} 
\includegraphics[width=.32\textwidth]{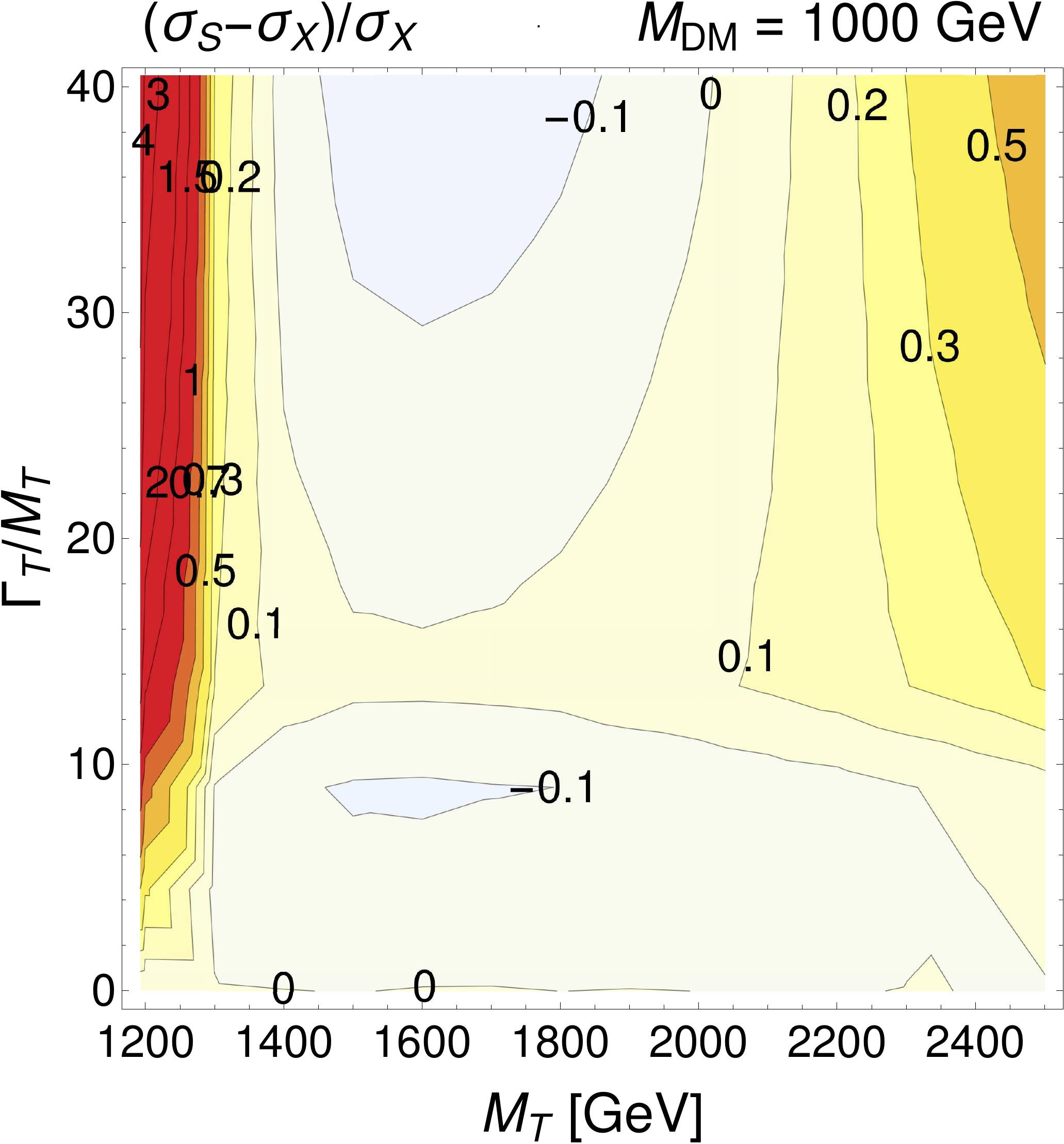}\\[5pt]  
\includegraphics[width=.32\textwidth]{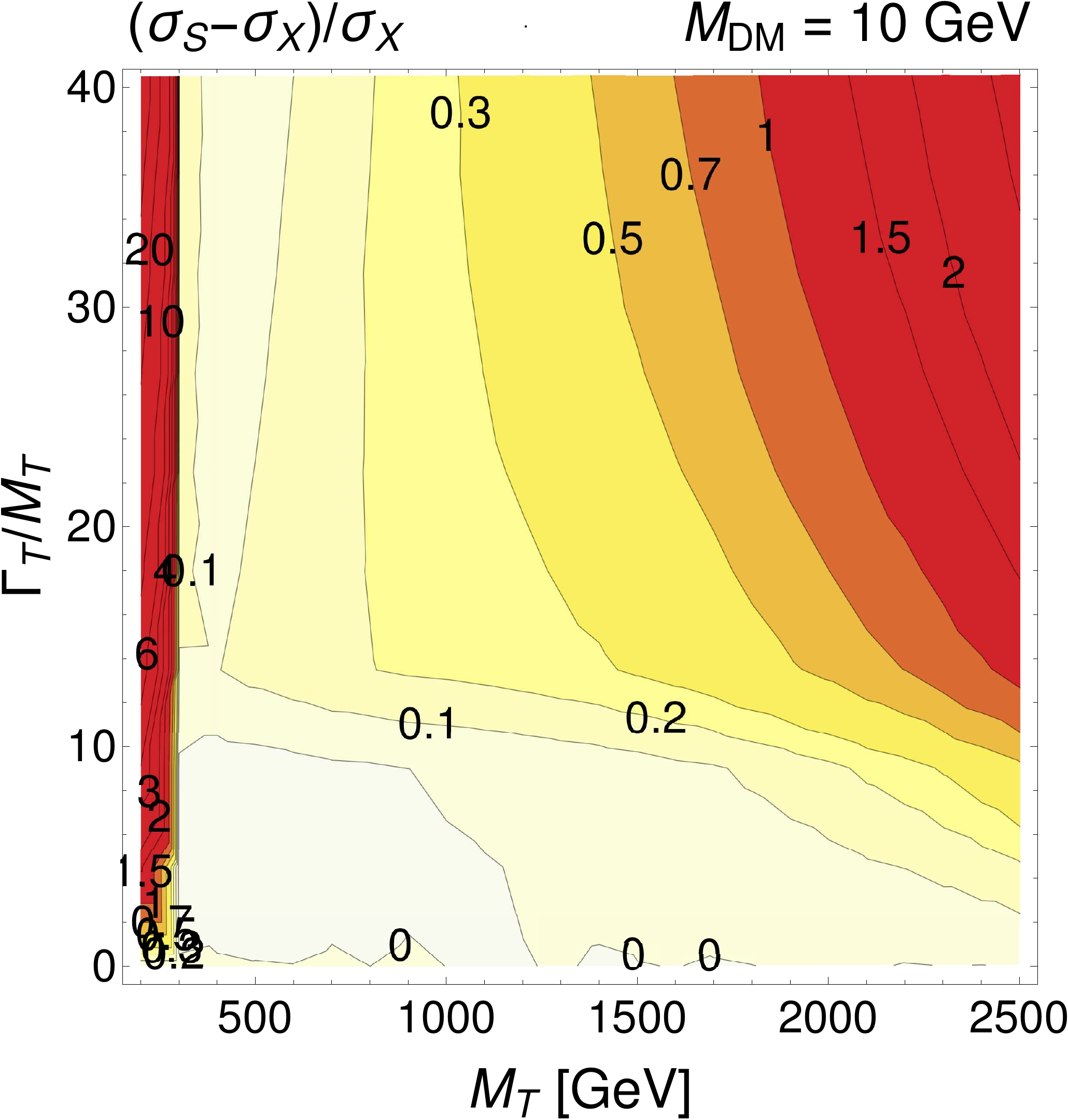} 
\includegraphics[width=.32\textwidth]{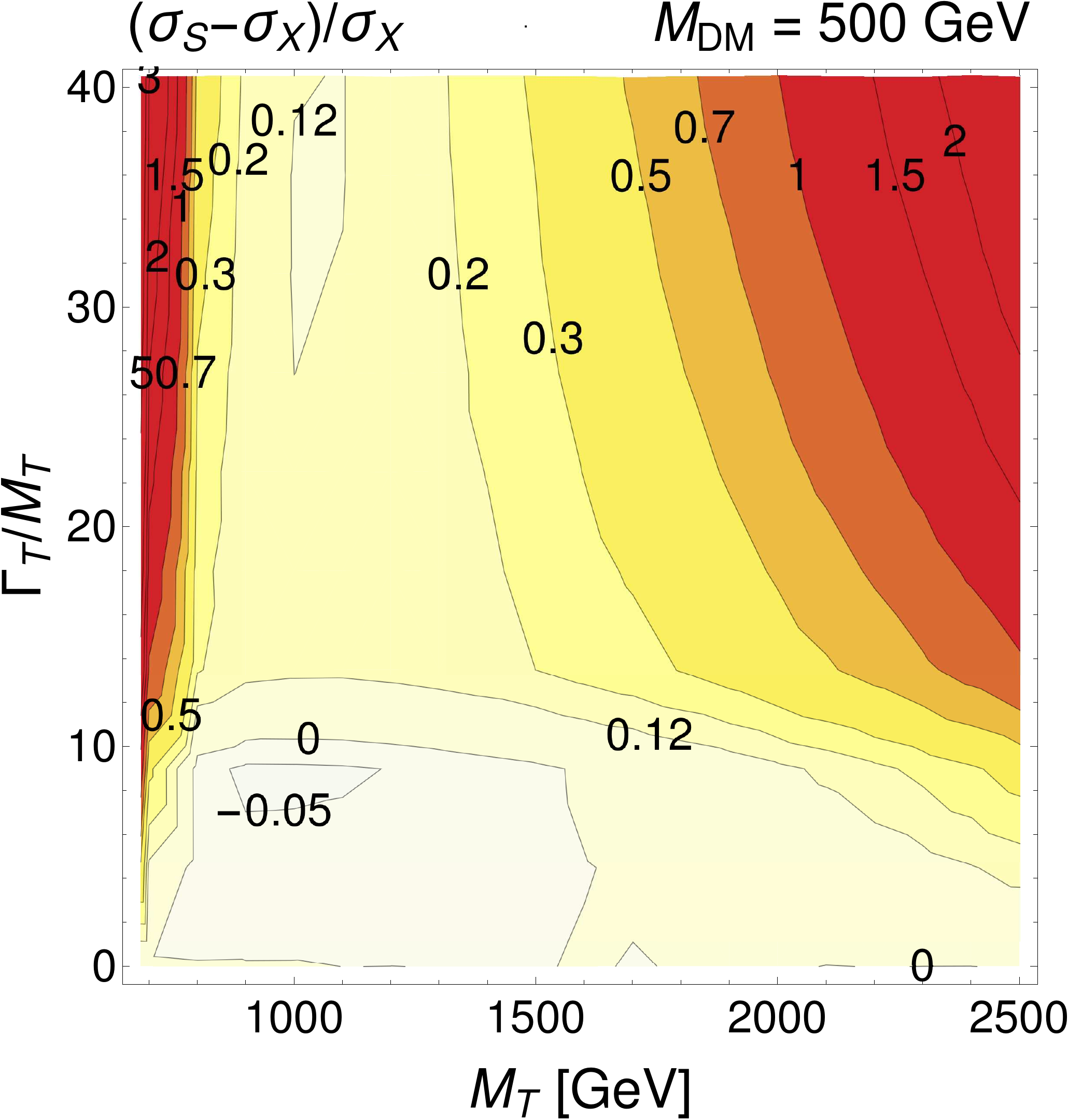} 
\includegraphics[width=.32\textwidth]{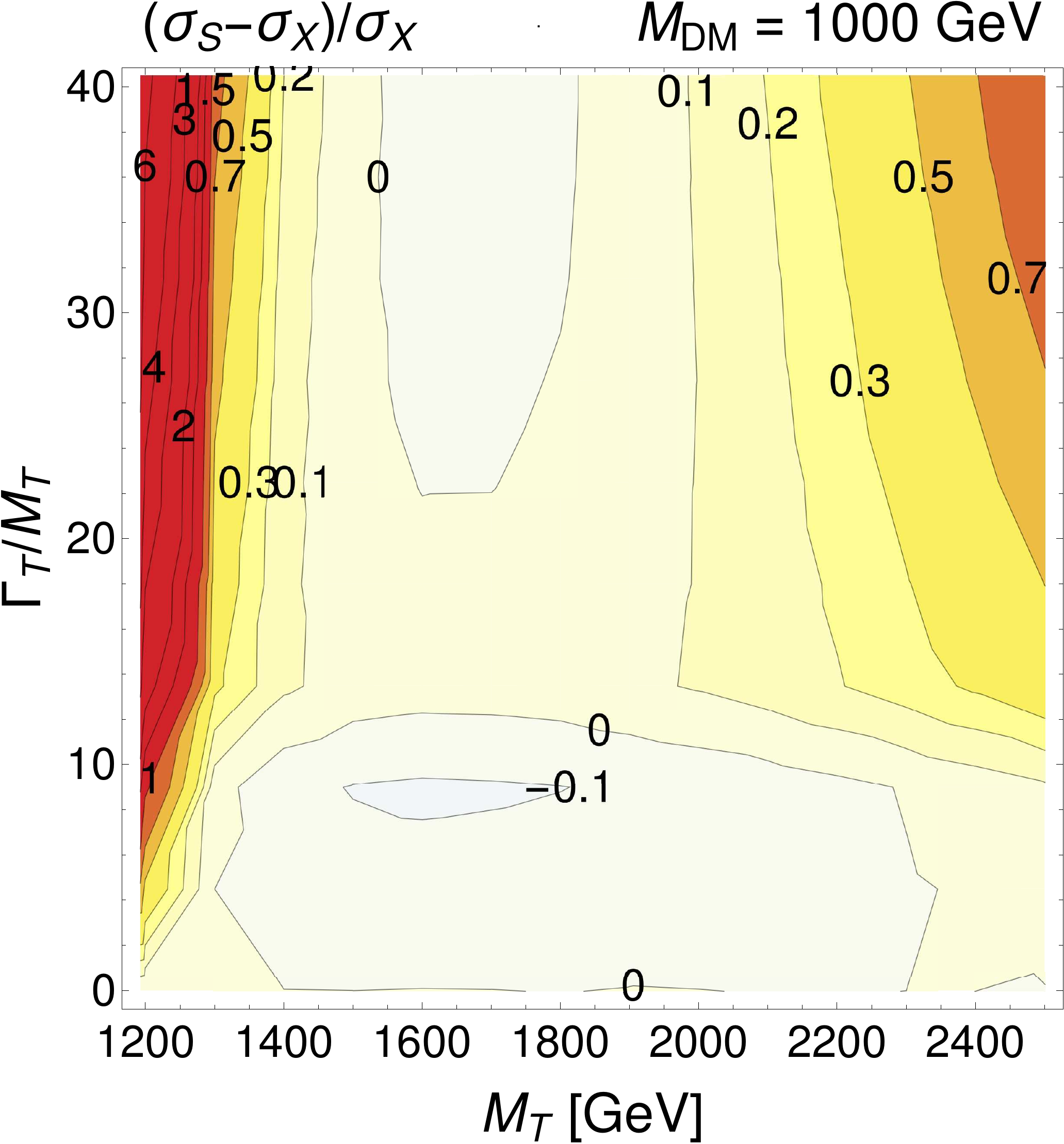} 
\caption{\label{fig:SXthird} Relative difference between the full signal and the QCD pair production cross sections for a $T$ coupling to a DM particle (coupling to third generation) of mass 10 GeV, 500 GeV and 1000 GeV. Top row: scalar DM; bottom row, vector DM.}
\end{figure}

In Fig.~\ref{fig:SXthird} the relative differences between the full signal and the QCD pair production cross sections $(\sigma_S - \sigma_X)/\sigma_X$ are plotted for an LHC energy of 13 TeV. Notice that here and in the following we do not apply cuts on \MET at parton level.

\begin{figure}[ht!]
\centering
\subfigure[Scalar DM: $M_{\rm DM}=1$ TeV, $M_T=2$ TeV]{
\includegraphics[width=.45\textwidth]{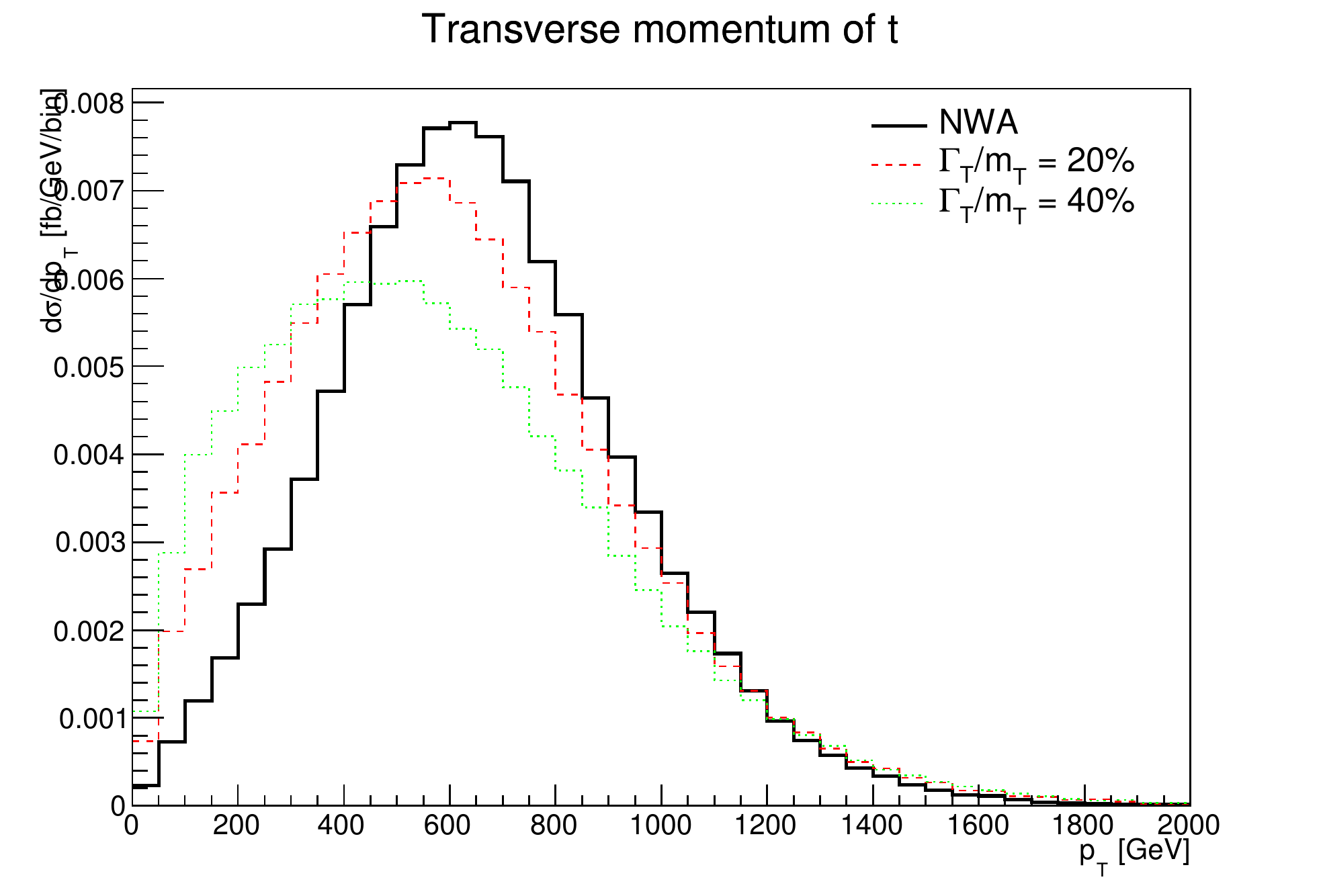} 
\includegraphics[width=.45\textwidth]{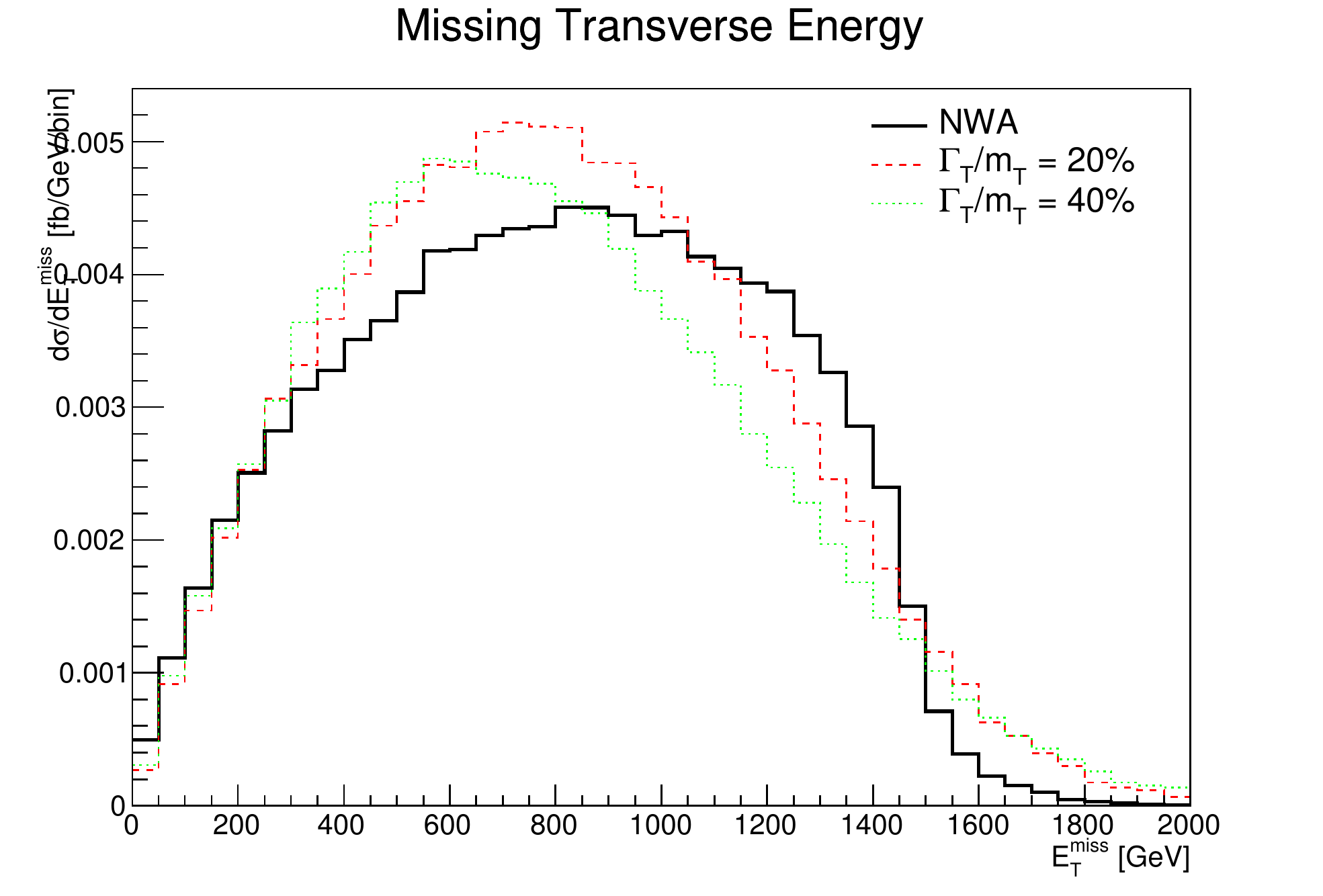}
}\\
\subfigure[Vector DM: $M_{\rm DM}=10$ GeV, $M_T=400$ GeV]{
\includegraphics[width=.45\textwidth]{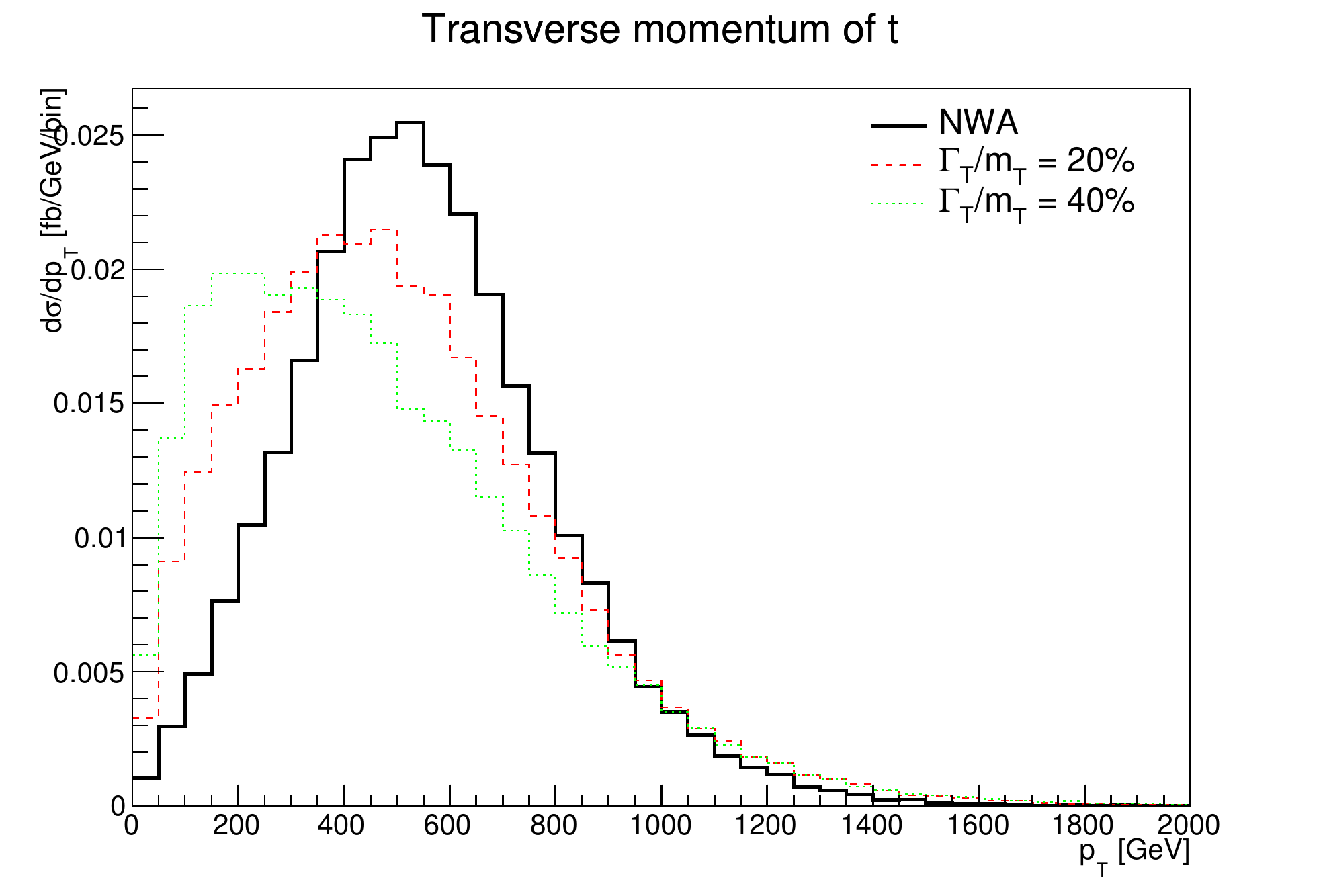} 
\includegraphics[width=.45\textwidth]{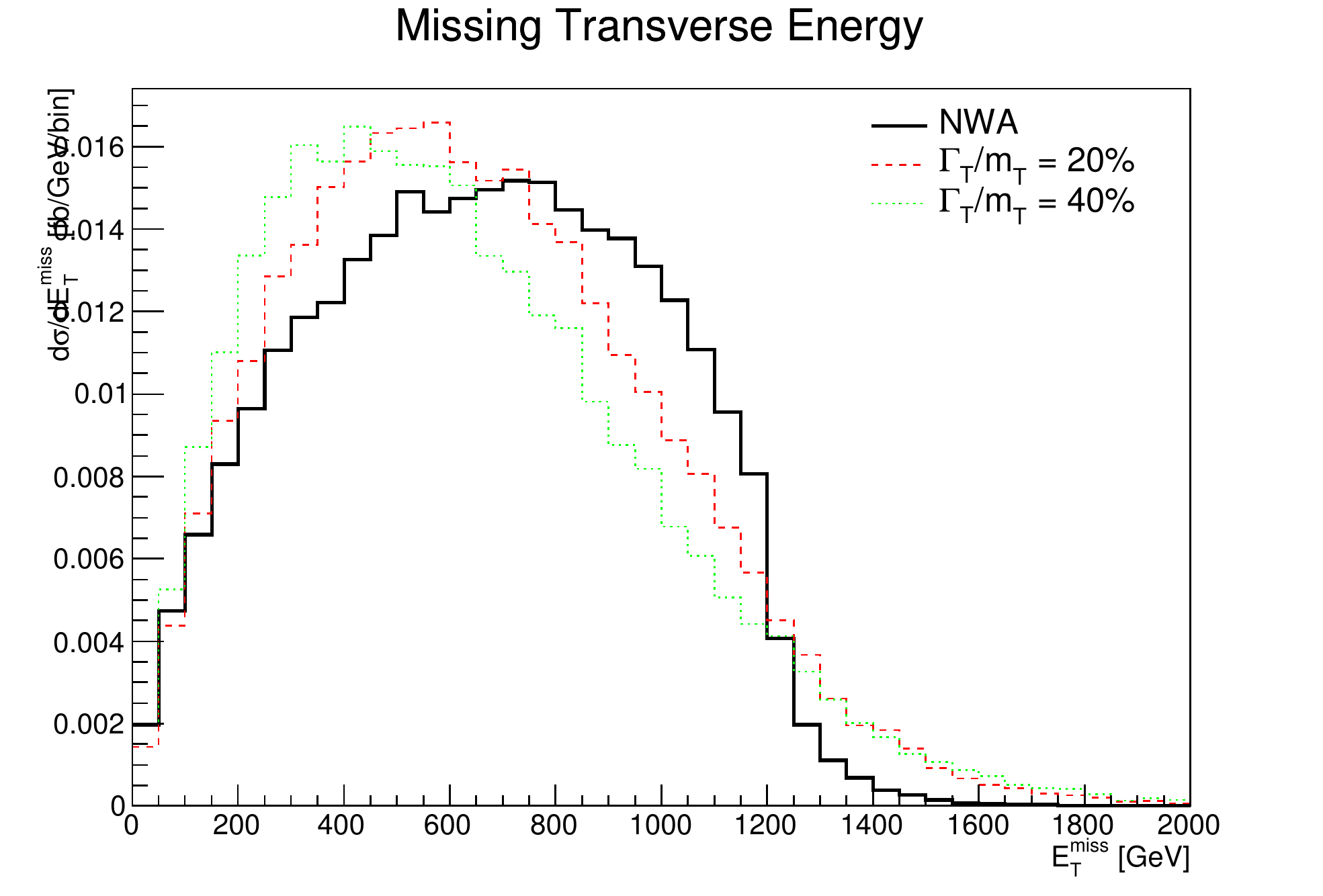}
}
\caption{\label{fig:DistributionCancellation}Differential distributions of transverse momentum of the top quark and \MET along the cancellation line for scalar and vector DM.}
\end{figure}

A number of conclusions can be derived from the observation of these results:
\begin{itemize}
\item As expected, and as a health check of our results, in the NWA limit ($\Gamma_T/M_T\to 0$) the QCD pair production channel is always an excellent approximation, as the off-shell and non-doubly-resonant contributions become negligible.
\item The effects of increasing the width becomes quickly relevant, independently of the DM spin, eventually becoming very large near the kinematics limit ($M_T=M_{DM}+m_t$) and for high $T$ masses, where the ratio can reach values above 100\% (represented by red regions in Fig.~\ref{fig:SXthird}). The increase near the kinematics limit can be explained by a non-trivial combination of factors, the most relevant being the fact that a larger width opens a larger phase space for the decay of the $T$, which is more limited (in the NWA) as the gap between the masses decreases. It is interesting to notice that the cross section for the full signal is large for values of $M_T$ beyond those ideally accessible in the NWA (see Fig.\ref{fig:Xsigma}). Therefore, even if the $T$ mass is too large to produce enough events in the NWA, if its width is sizable it might still be possible to detect it, unless the experimental acceptances drop with a comparable rate with respect to the NWA values. In this respect, the performance of the aforementioned experimental searches will be discussed in the following section.
\item For all channels, and in specific regions, a cancellation of effects takes place. Such cancellation makes the QCD pair production cross section similar to the cross section of the full signal even for large values of the width. The cancellation appears at different values of the $T$ mass depending on the mass of the DM and of its spin and becomes stronger  when the value of $M_{\rm DM}$ increases. Yet this cancellation does not mean that results in the NWA approximation are valid also for larger widths, as the cancellation is an accidental result due to the different scaling of the cross sections in NWA and large width regime. The differences between NWA and large width results are clearer at differential level. In Fig.~\ref{fig:DistributionCancellation} we show the differential distributions of the missing transverse energy and of the transverse momentum of the top quark along the cancellation line for a scalar DM particle of mass 1000 GeV and for a vector DM particle of mass 10 GeV. A similar effect was already observed in \cite{Moretti:2016gkr}, considering XQ decaying to SM particles instead of DM.
\end{itemize}

%%%%%%%%%%%%%%%%%%%%%%%%%%%%%%%%%%%%%%%%%%%%%%%%%%%%%%%%%%%%%%%%%%%%%%%%%

\subsection{Large width effects at detector level \label{sec:Detector3}}

\begin{figure}[ht!]
\centering
\includegraphics[width=.32\textwidth]{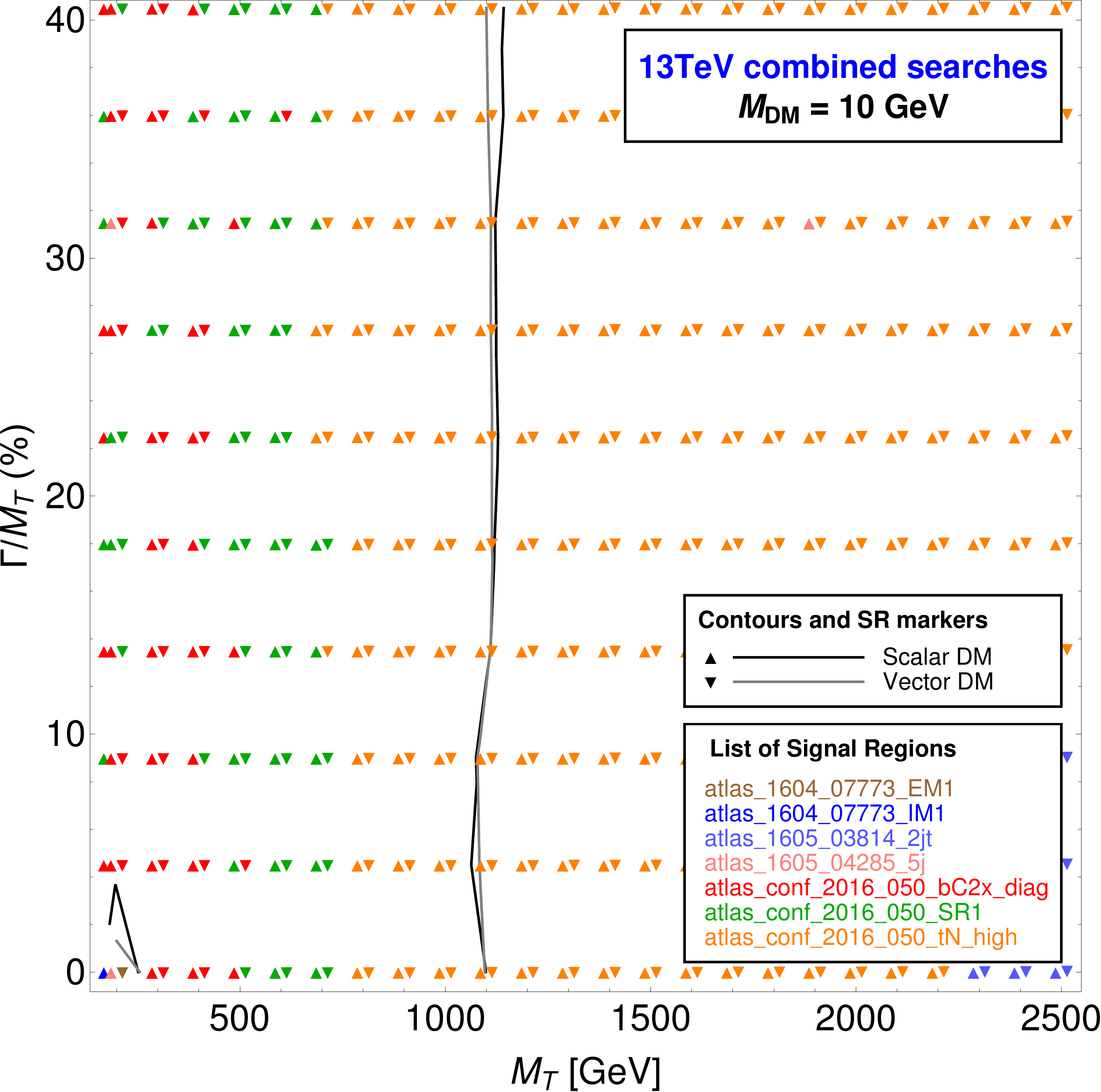} 
\includegraphics[width=.32\textwidth]{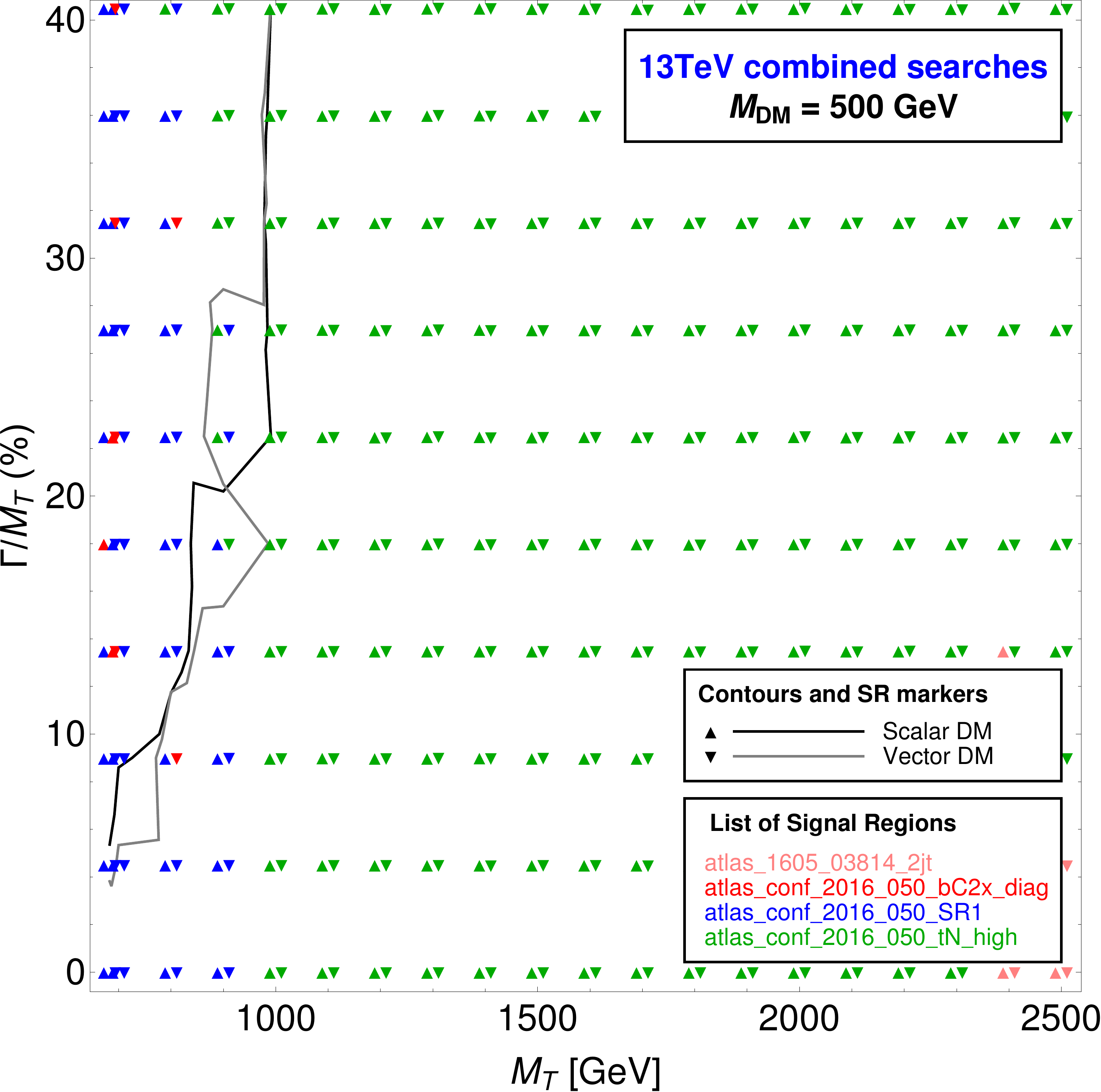} 
\includegraphics[width=.32\textwidth]{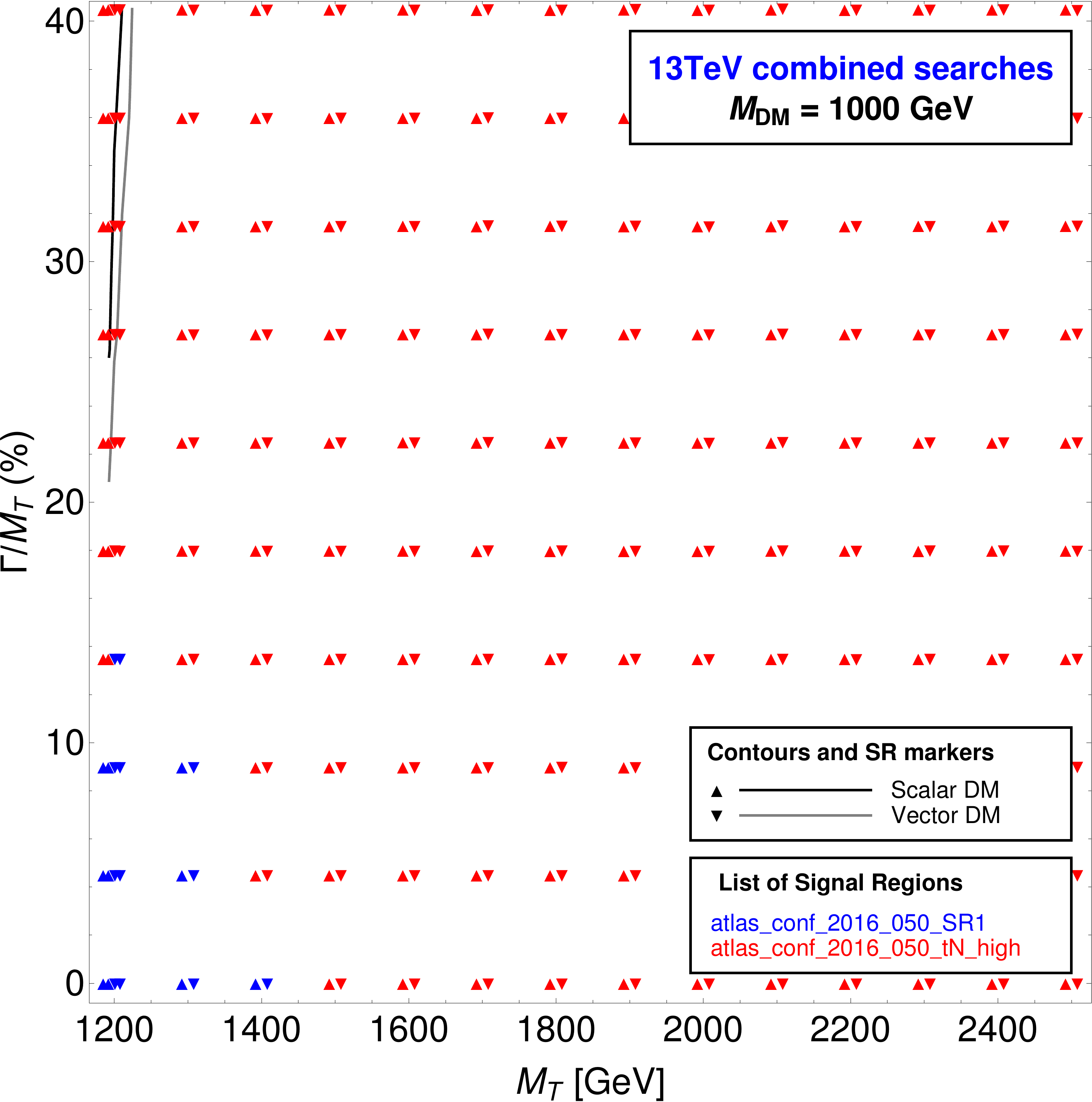} 
\caption{\label{fig:Exclusion3}{\sc CheckMATE} results for a $T$ coupling to a DM particle (coupling to third generation) of mass 10 GeV, 500 GeV and 1500 GeV. The black (grey) line show which part of the parameter space is excluded for the scalar (vector) DM scenario. 
}
\end{figure}

In this section we consider the effects of large widths on the exclusion limits for the $T$ mass. We show in Fig~\ref{fig:Exclusion3} the exclusion limit, corresponding to $r_{\rm max} (\equiv {S-1.96 \Delta S\over S_95^{\rm exp}})= 1$ as defined in \cite{Drees:2013wra}, in the $(M_T, \Gamma_T / M_T)$ plane for both scalar and vector DM scenarios and for the same values of the DM mass previously considered.
For each simulated point the best Signal Region (SR) is also shown using a color code. 

\begin{figure}[ht!]
\centering
\includegraphics[width=.32\textwidth]{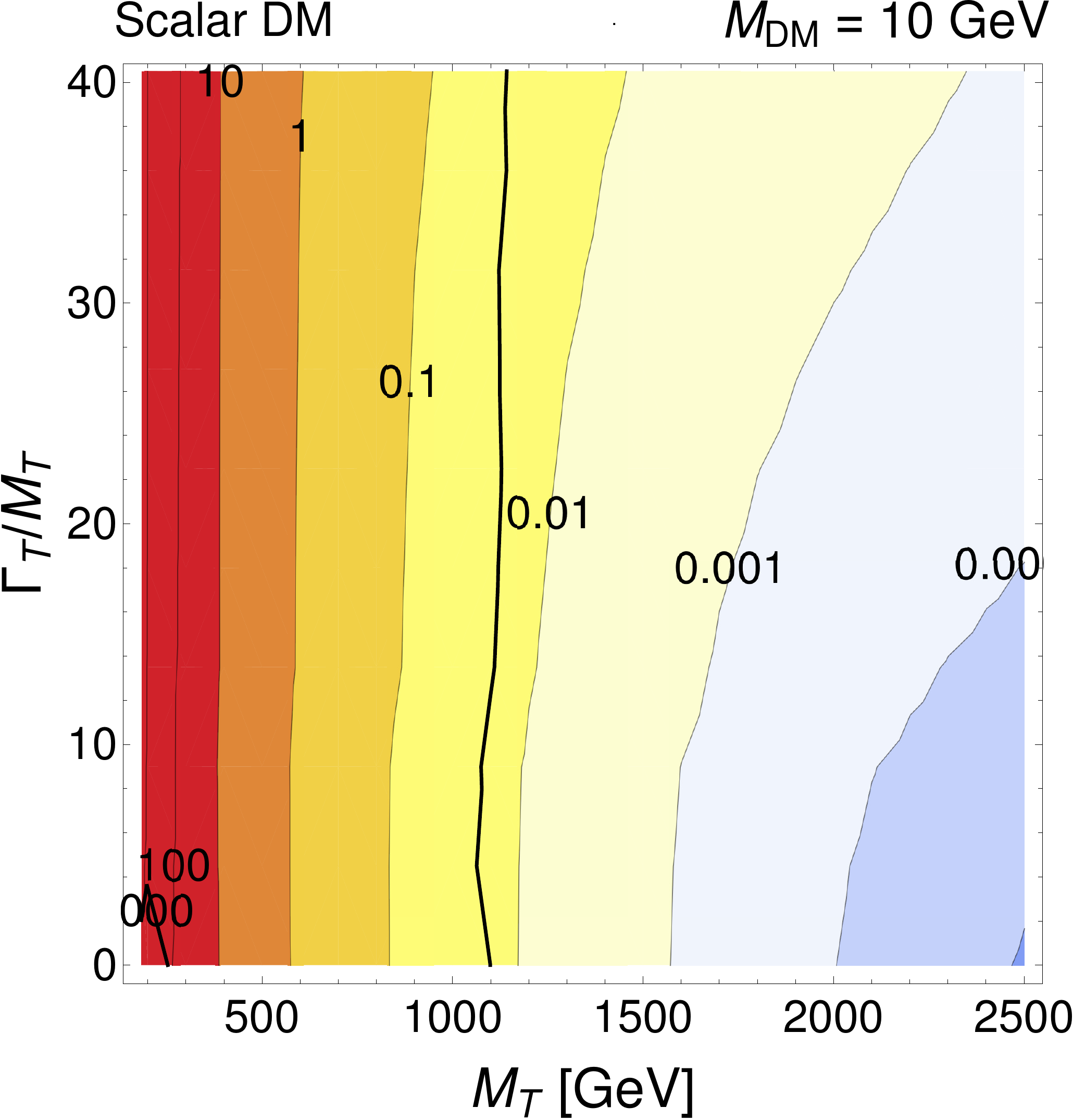} 
\includegraphics[width=.32\textwidth]{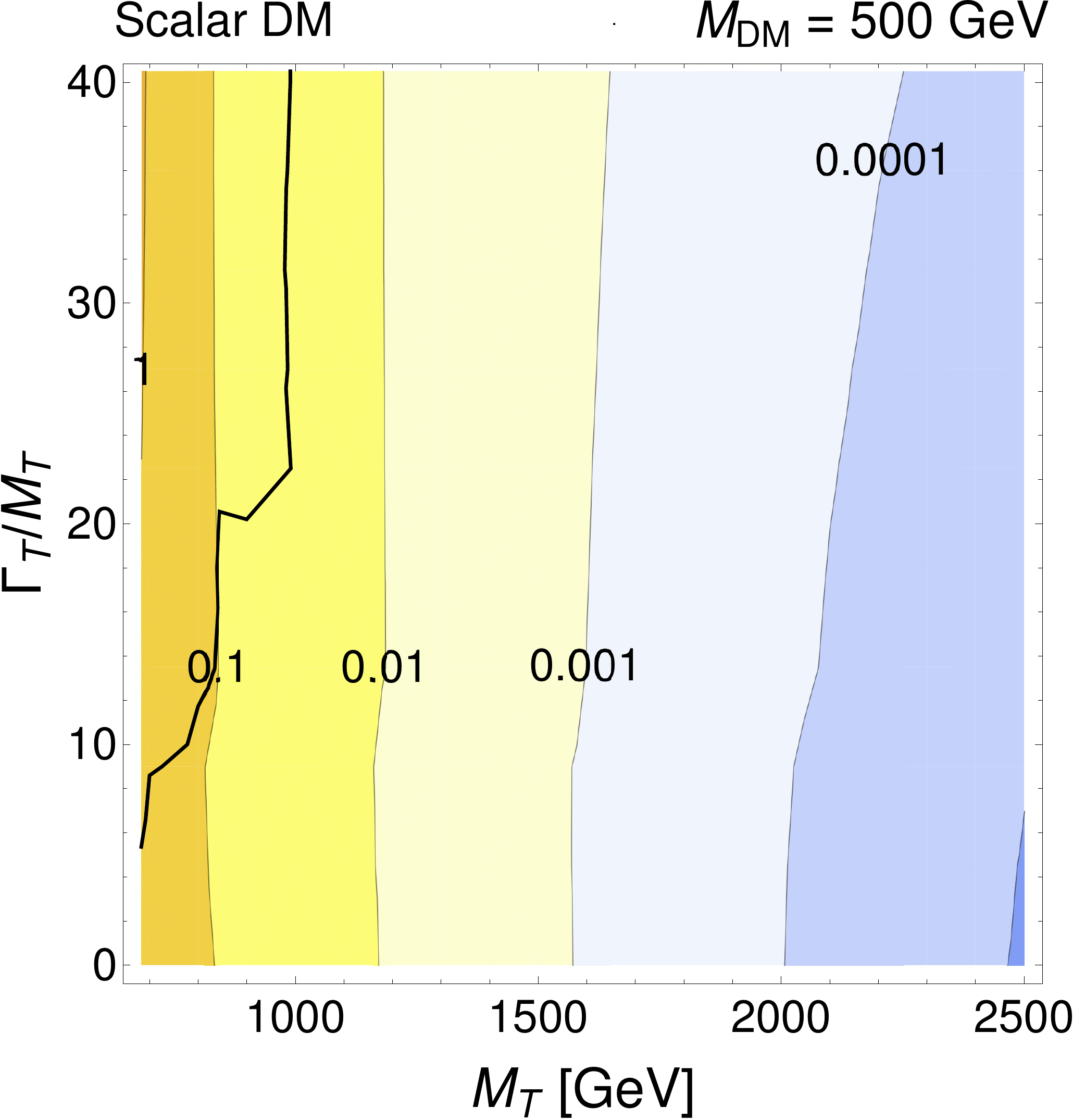} 
\includegraphics[width=.32\textwidth]{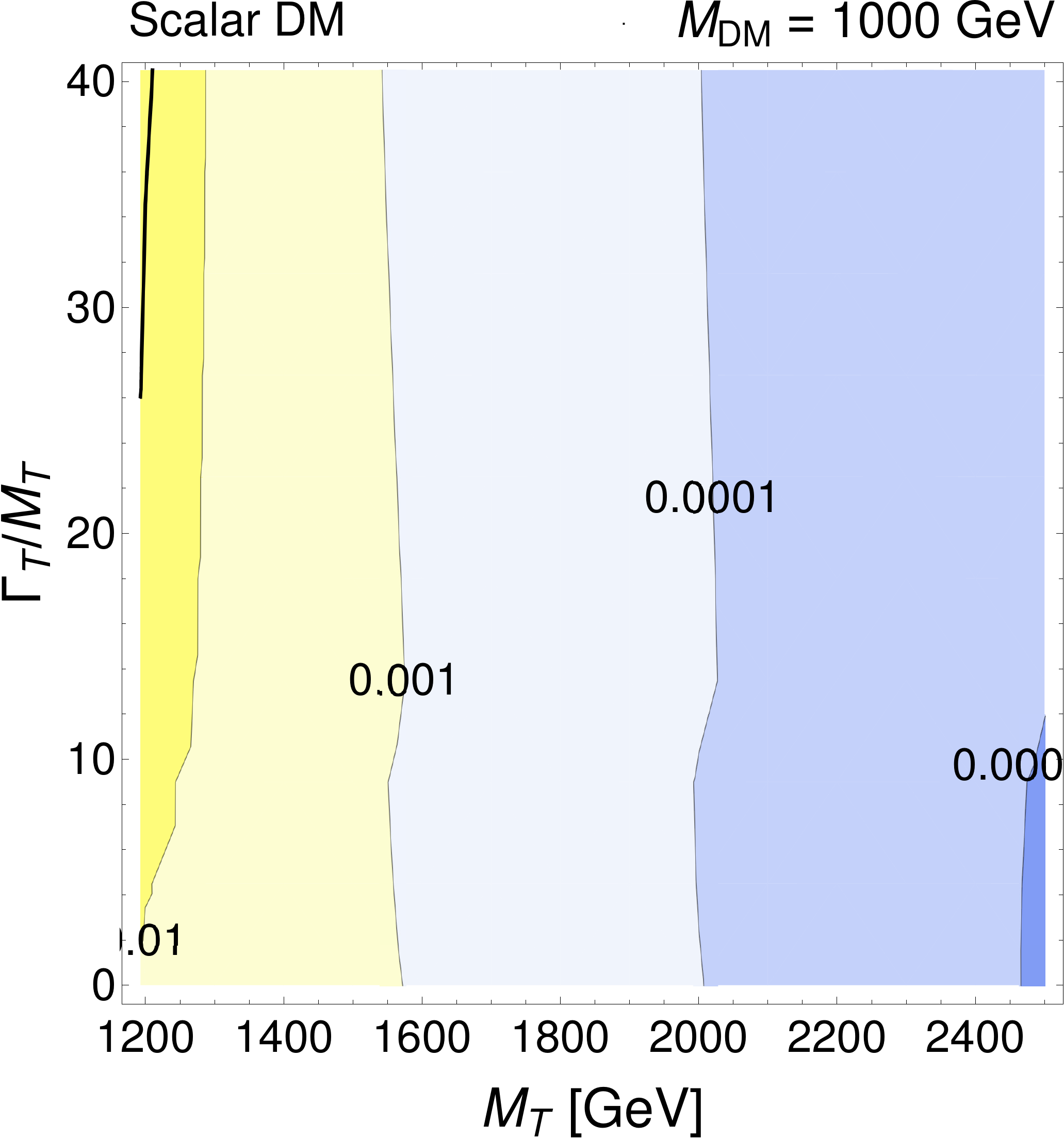}\\ 
\includegraphics[width=.32\textwidth]{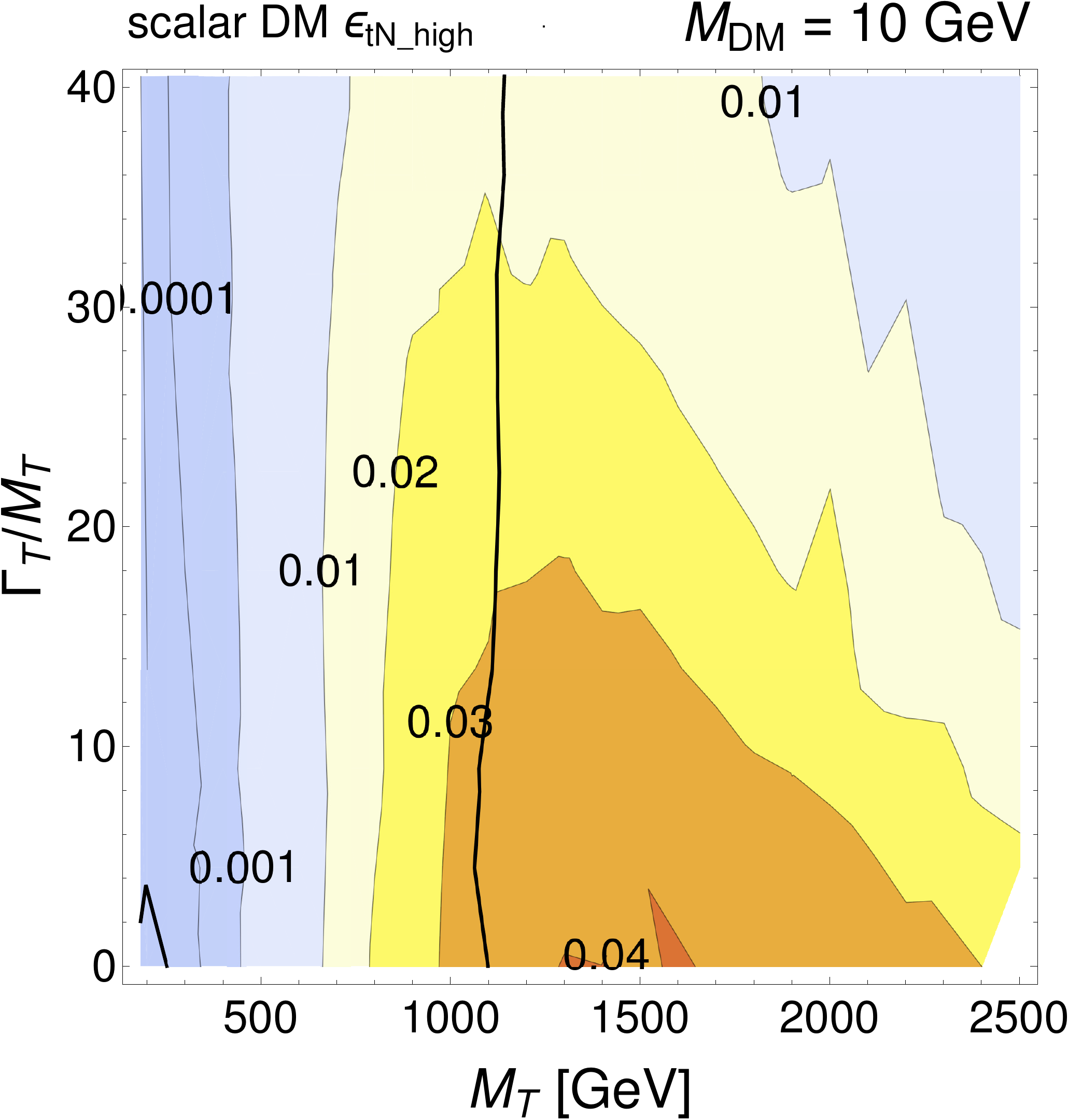} 
\includegraphics[width=.32\textwidth]{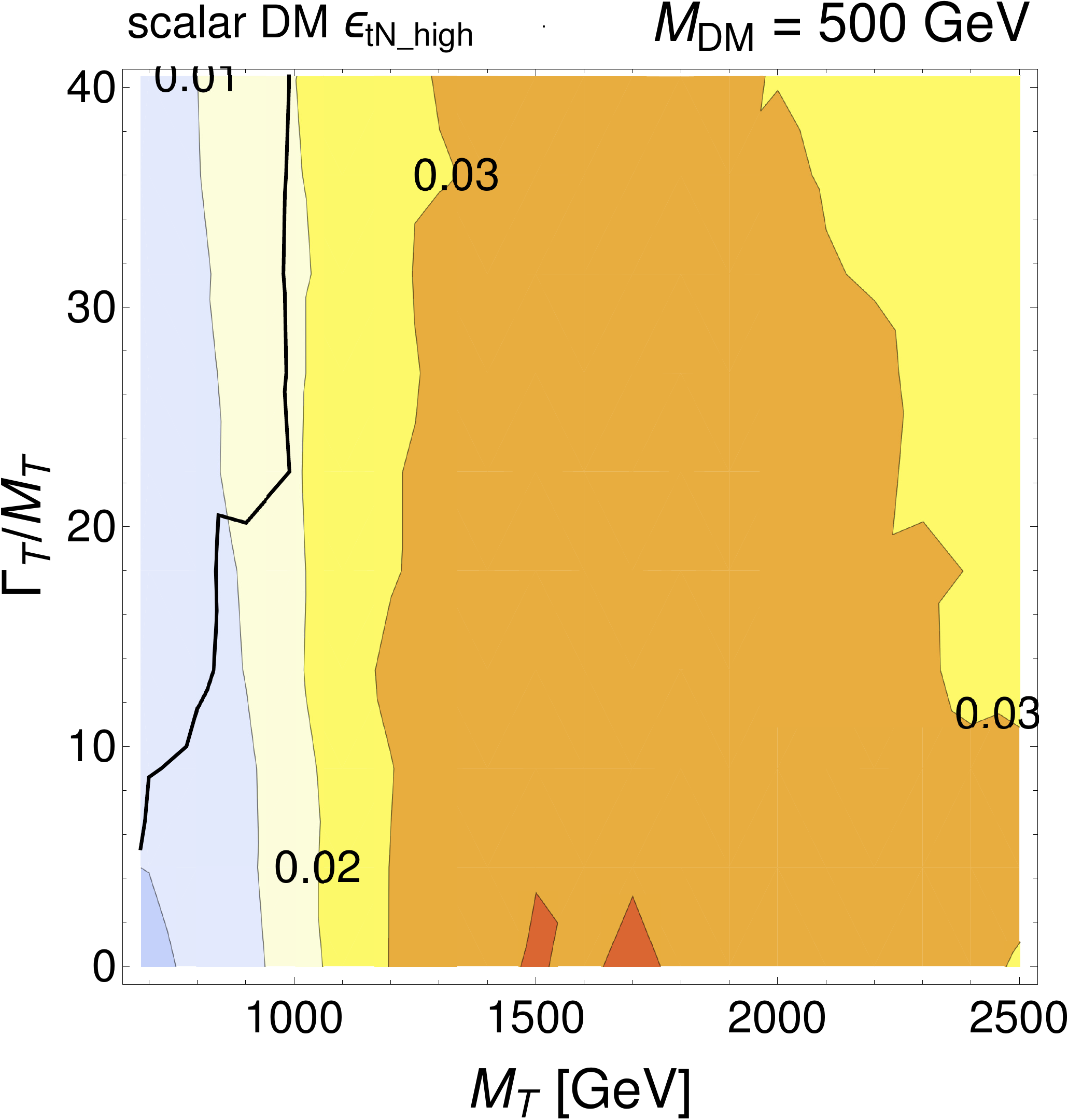} 
\includegraphics[width=.32\textwidth]{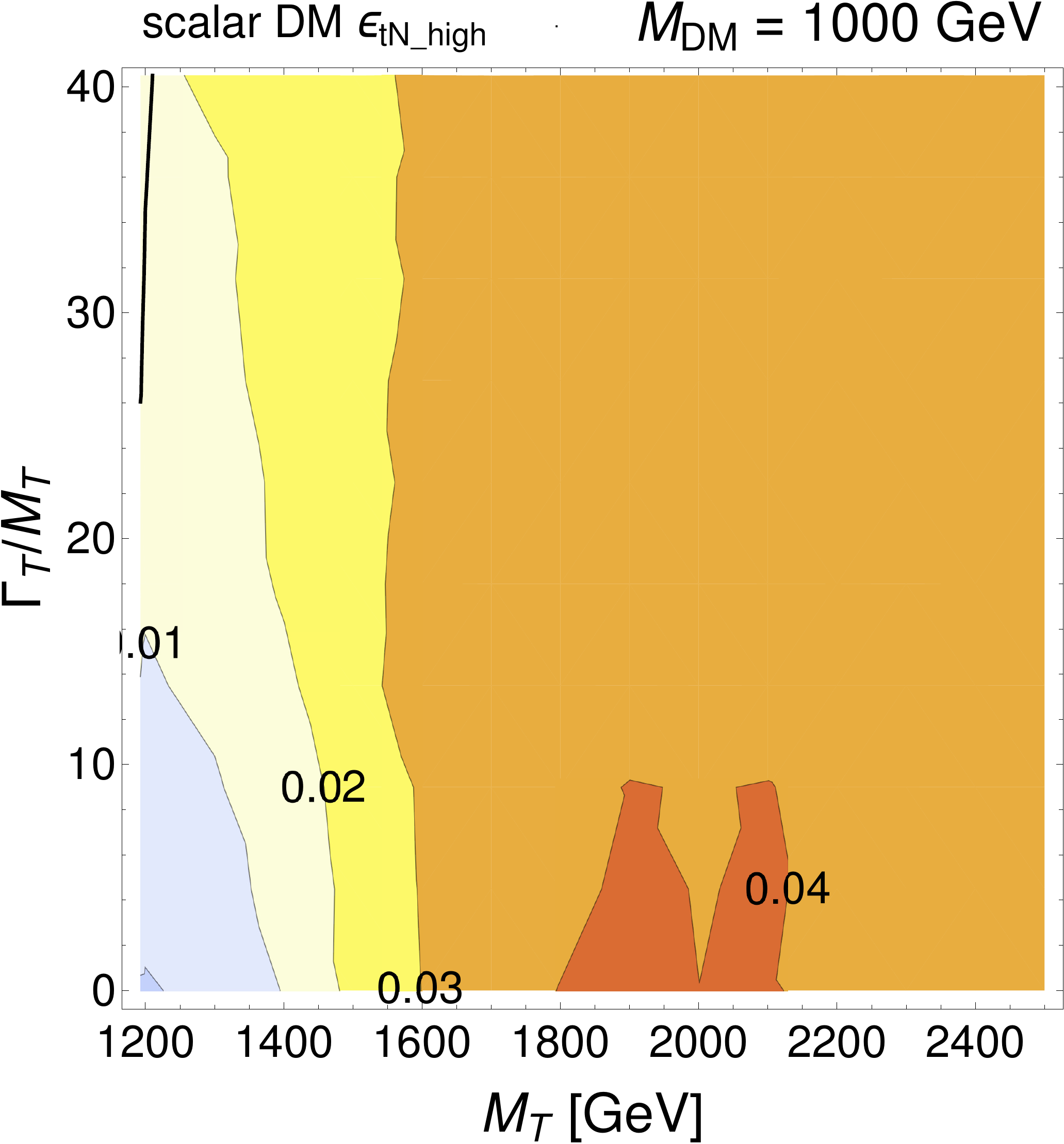}
\caption{\label{fig:sigmaEffs3} Top row: full signal cross sections for the scalar DM case. Bottom row: efficiencies of the SR tN\_high from the analysis ATLAS-CONF-2016-050~\cite{ATLAS-CONF-2016-050} for different scalar DM masses.}
\end{figure}
\begin{figure}[ht!]
\centering
\includegraphics[width=.32\textwidth]{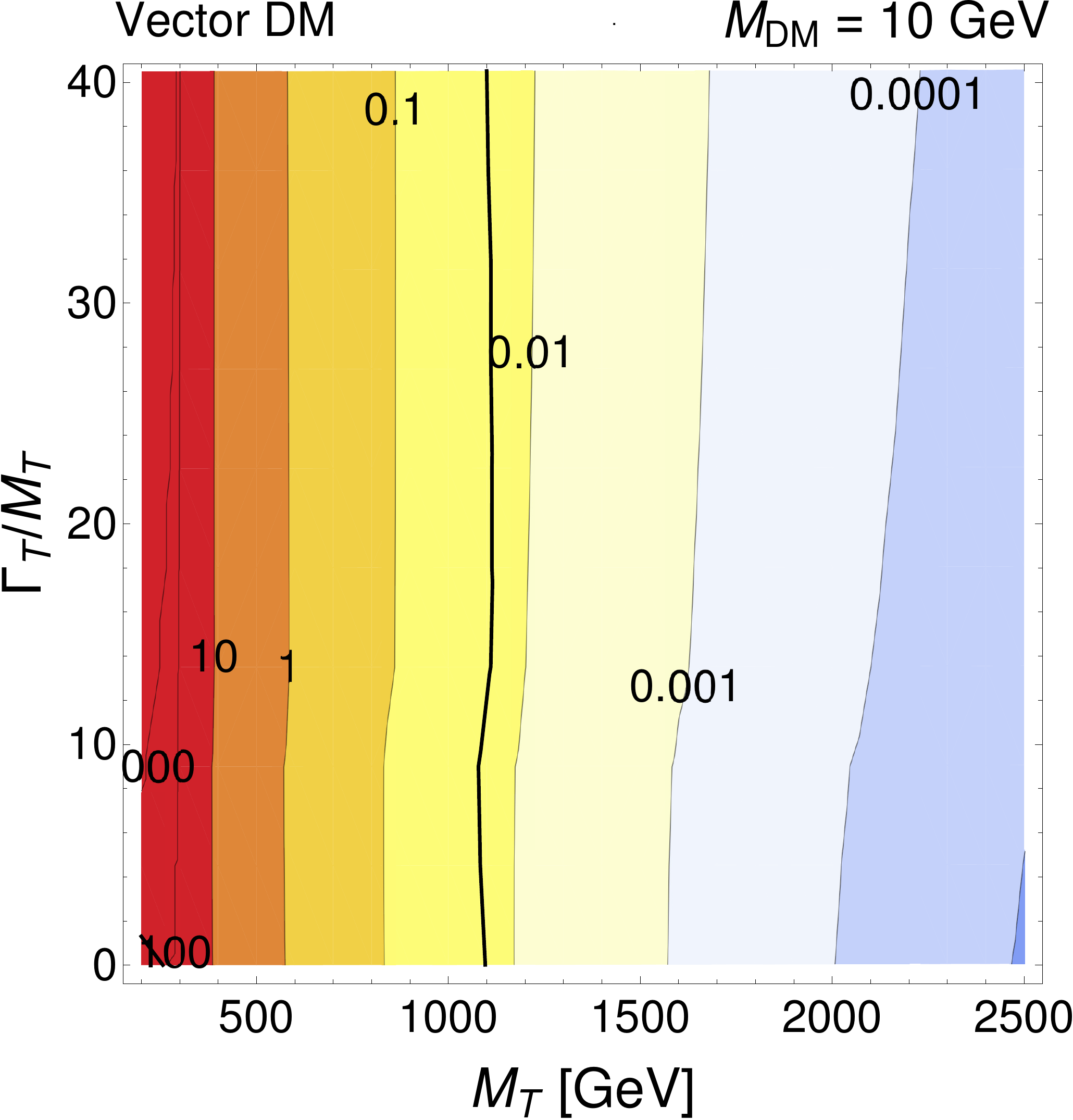} 
\includegraphics[width=.32\textwidth]{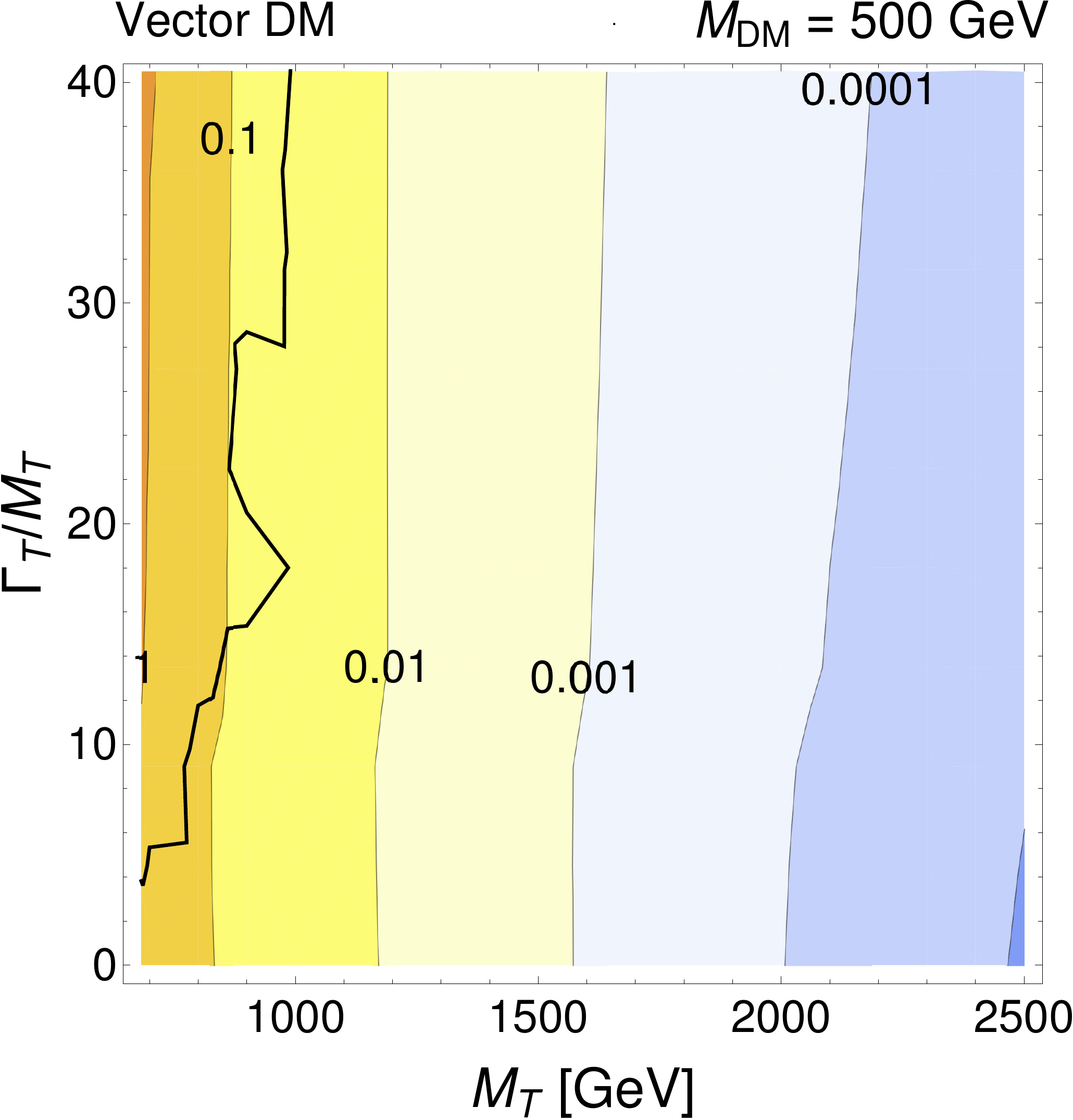} 
\includegraphics[width=.32\textwidth]{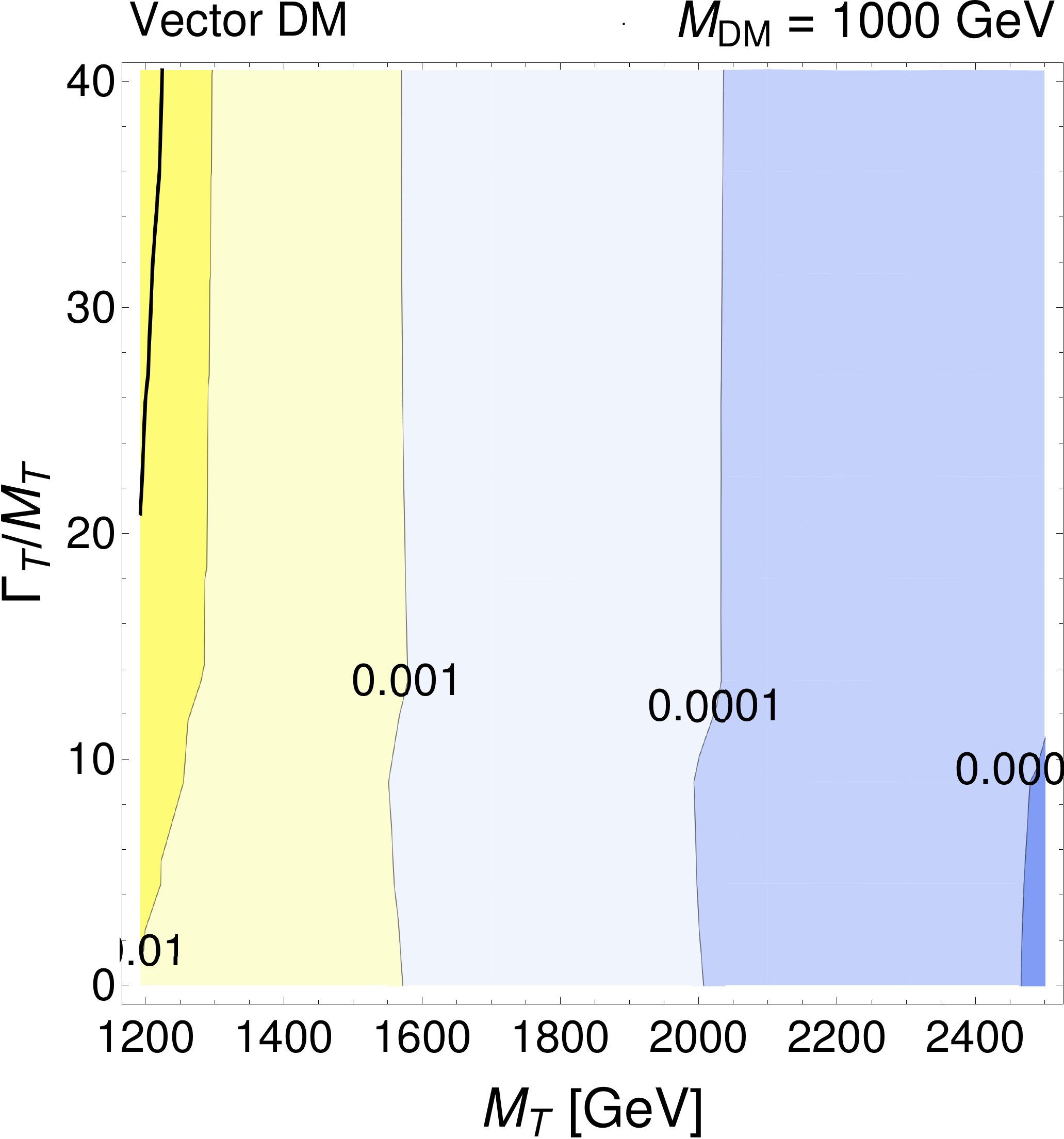}\\ 
\includegraphics[width=.32\textwidth]{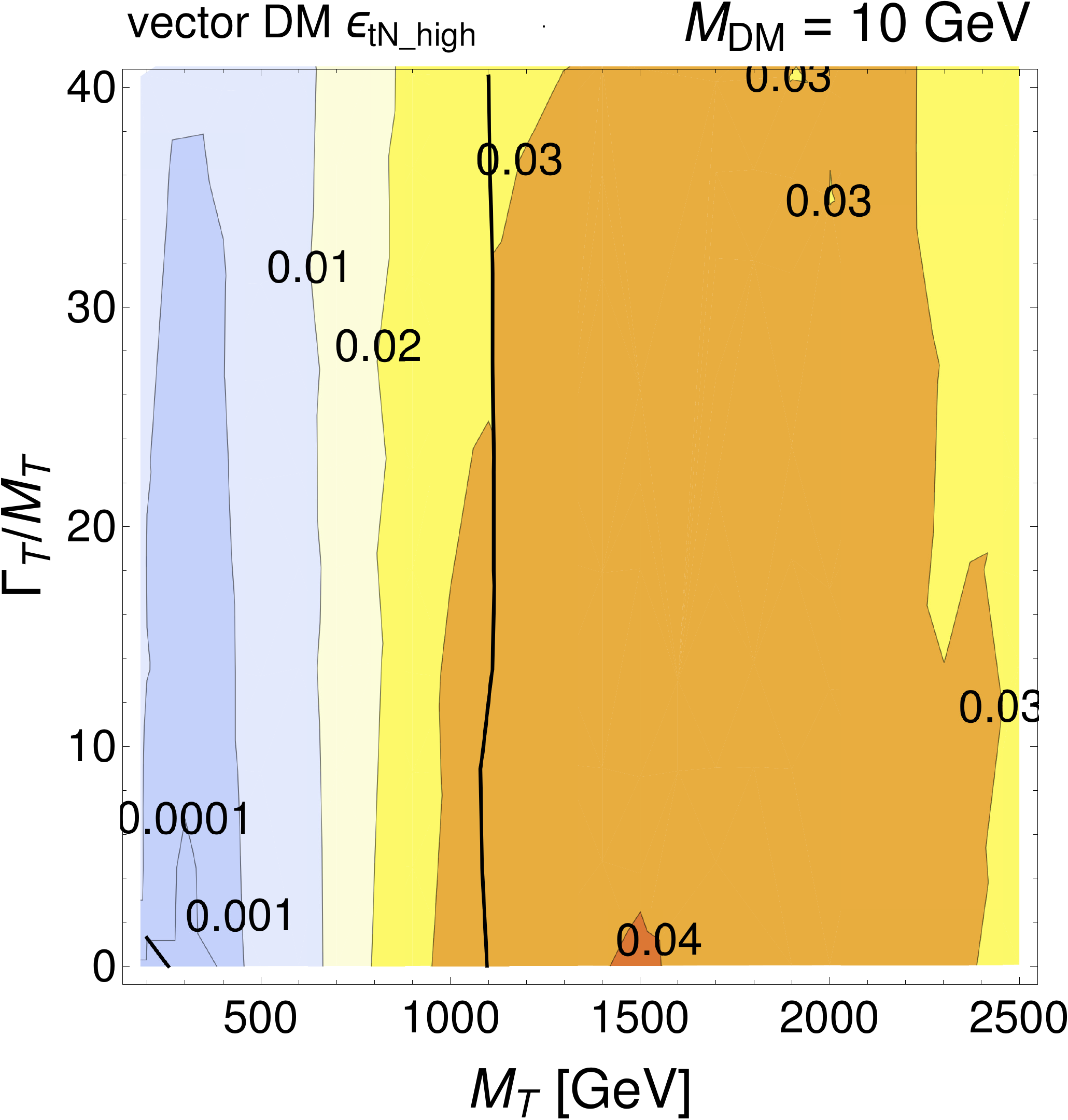} 
\includegraphics[width=.32\textwidth]{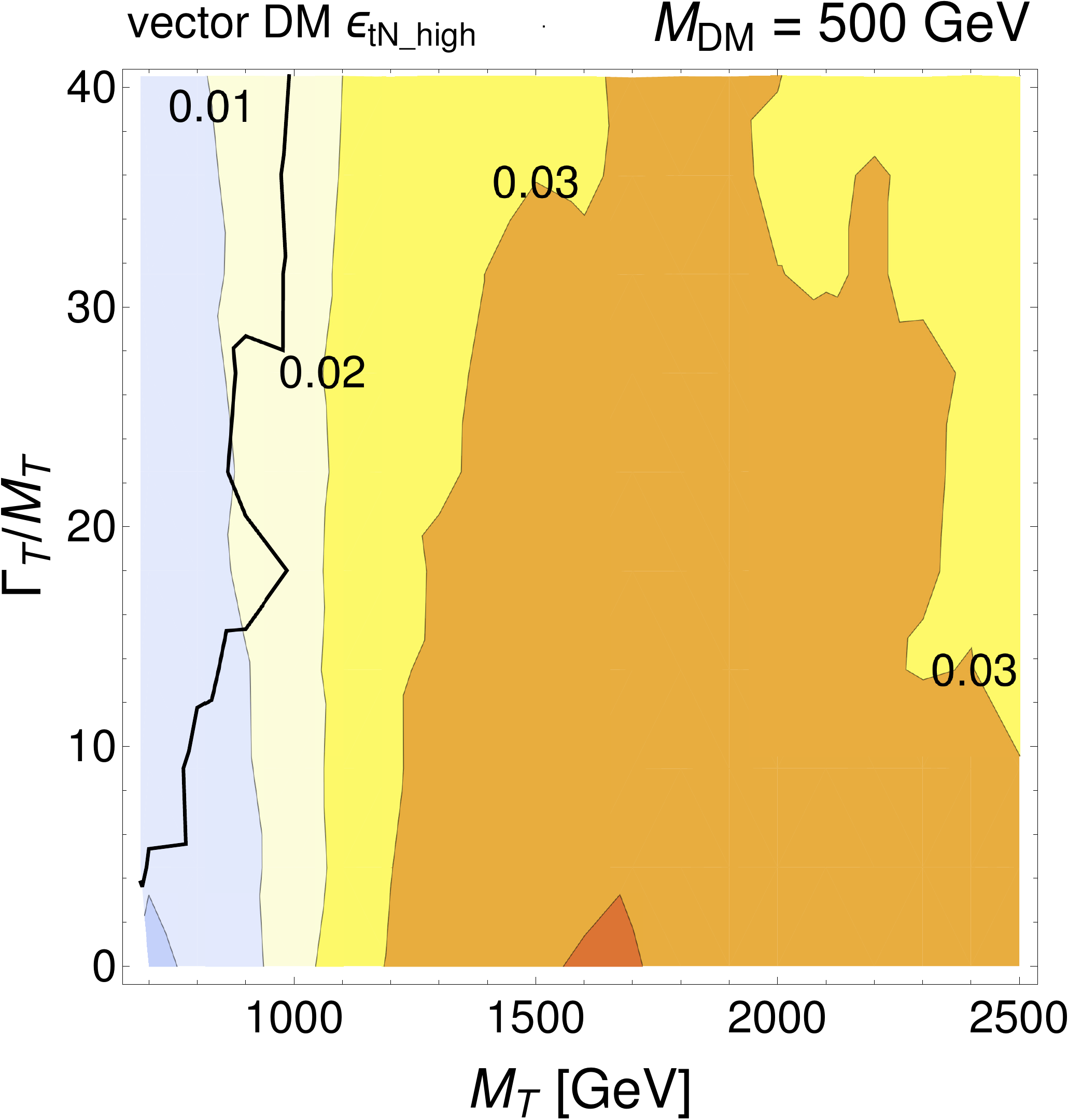} 
\includegraphics[width=.32\textwidth]{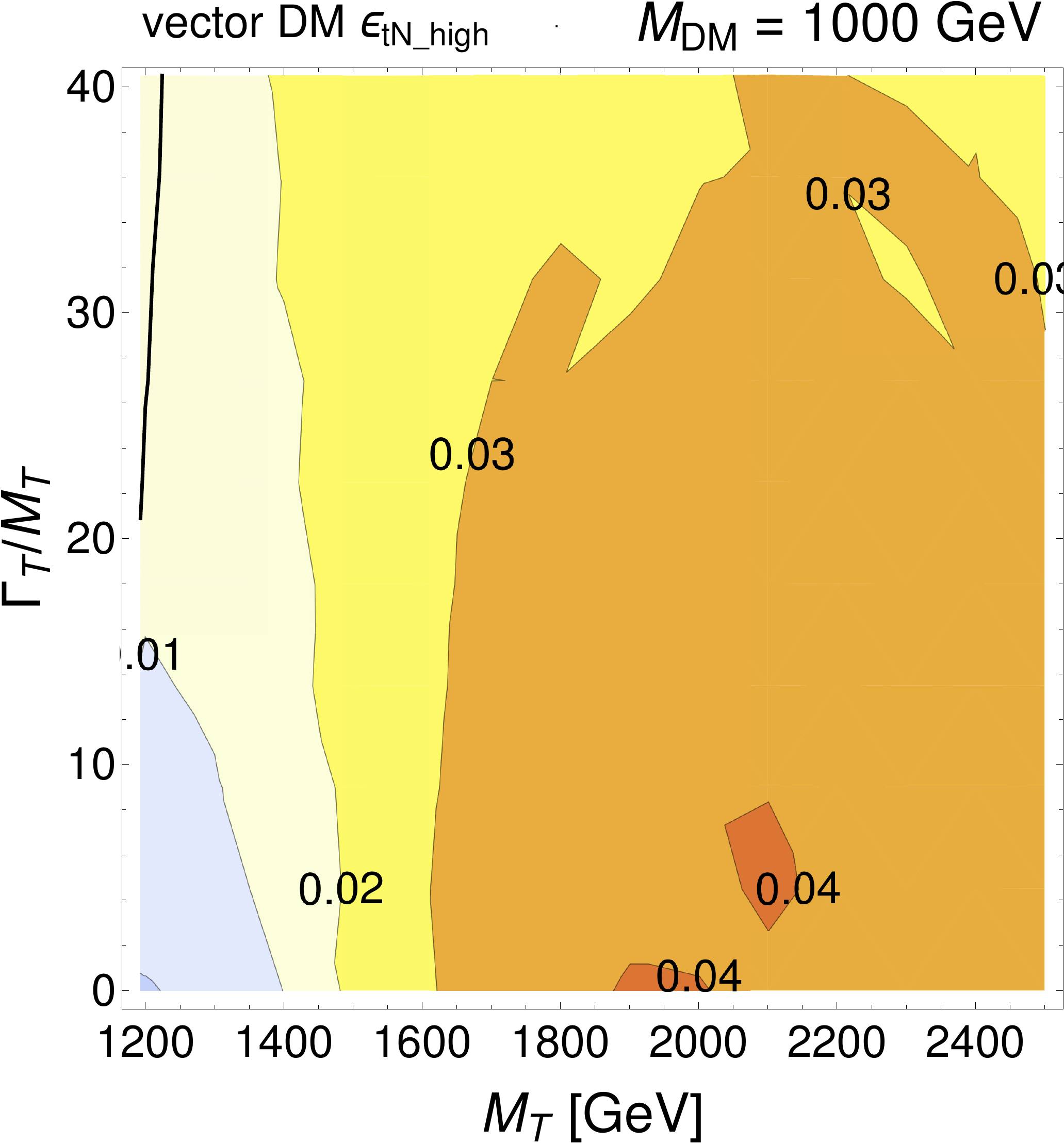}
\caption{\label{fig:sigmaEffv3} Top row: full signal cross sections for the vector DM case. Bottom row: efficiencies of the SR tN\_high from the analysis ATLAS-CONF-2016-050~\cite{ATLAS-CONF-2016-050} for different vector DM masses.}
\end{figure}
\begin{figure}[ht!]
\centering
\includegraphics[width=.45\textwidth]{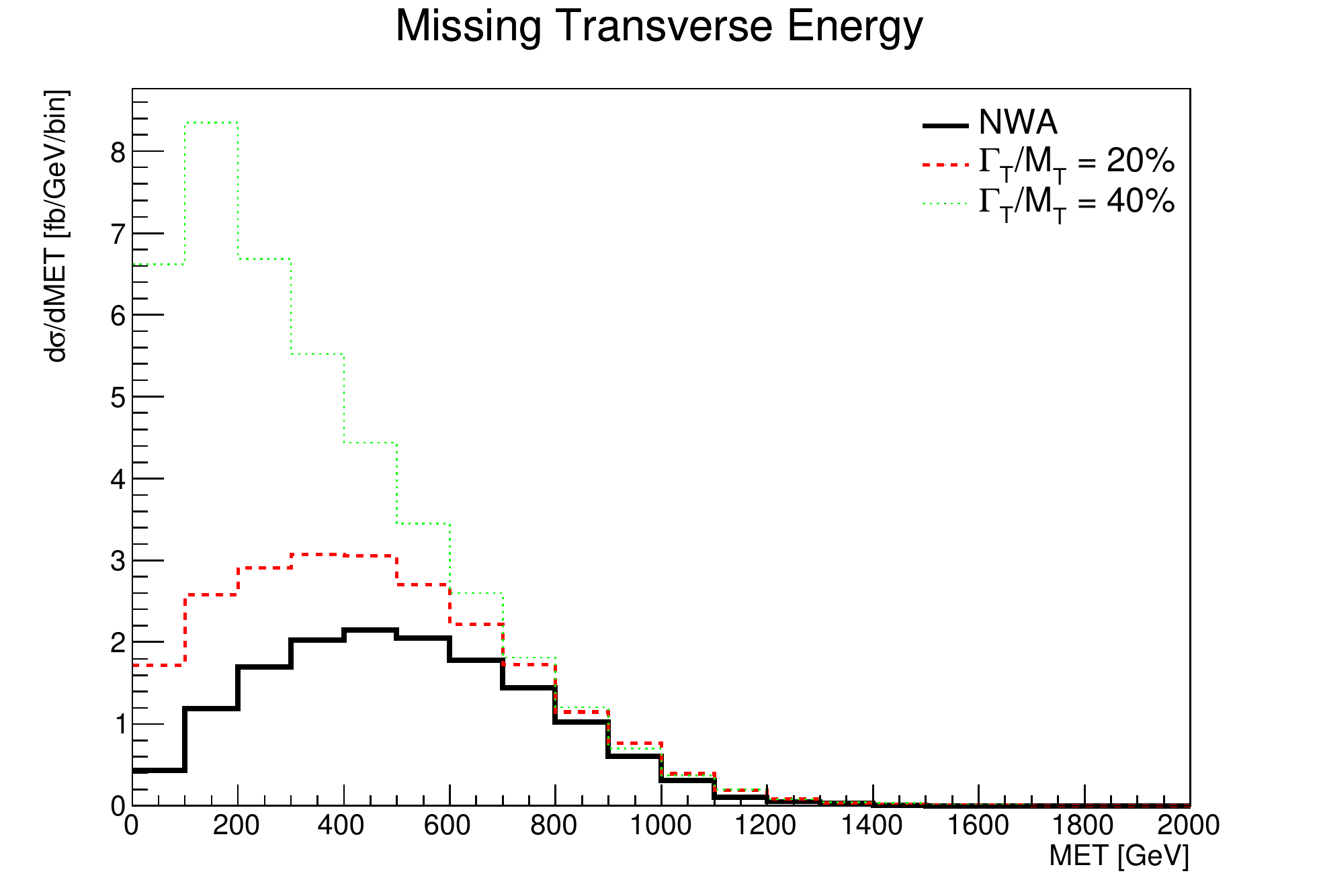} 
\includegraphics[width=.45\textwidth]{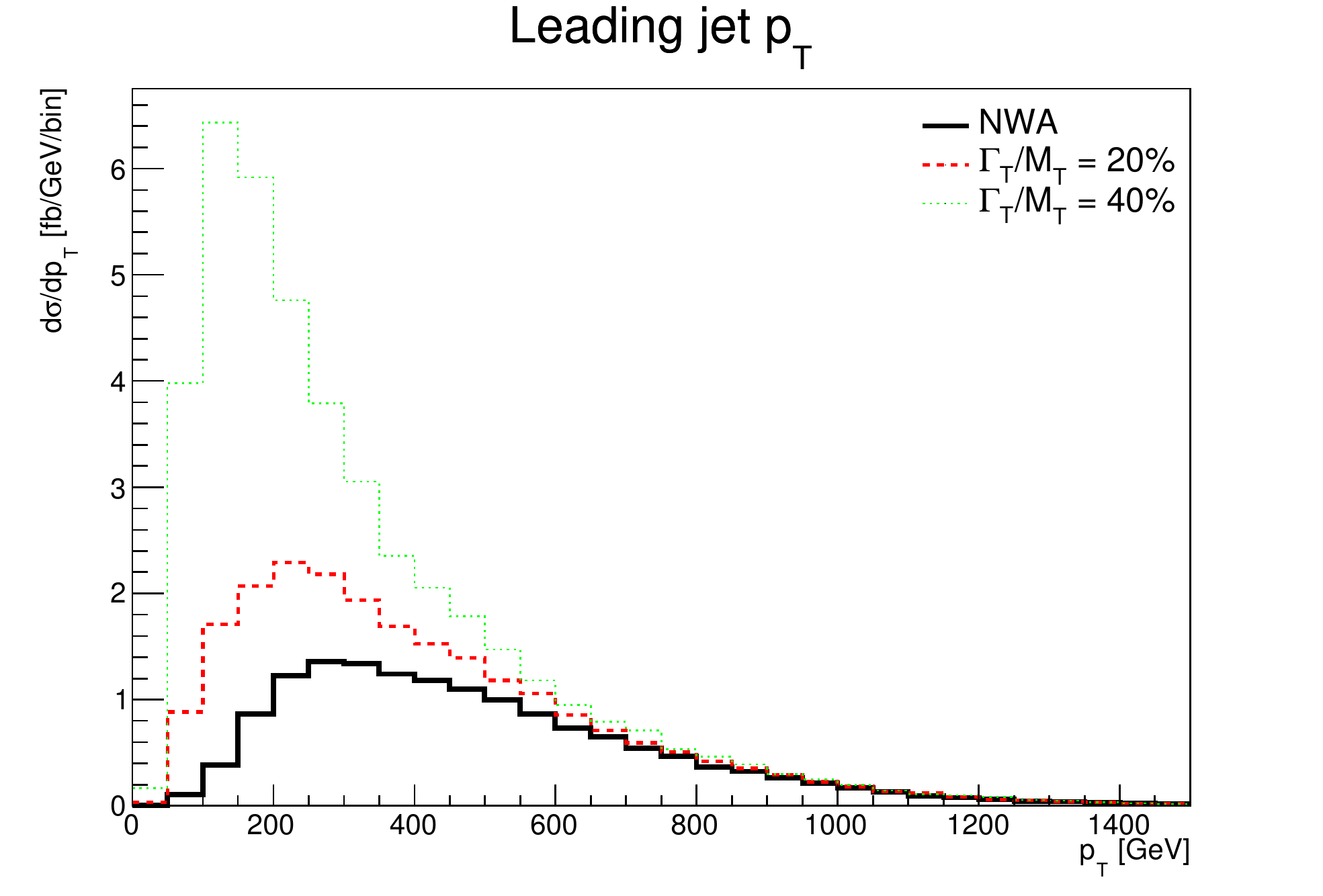} 
\caption{\label{fig:cut3G10}Differential distributions along the bound for a $T$ with mass $M_T=1100$ GeV coupling to the top quark and scalar DM with mass $M_{DM}=10$ GeV.}
\end{figure}

The main conclusions which can be derived are the following:
\begin{itemize}
\item For all values of the DM mass the bounds for scalar and vector DM do not show sizable differences. The most sensitive SR is almost always tN\_high from the analysis ATLAS-CONF-2016-050, which is optimised for ``high mass splitting, leading to very boosted top quarks where the decay products are close-by and can be reconstructed within a single large-R jet''~\cite{ATLAS-CONF-2016-050}. It is characterised by the requirement of at least 4 jets, at least 1 b-jet and $\MET$ larger than 450 GeV, and by further kinematical requirements on combination of variables, such as $E_{T,\bot}^{\rm miss}$ for the definition of which we refer to~\cite{ATLAS-CONF-2016-050}, to optimize the signal-over-background ratio for boosted top quarks produced in association with $\MET$, as is the case in both stops decaying to top and neutralino and $T$ quarks decaying to top and bosonic DM candidates. Therefore, this SR is dominantly sensitive to topologies of resonant production, which depend weakly on the spin of the DM particle. 
\item For $M_{\rm DM}$ = 10 GeV the exclusion bound is around $M_T = 1100$ GeV and has basically no width dependence. It is therefore instructive compare the width dependence of the full signal cross section and of the efficiency for the tN\_high signal region, shown in Figs.\ref{fig:sigmaEffs3} and \ref{fig:sigmaEffv3}. Clearly, the increase in the cross section is compensated by an analogous decrease in the efficiency of this SR, and this compensation accounts for the fact that the bound is almost independent of the width. The reduction of the efficiency between small and large widths in the bound region is mostly due to the cuts on the \MET and on the $p_T$ of the 4 jets, respectively 450 GeV and \{120,80,50,25\} GeV in this SR~\cite{ATLAS-CONF-2016-050}. In figure \ref{fig:cut3G10} we plot the distributions of these observables at detector level, where it is possible to see that cutting on these variables has a stronger effect for the large width scenarios. It is worth noticing that points where the $T$ mass is close to the top mass and its width approaches the NWA are not excluded: in such region the top background hides the XQ signal and makes it undetectable.
\item For $M_{\rm DM}$ = 500 GeV and 1000 GeV the bound shows a slight dependence on the width: the larger the width, the stronger the exclusion. This could be understood looking again at the relation between the efficiencies of the most sentitive SR and the full signal cross section. It's also worth noticing that for these DM masses the NWA region is never excluded, only XQ with a large width can be excluded, and only up to mass of $M_T \sim$ 1000 (1200) GeV for $M_{\rm DM}$ = 500 (1000) GeV. 
\item For higher DM masses the exclusion contour is gradually pushed to the kinematics limit and above the maximum value of the width-over-mass ratio we have tested (40\%), and eventually disappears due to the limited sensitivity of the detector for small mass splittings between $T$ and $DM$. 
\end{itemize} 

%\vspace{\baselineskip}
\subsubsection{Dependence on the chirality of the couplings}

% \ \vspace{\baselineskip}

\begin{figure}[ht!]
\centering
\includegraphics[width=.45\textwidth]{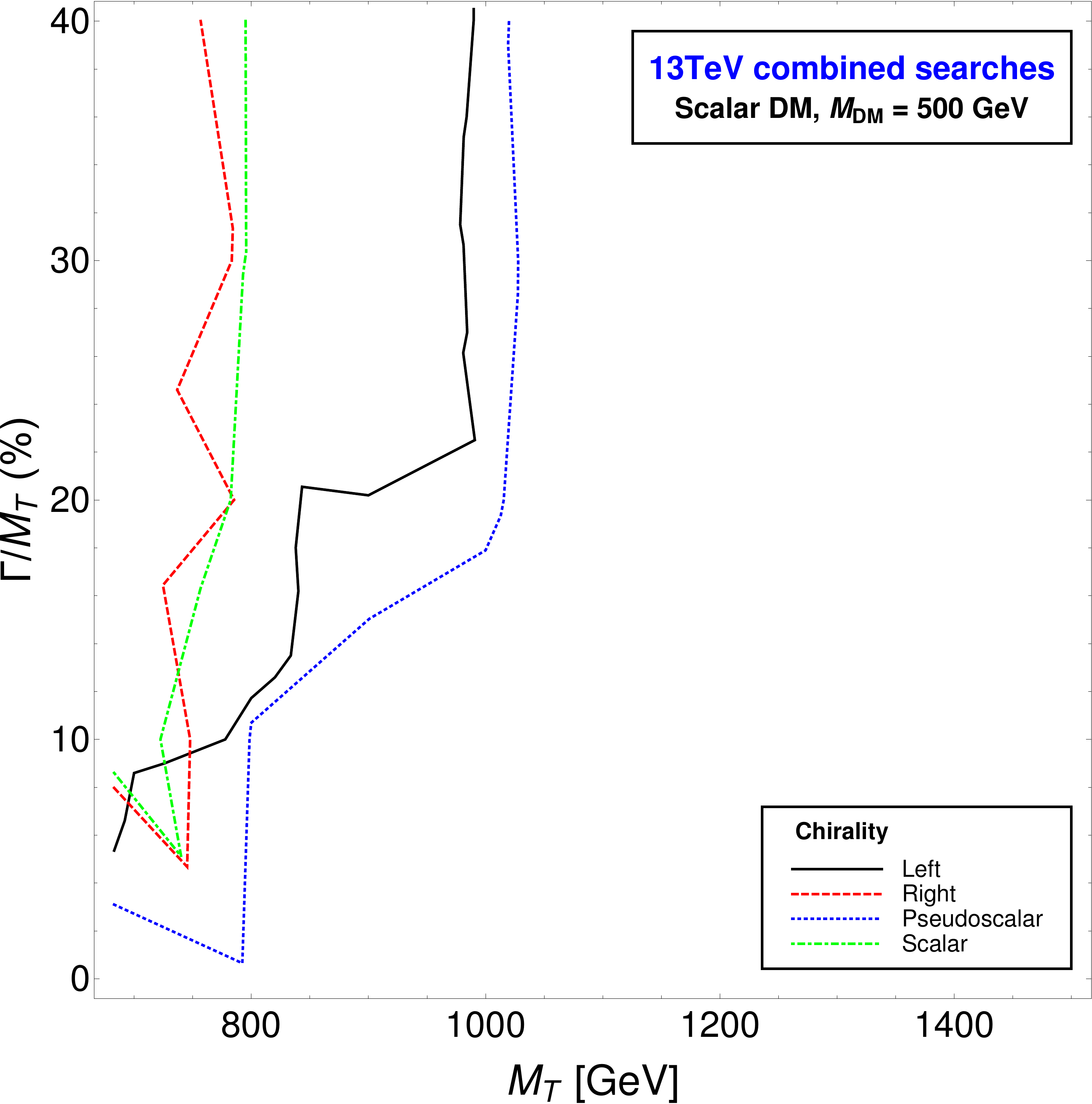} 
\includegraphics[width=.45\textwidth]{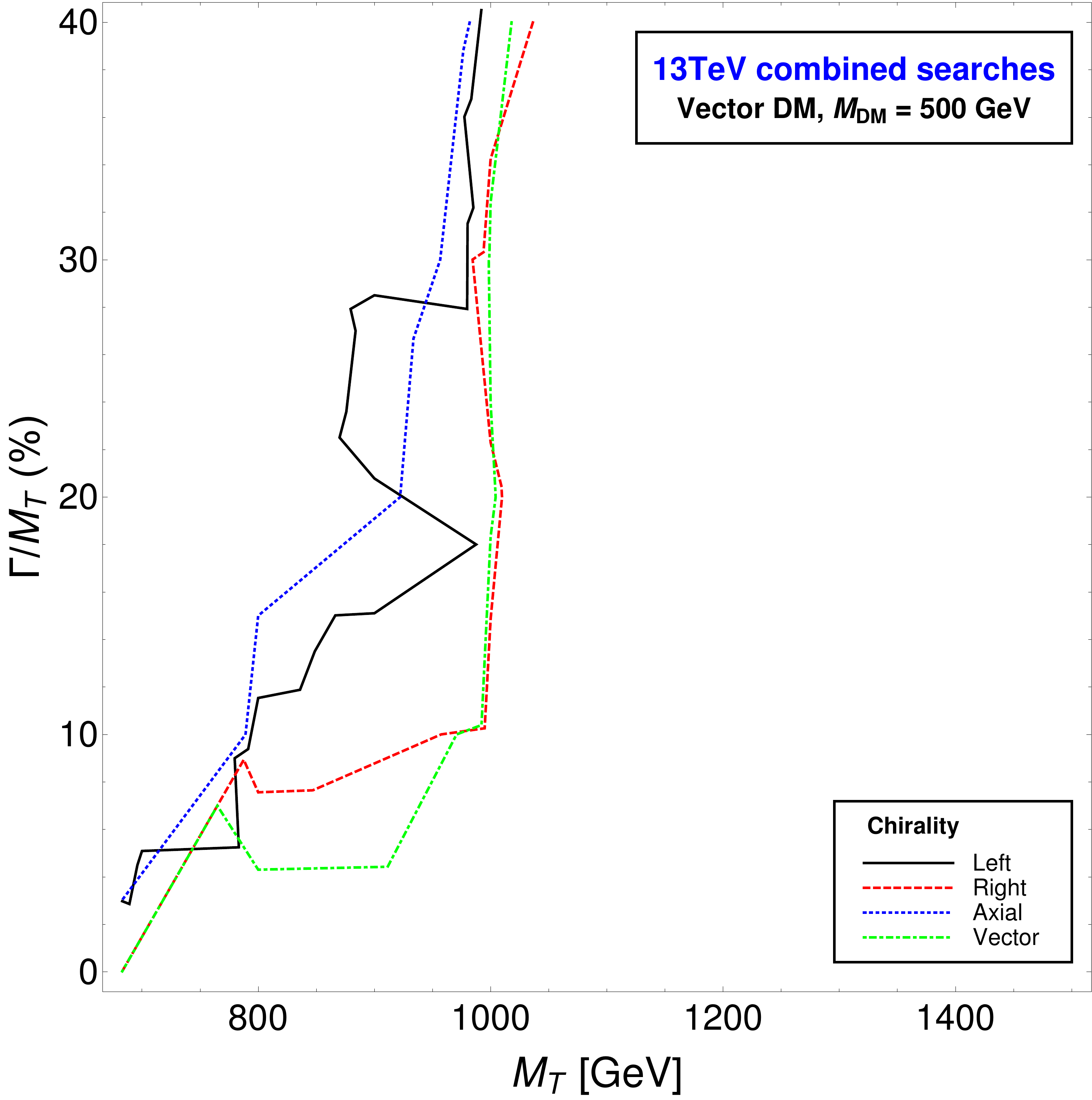} 
\caption{\label{fig:3Gchirality}Exclusion bounds for a $T$ interacting with the SM top quark and DM for different hypotheses on the chirality of the couplings: for a VLQ $T$ pure left-handed and pure right-handed couplings, and for a ChQ $T$ pure scalar (vector) or pseudoscalar (axial-vector) couplings if $T$ interacts with scalar (vector) DM.}
\end{figure}

To conclude the analysis of XQs interacting with DM states and third generation SM quarks, we consider how the bounds change if the $T$ quark is a VLQ singlet (pure right-handed couplings) or a ChQ (where we consider either pure scalar or pseudoscalar couplings if the DM is a scalar or pure vector or axial-vector couplings if the DM is a vector). In Fig.~\ref{fig:3Gchirality} the bounds are shown for all the aforementioned scenarios: keeping in mind that the uncertainty due to the use of a recasting tool is quite large, it is possible to see that with the set of experimental searches considered in this study, the differences between various chiralities are not significant for the vector DM scenario, while there are visible differences if the DM is scalar. The more pronounced effects for scalar DM with respect to vector DM can be due to the fact that the transverse components of vector DM are less sensitive to the coupling chiralities than its (pseudo)scalar one, and dilute the effects of different couplings. Therefore, with the set of cuts currently used to optimise the discovery of new physics in the $t \bar t + \MET$ channels, a characterisation of the couplings of a $T$ interacting with a vector DM and the top quark would be challenging even in the large width regime. If the DM is scalar there could be more room for a characterisation of the properties of the $T$. Designing signal regions optimised for the discrimination of different coupling hypotheses and for different $\Gamma/M$ regimes would be advisable in case of discovery of a signal in this channel, but this goes beyond the scope of the present analysis and we defer this to a future study.

%%%%%%%%%%%%%%%%%%%%%%%%%%%%%%%%%%%%%%%%%%%%%%%%%%%%%%%%%%%%%%%%%%%%%%%%%
%%%%%%%%%%%%%%%%%%%%%%%%%%%%%%%%%%%%%%%%%%%%%%%%%%%%%%%%%%%%%%%%%%%%%%%%%
% \clearpage
%\newpage

\section{Extra $T$ quark interacting with Dark Matter and the SM up quark}
\label{sec:Tu}

In this section we will study the case of XQs coupling to first generation SM quarks and a DM candidate. The possible final states are therefore $S^0_{DM}u \; S^0_{DM}\bar u$ and $\ V^0_{DM}u \; V^0_{DM}\bar u$.

%%%%%%%%%%%%%%%%%%%%%%%%%%%%%%%%%%%%%%%%%%%%%%%%%%%%%%%%%%%%%%%%%%%%%%%%%

\subsection{Large width effects at parton level}

\begin{figure}[ht!]
% \begin{center}
% \begin{picture}(140,100)(0,-10)
% \SetWidth{1}
% \Line[arrow](0,0)(50,0)
% \Text(-2,0)[rc]{\large $u$}
% \Line[arrow](50,80)(0,80)
% \Text(-2,80)[rc]{\large $\bar u$}
% \Photon(50,0)(120,0){3}{9}
% \Line[dash](50,0)(120,0)
% \Text(122,0)[lc]{\large $S^0_{DM}, V^0_{DM}$}
% \SetColor{Red}\SetWidth{1.5}
% \Line[arrow](50,0)(50,40)
% \Text(45,20)[rc]{\large \Red{$T$}}
% \Line[arrow](50,40)(50,80)
% \Text(45,60)[rc]{\large \Red{$T$}}
% \SetColor{Black}\SetWidth{1}
% \Photon(50,80)(120,80){3}{9}
% \Line[dash](50,80)(120,80)
% \Text(122,80)[lc]{\large $S^0_{DM}, V^0_{DM}$}
% \Gluon(50,40)(90,40){3}{6}
% \Line[arrow](90,40)(120,60)
% \Text(122,60)[lc]{\large $u$}
% \Line[arrow](120,20)(90,40)
% \Text(122,20)[lc]{\large $\bar u$}
% \end{picture}\hskip 60pt
% \begin{picture}(140,100)(0,-10)
% \SetWidth{1}
% \Line[arrow](0,0)(50,0)
% \Text(-2,0)[rc]{\large $u$}
% \Line[arrow](50,80)(0,80)
% \Text(-2,80)[rc]{\large $\bar u$}
% \Photon(50,40)(120,40){3}{9}
% \Line[dash](50,40)(120,40)
% \Text(122,40)[lc]{\large $S^0_{DM}, V^0_{DM}$}
% \Line[arrow](50,0)(50,40)
% \Text(45,20)[rc]{\large $u$}
% \SetColor{Red}\SetWidth{1.5}
% \Line[arrow](50,40)(50,80)
% \Text(45,60)[rc]{\large \Red{$T$}}
% \SetColor{Black}\SetWidth{1}
% \Photon(50,80)(120,80){3}{9}
% \Line[dash](50,80)(120,80)
% \Text(122,80)[lc]{\large $S^0_{DM}, V^0_{DM}$}
% \Gluon(50,0)(90,0){3}{6}
% \Line[arrow](90,0)(120,10)
% \Text(122,10)[lc]{\large $u$}
% \Line[arrow](120,-10)(90,0)
% \Text(122,-10)[lc]{\large $\bar u$}
% \end{picture}
% \end{center}
\centering
\includegraphics[width=.8\textwidth]{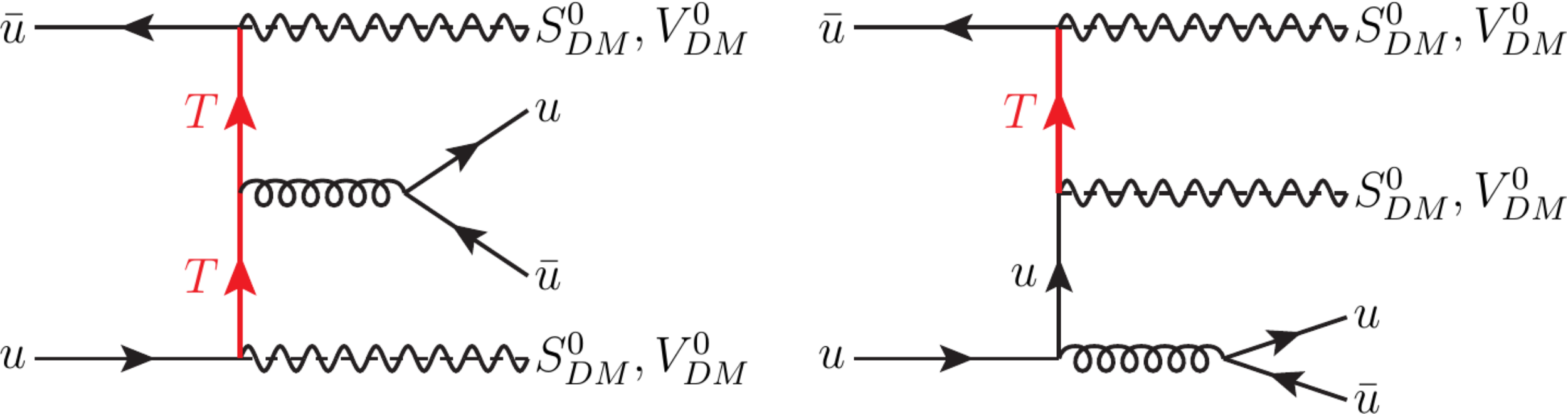}
\caption{Examples of topologies which are peculiar to scenarios with heavy quarks coupling to first generation.}
\label{fig:firstgentopologies}
\end{figure}

When the $T$ quark couples to quarks of the first generation, the $2\to 4$ process contains topologies where the initial state partons interact directly with the $T$ (examples are shown in Fig.~\ref{fig:firstgentopologies}) which are absent in the case of coupling to third generation. 

\begin{figure}[ht!]
\centering
\includegraphics[width=.32\textwidth]{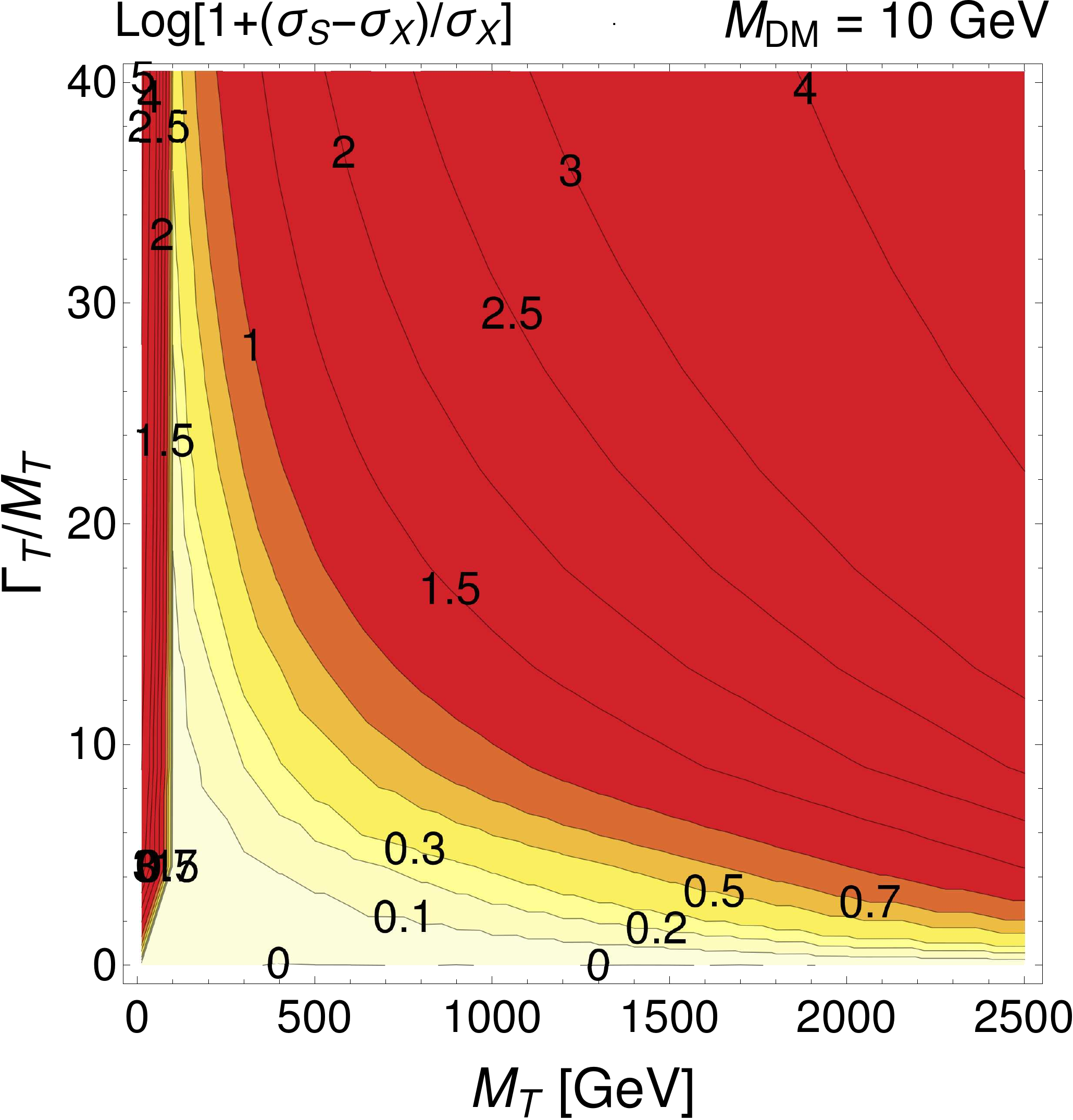} 
\includegraphics[width=.32\textwidth]{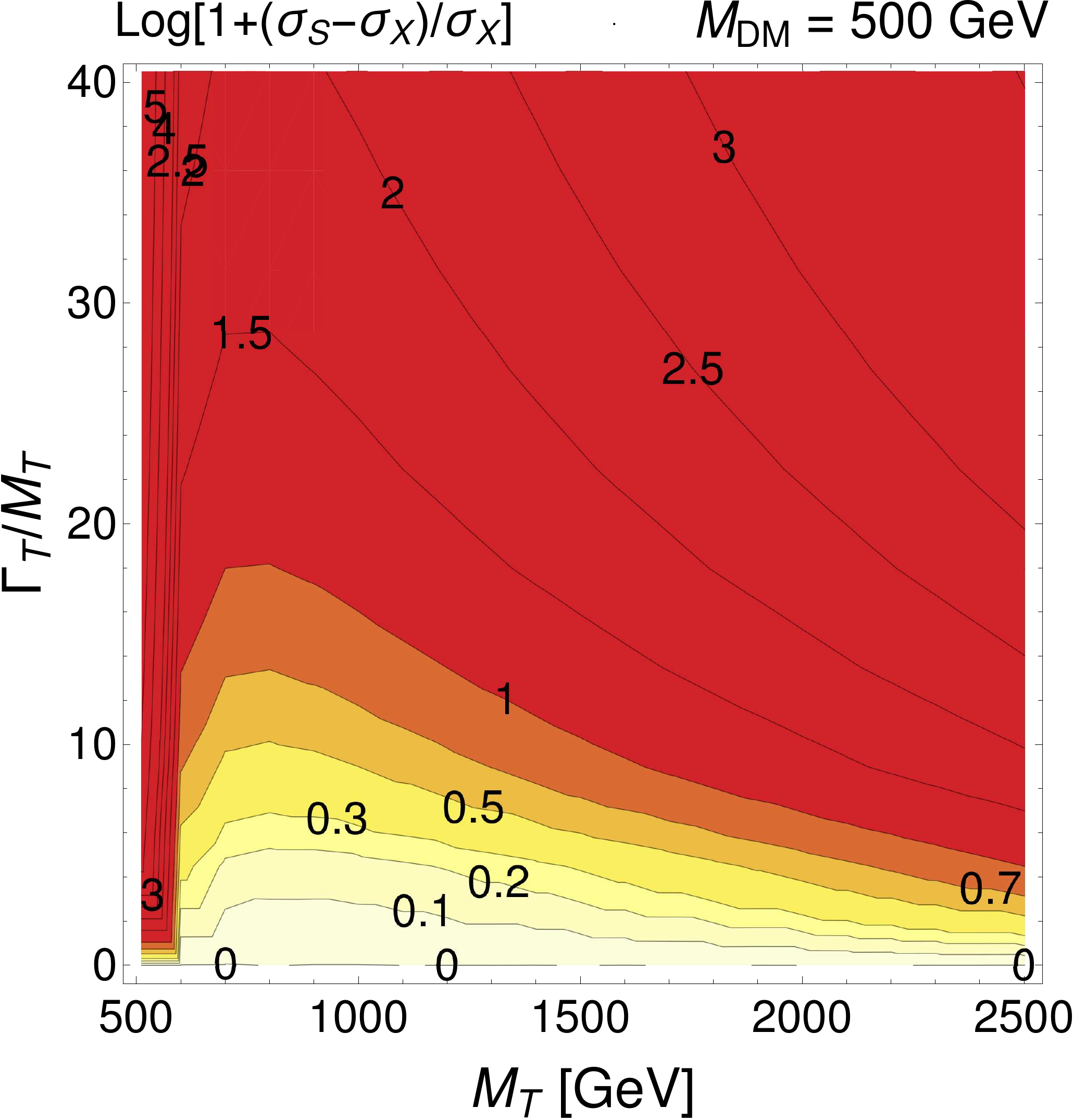} 
\includegraphics[width=.32\textwidth]{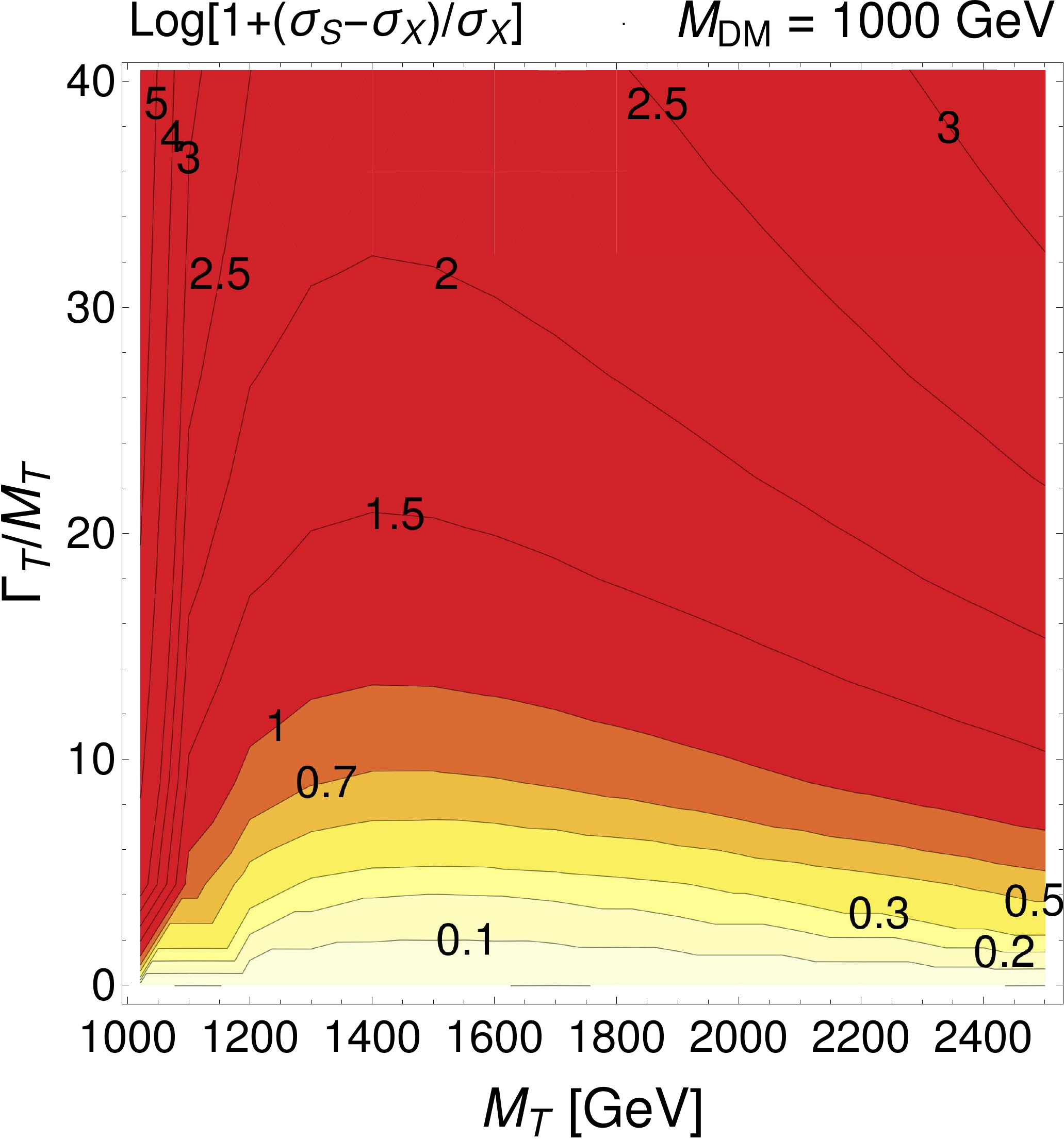}\\[5pt] 
\includegraphics[width=.32\textwidth]{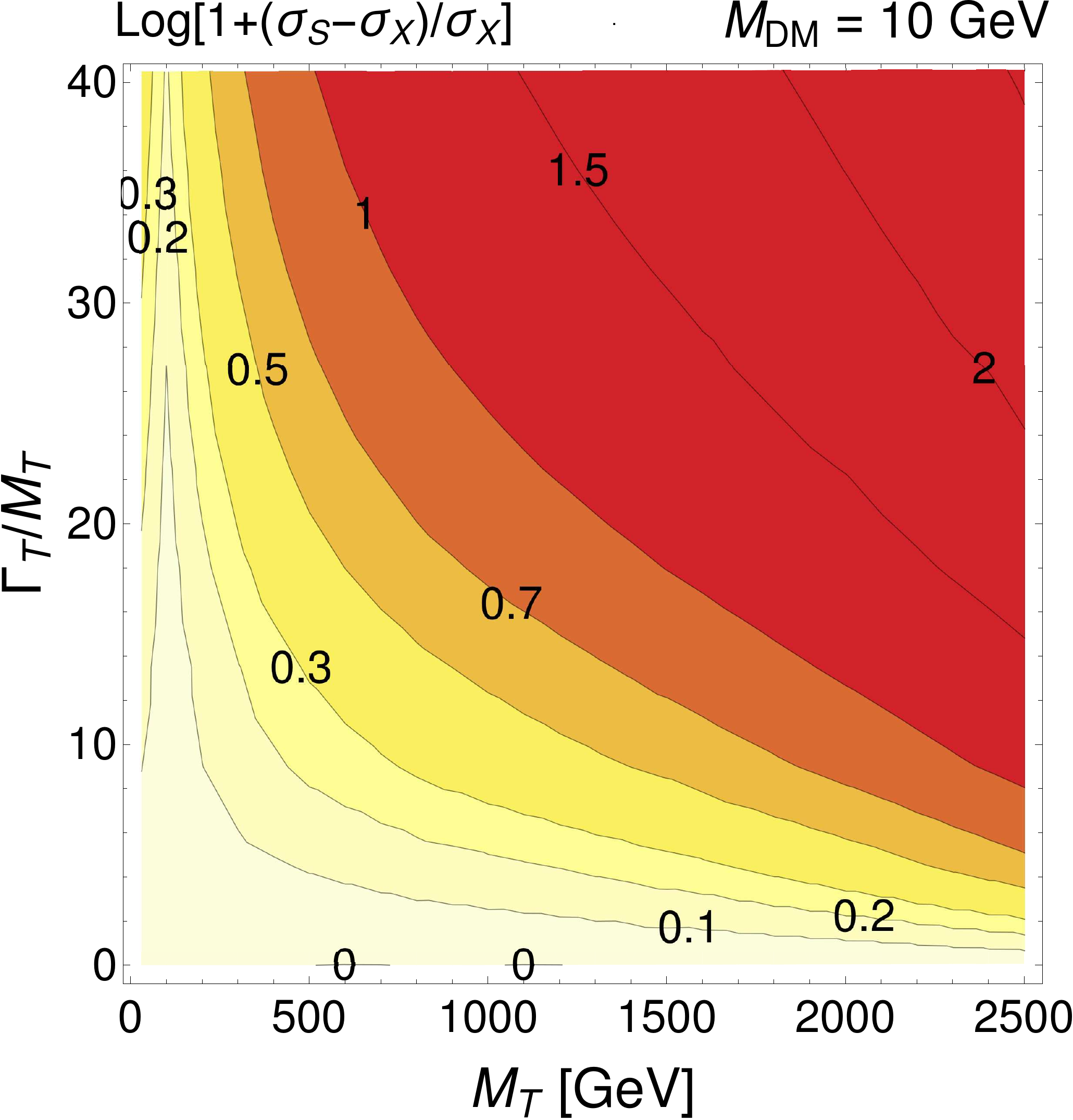} 
\includegraphics[width=.32\textwidth]{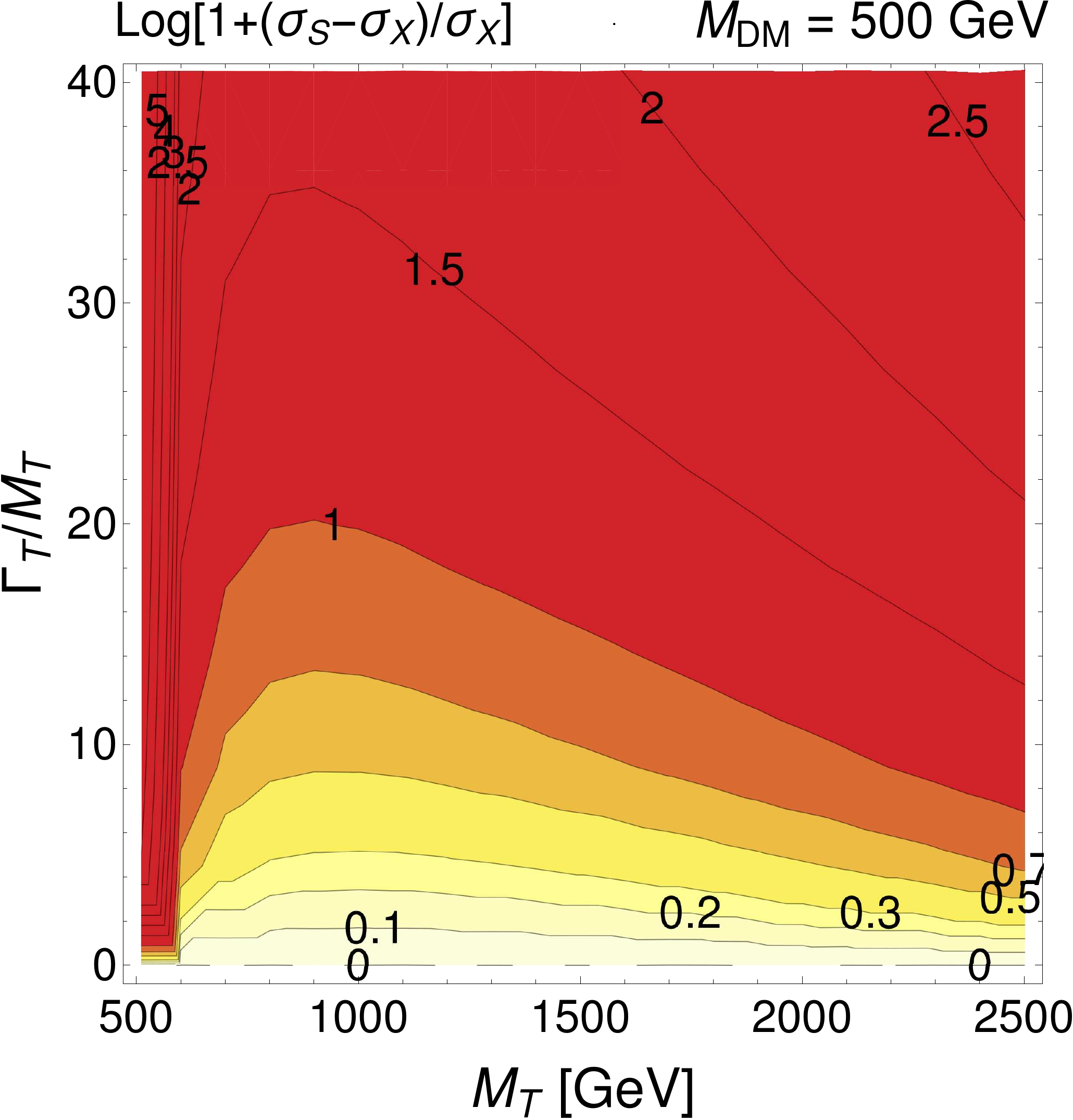} 
\includegraphics[width=.32\textwidth]{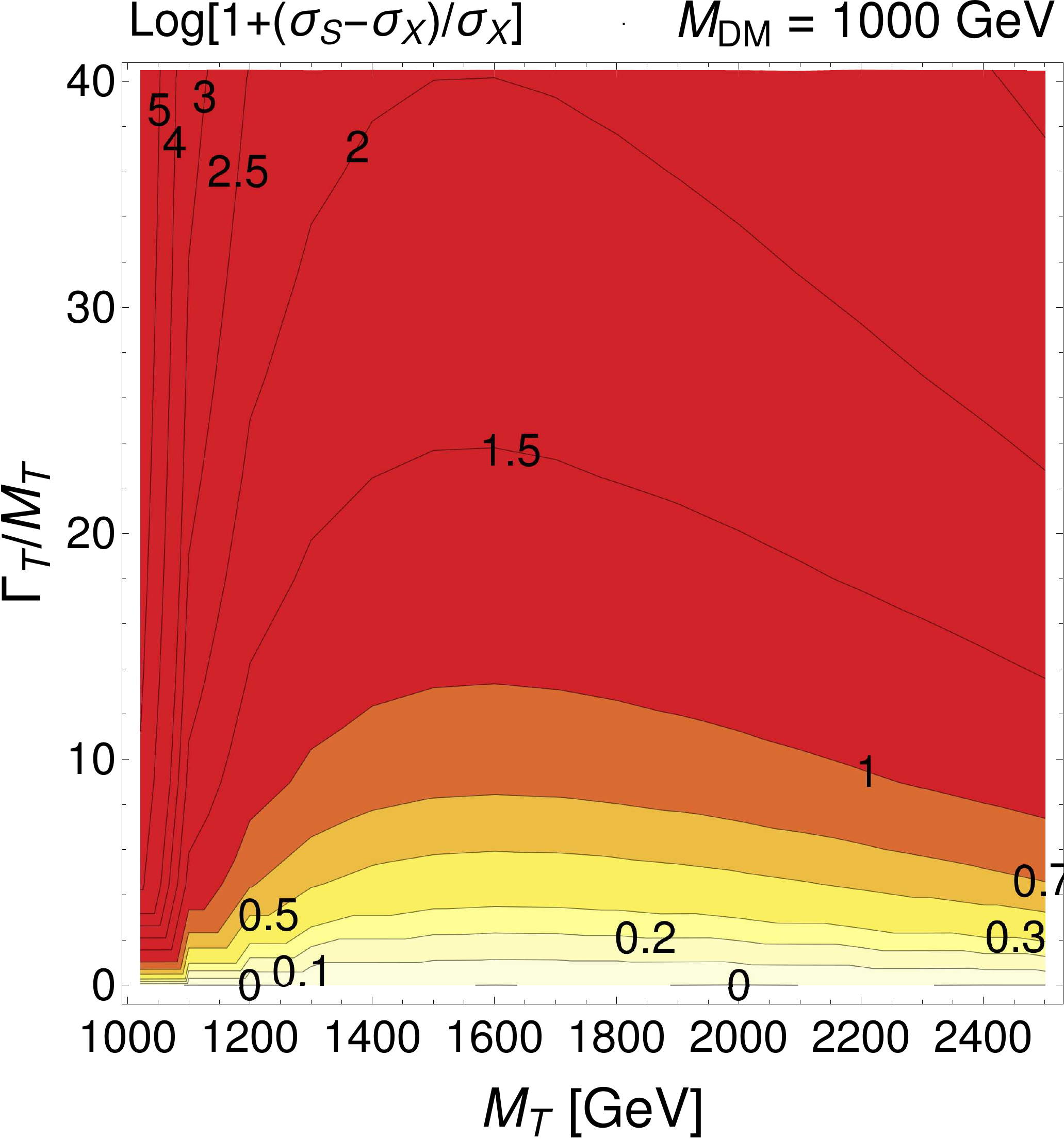} 
\caption{\label{fig:SXfirst}Relative difference between the full signal and the QCD pair production cross sections for a $T$ coupling to a DM particle (coupling to first generation) of mass 10 GeV, 500 GeV and 1000 GeV. Due to the large differences between cross sections, the ratio is plotted as $\log [1+(\sigma_S - \sigma_X)/\sigma_X]$ instead of $(\sigma_S - \sigma_X)/\sigma_X$. Notice that in that case the contours at 0.1, 0.2, 0.3, 0.5 and 1 respectively correspond to a value of $(\sigma_S - \sigma_X)/\sigma_X$ equal to 26\%, 58\%, 100\%, 216\% and 900\%. Top row: scalar DM; bottom row, vector DM.}
\end{figure}

These topologies contain collinear divergences, due to the gluon splitting, which drastically enhance the full signal cross section with respect to QCD pair-production. In Fig.~\ref{fig:SXfirst} the logarithm of the relative differences between the full signal cross section and the QCD pair production cross section are plotted for an LHC energy of 13 TeV. Notice that to allow a consistent comparison with the NWA case no cuts have been applied on the light jet at parton level. 

The main conclusions which can be derived from our results are the following:
\begin{itemize}
\item In the NWA the full signal and the QCD pair production topologies become equivalent, as expected. The latter topologies describe the process in an excellent way in the NWA, as subleading topologies and off-shell contributions are indeed negligible.
\item The contributions of new topologies and of off-shell $T$ become more and more relevant as the width of the $T$ increases, quickly becoming extremely relevant for the determination of the cross section, especially when the mass of the XQ and of the DM particle are close. 
\item The cancellation of effects which makes the $\sigma_S$ similar to $\sigma_X$ as in the case of coupling to third generation is not observed in this case. However, a minimum of the cross section ratio (for fixed $\Gamma_T/M_T$) appears for all value of the DM mass and spin in regions that are very similar to the cancellation region observed in section \ref{sec:Parton3}. This decrease is due again to a different scaling of the phase space in the NWA and large width regimes, but due to the additional diagrams in the case of coupling with first generation, the cancellation only lowers the cross section ratio and does not bring it to zero as it was the case for third generation coupling.
\end{itemize}

%%%%%%%%%%%%%%%%%%%%%%%%%%%%%%%%%%%%%%%%%%%%%%%%%%%%%%%%%%%%%%%%%%%%%%%%%

\subsection{Large width effects at detector level}

\begin{figure}[ht!]
\centering
\includegraphics[width=.32\textwidth]{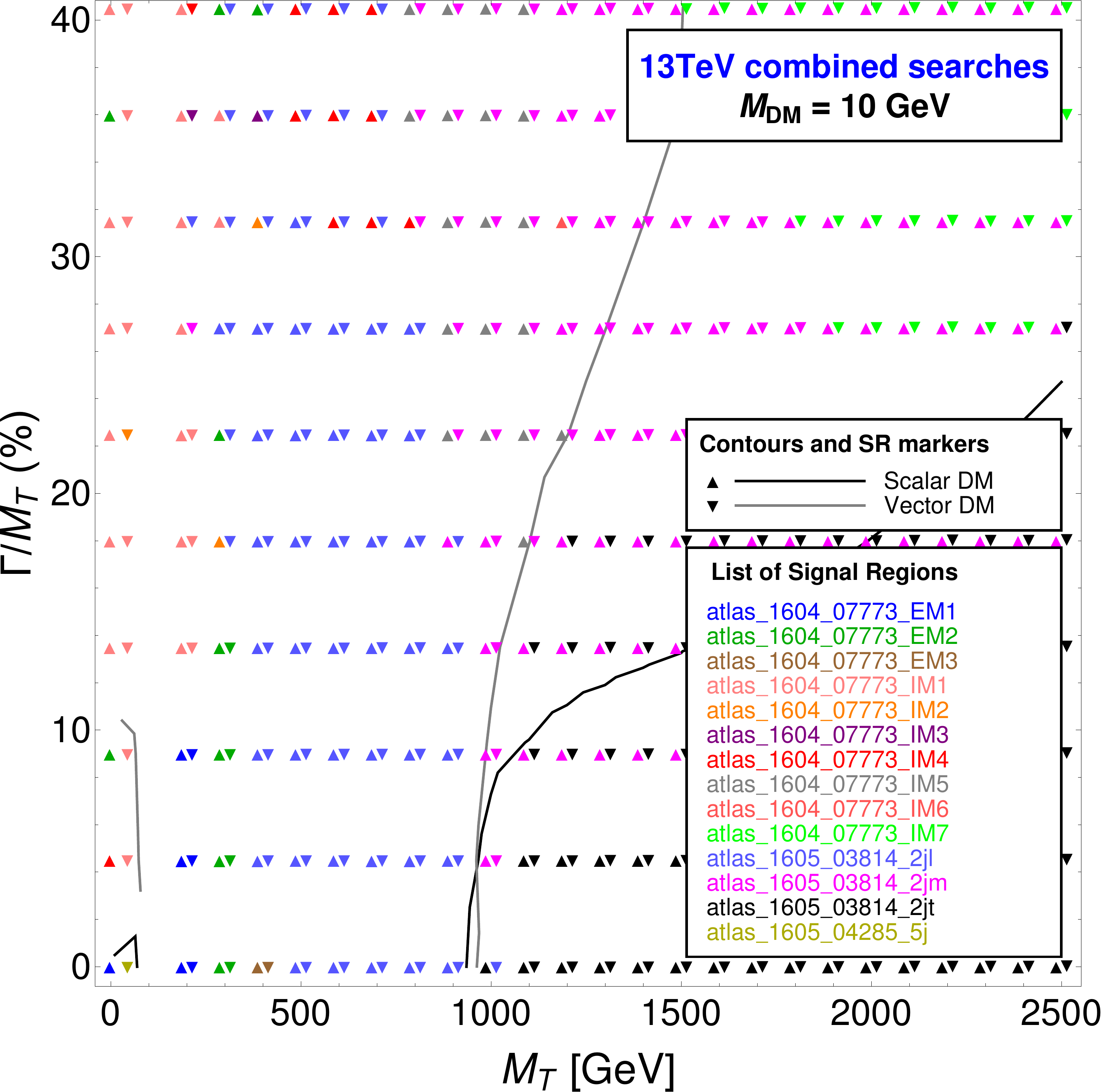} 
\includegraphics[width=.32\textwidth]{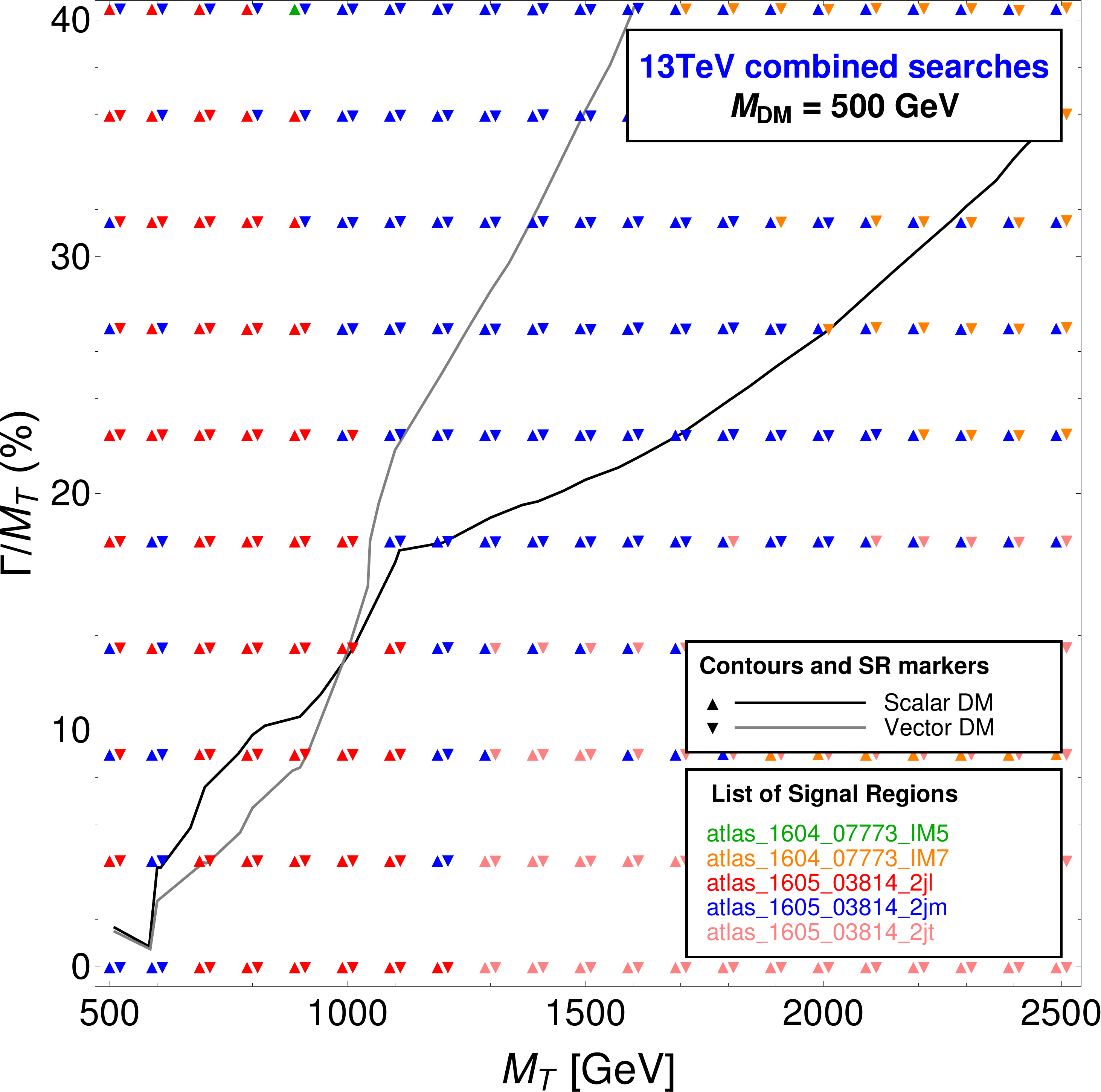} 
\includegraphics[width=.32\textwidth]{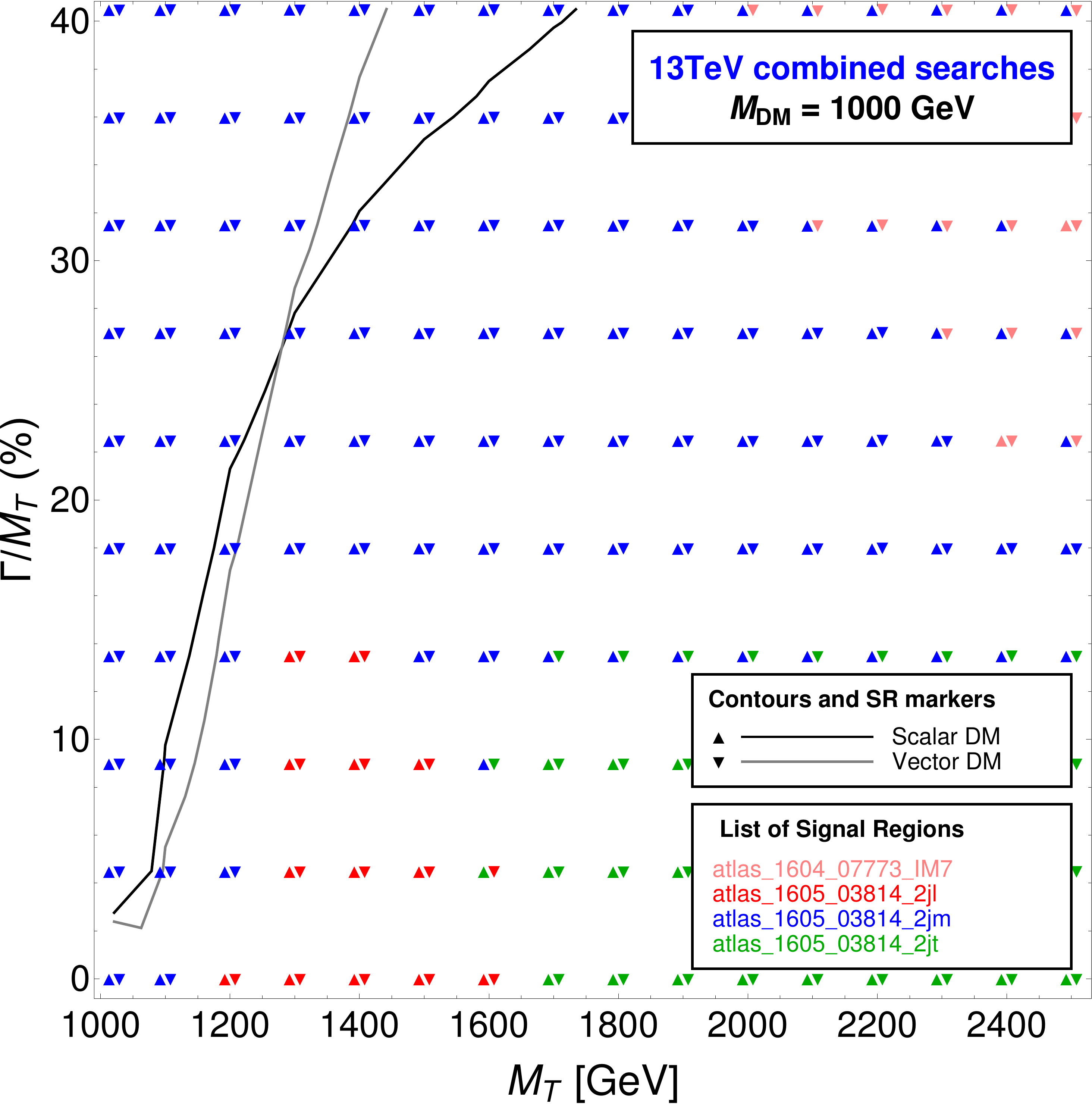} 
\caption{\label{fig:Exclusion1}{\sc CheckMATE} results for a $T$ coupling to a DM particle (coupling to first generation) of mass 10 GeV, 500 GeV and 1000 GeV. The black (grey) line shows which part of the parameter space is excluded in the scalar (vector) DM scenario.}
\end{figure}

\begin{figure}[ht!]
\centering
\includegraphics[width=.32\textwidth]{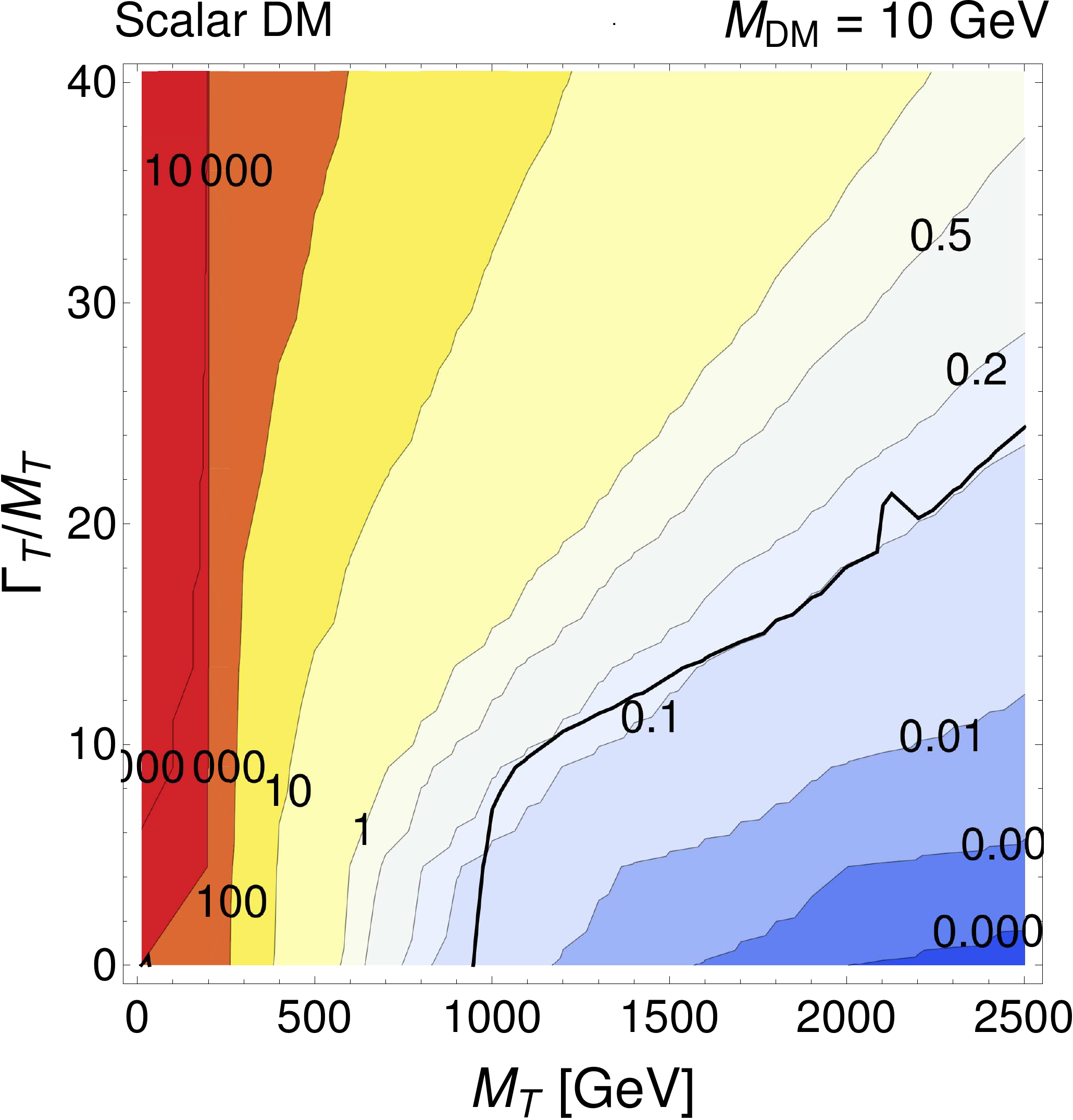} 
\includegraphics[width=.32\textwidth]{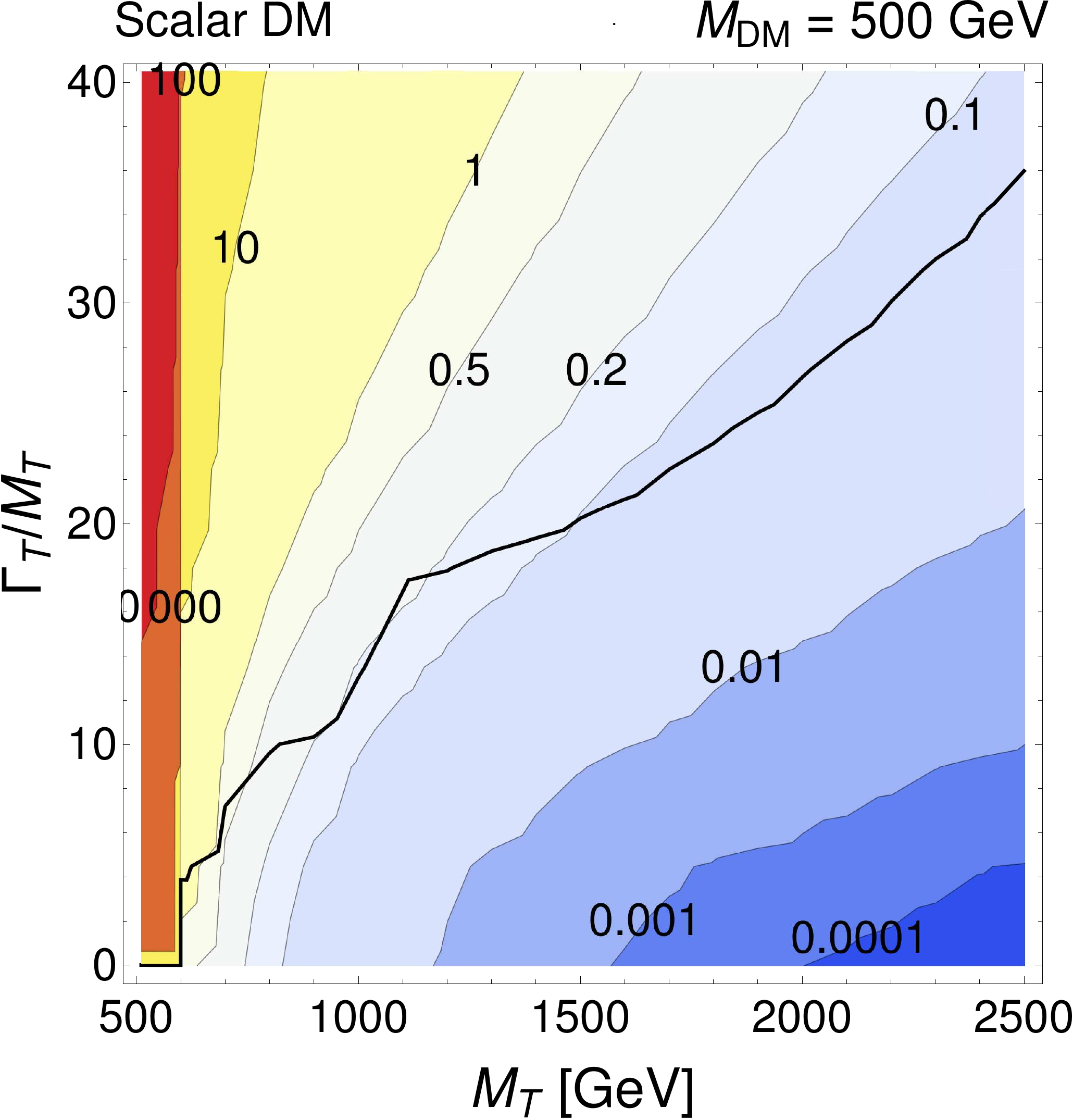} 
\includegraphics[width=.32\textwidth]{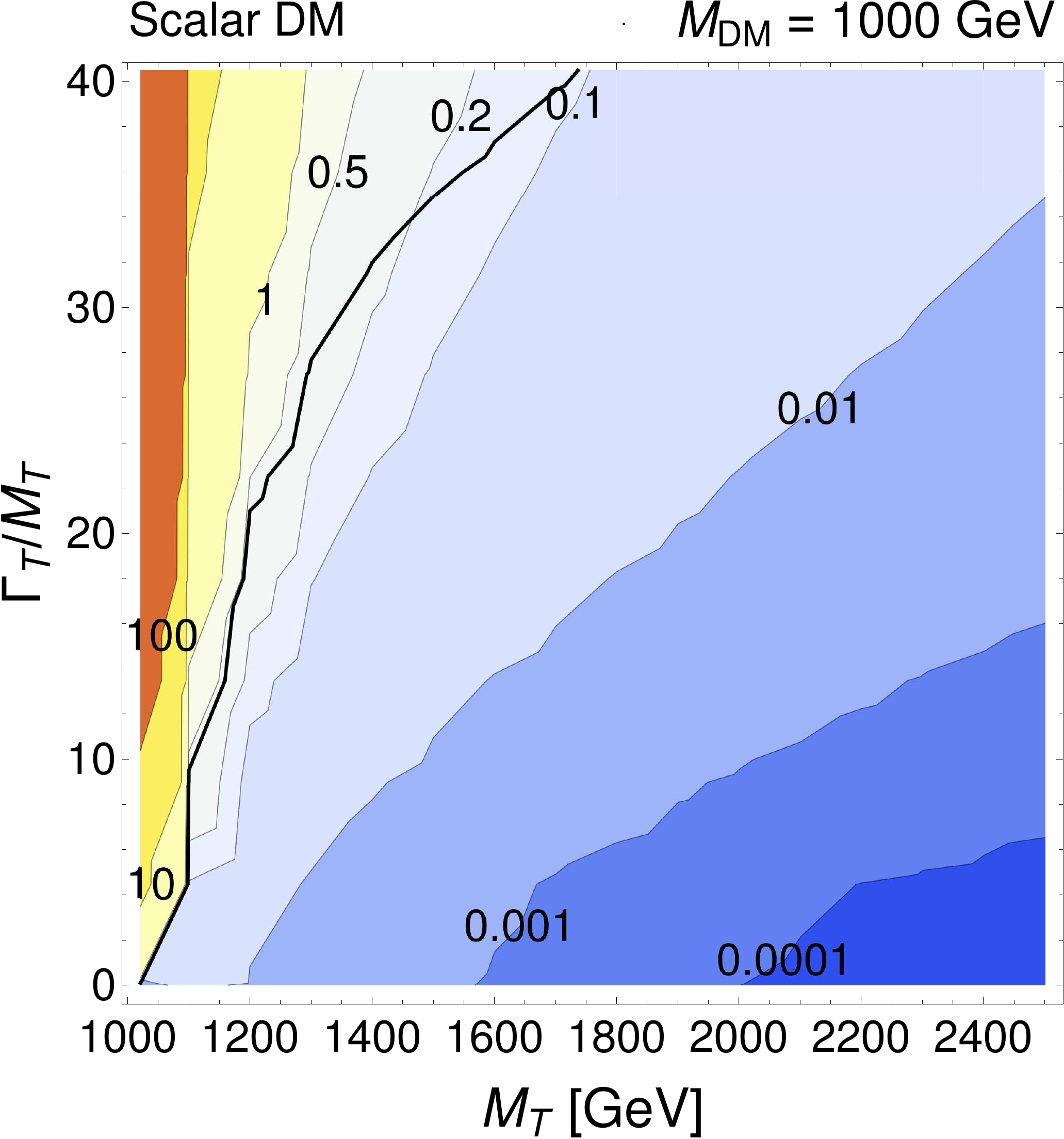}\\ 
\includegraphics[width=.32\textwidth]{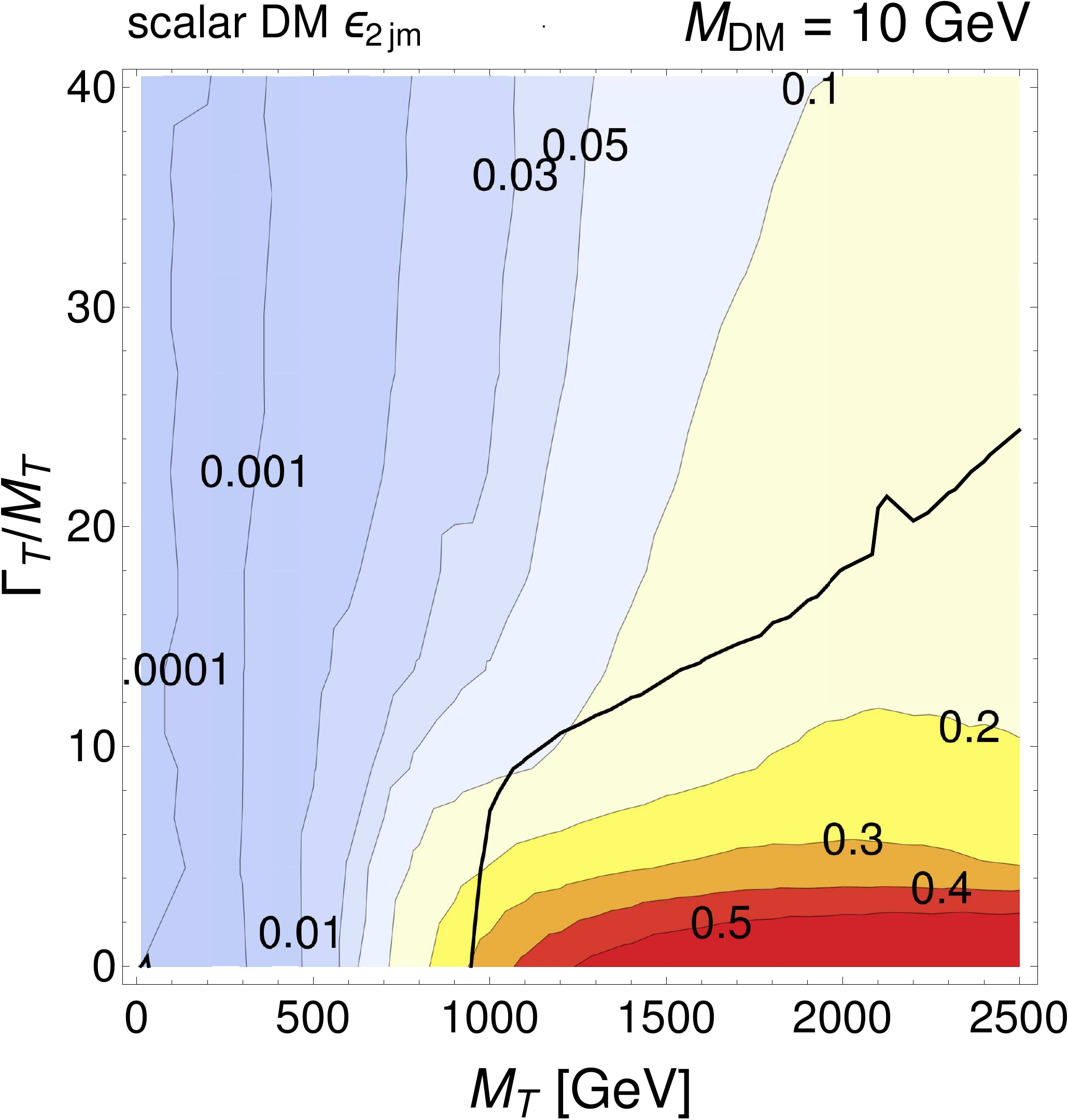} 
\includegraphics[width=.32\textwidth]{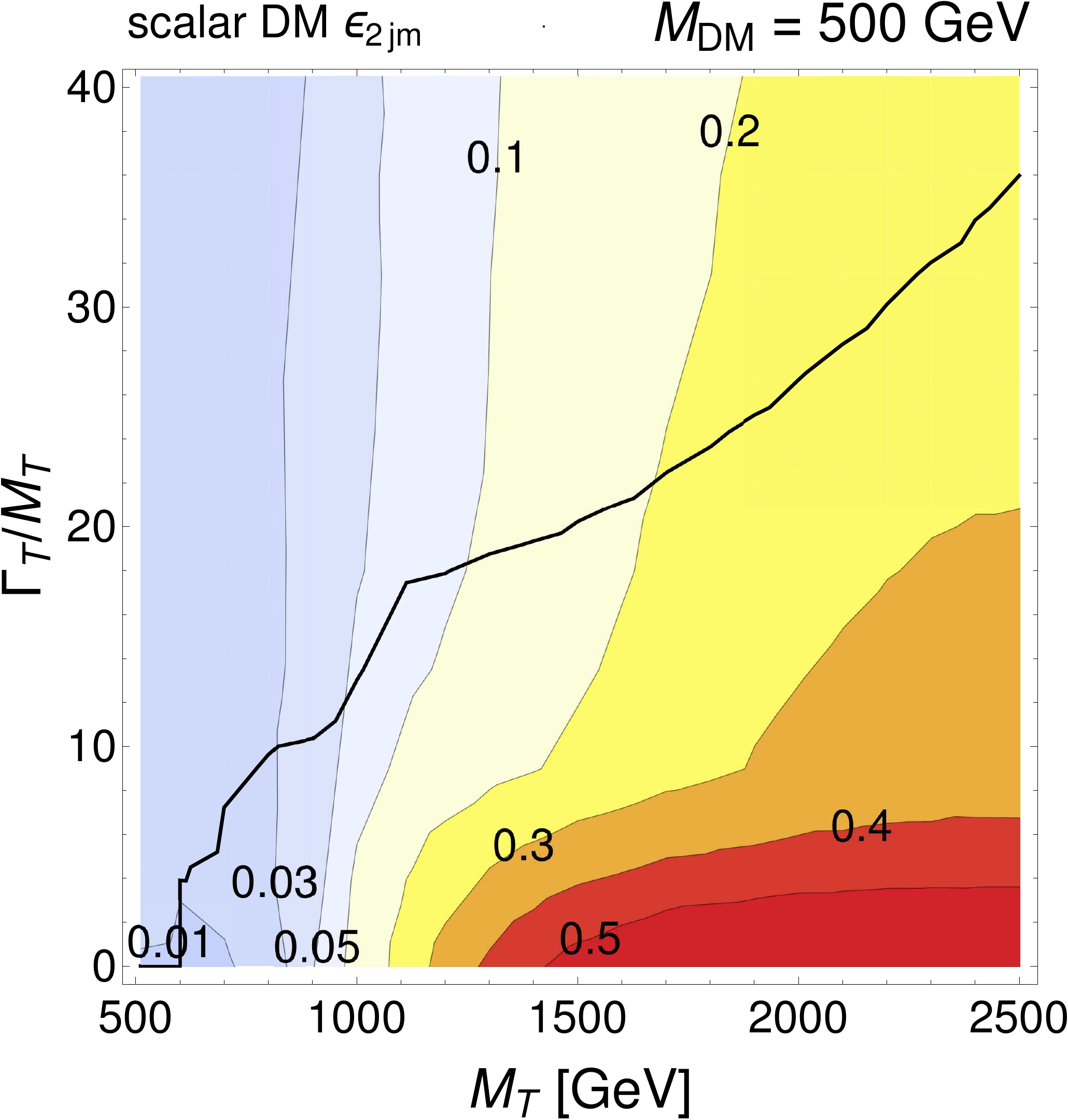} 
\includegraphics[width=.32\textwidth]{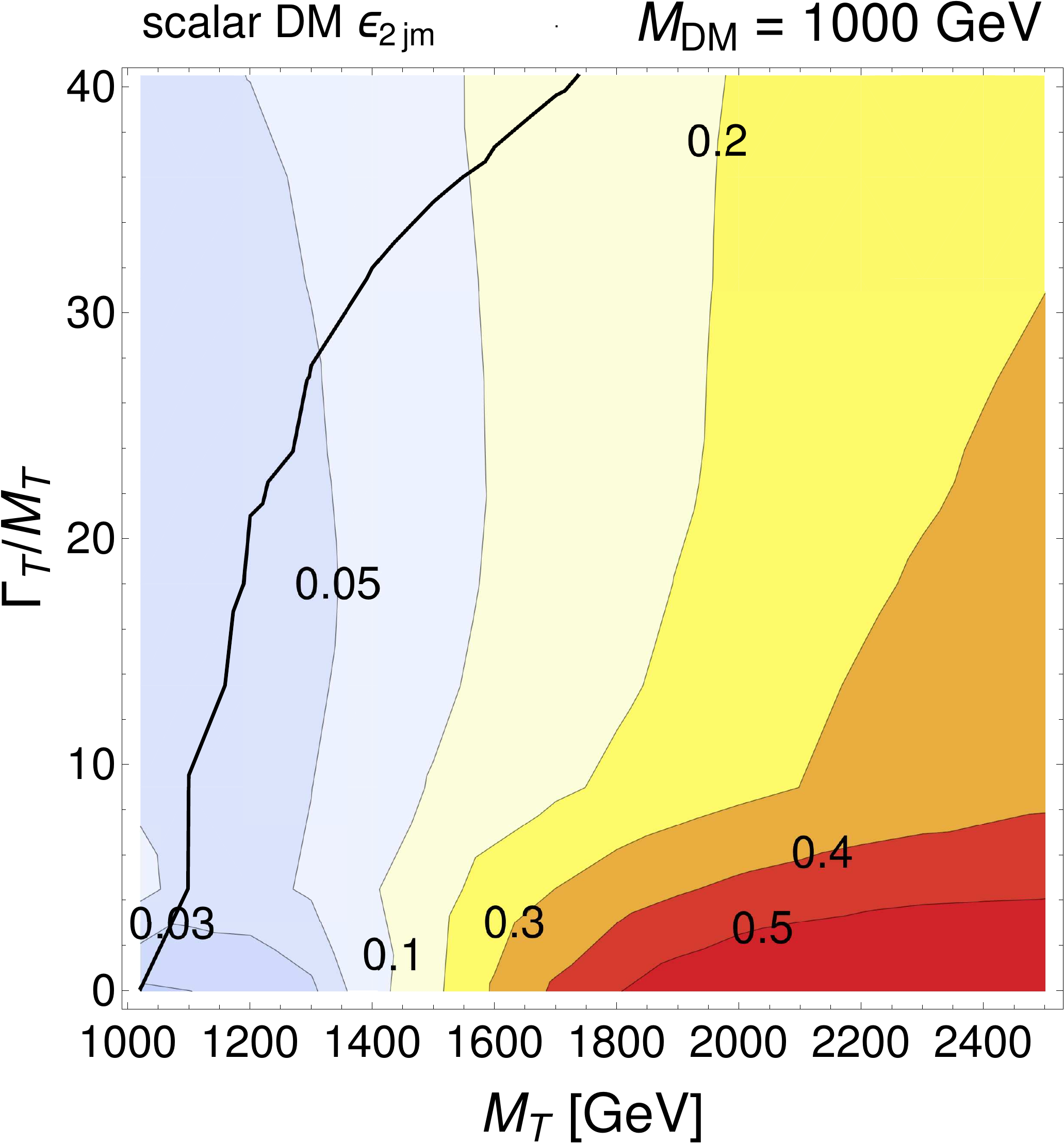}
\caption{\label{fig:sigmaEffs1} Top row: full signal cross sections for the scalar DM case. Bottom row: efficiencies of the SR 2jm from the ATLAS search~\cite{Aaboud:2016zdn} for different scalar DM masses.}
\end{figure}

\begin{figure}[ht!]
\centering
\includegraphics[width=.32\textwidth]{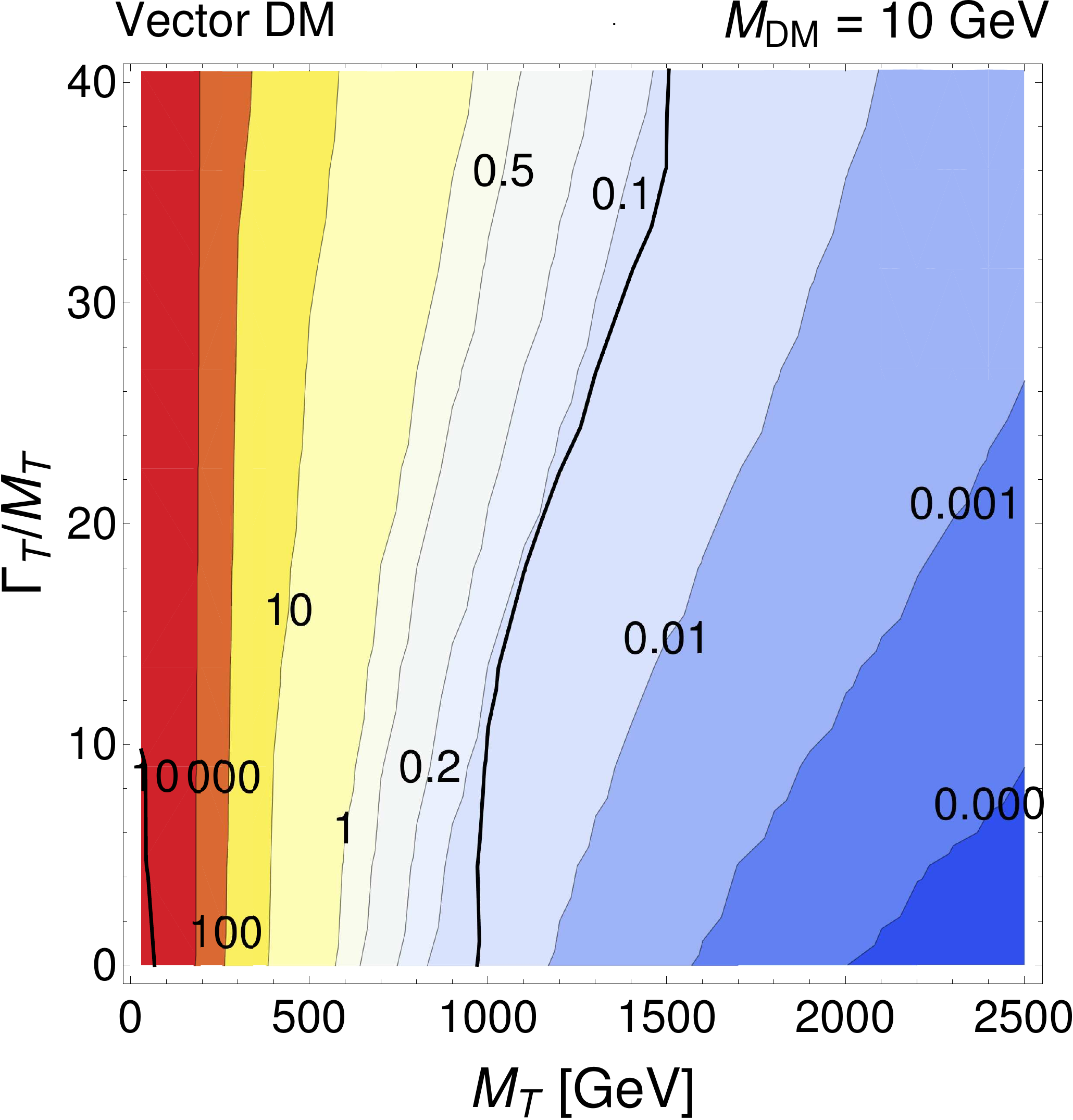} 
\includegraphics[width=.32\textwidth]{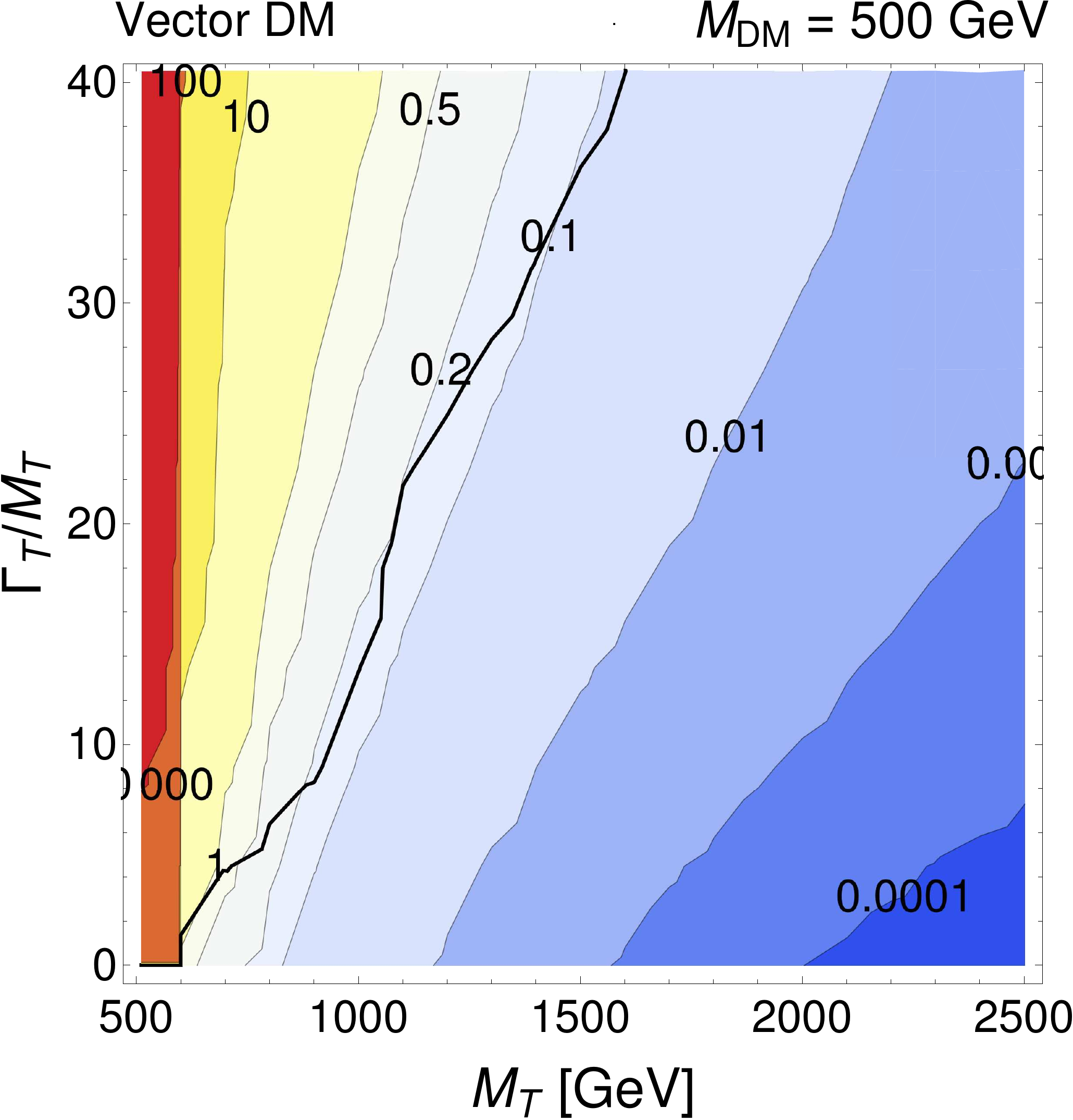} 
\includegraphics[width=.32\textwidth]{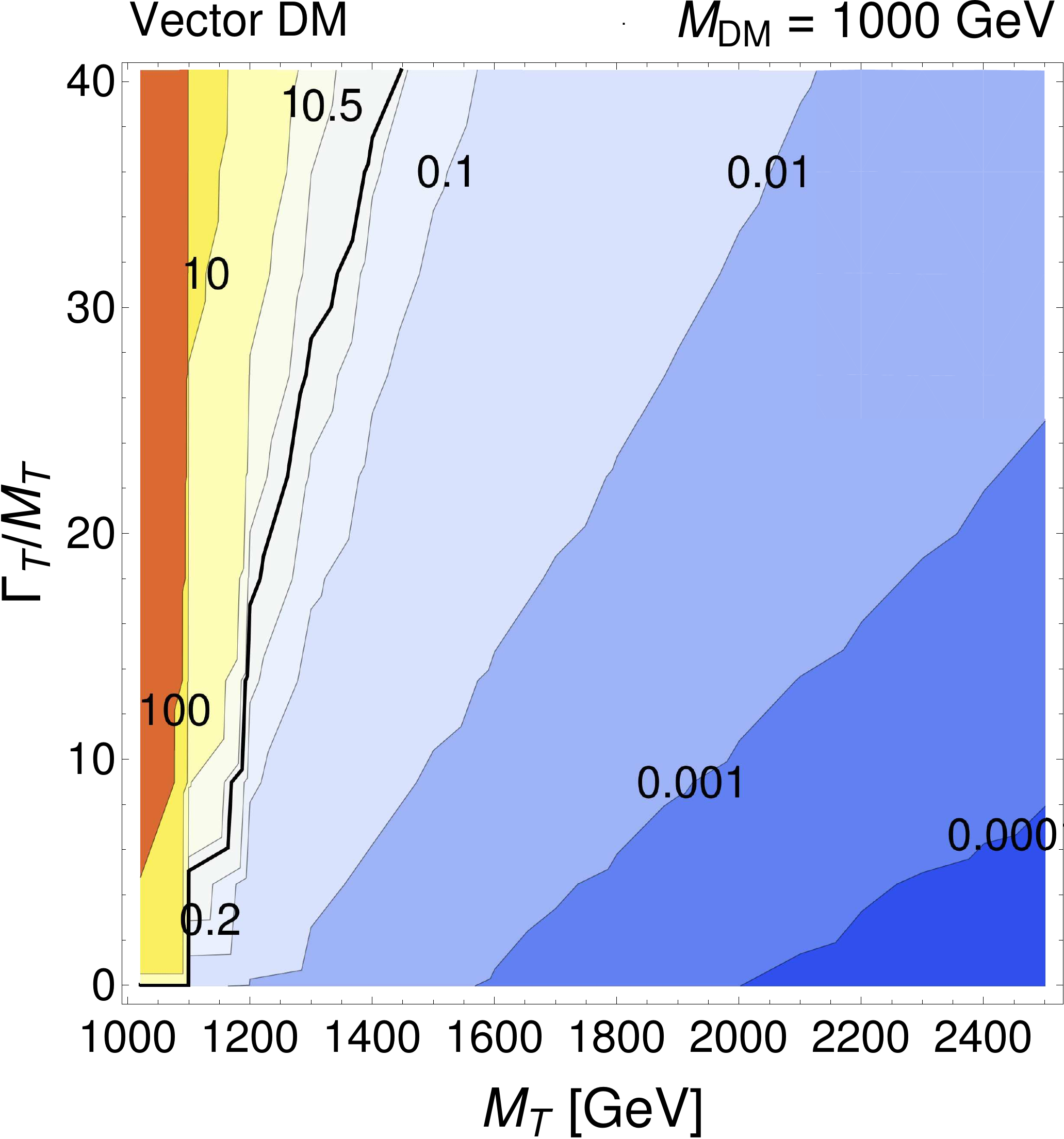}\\ 
\includegraphics[width=.32\textwidth]{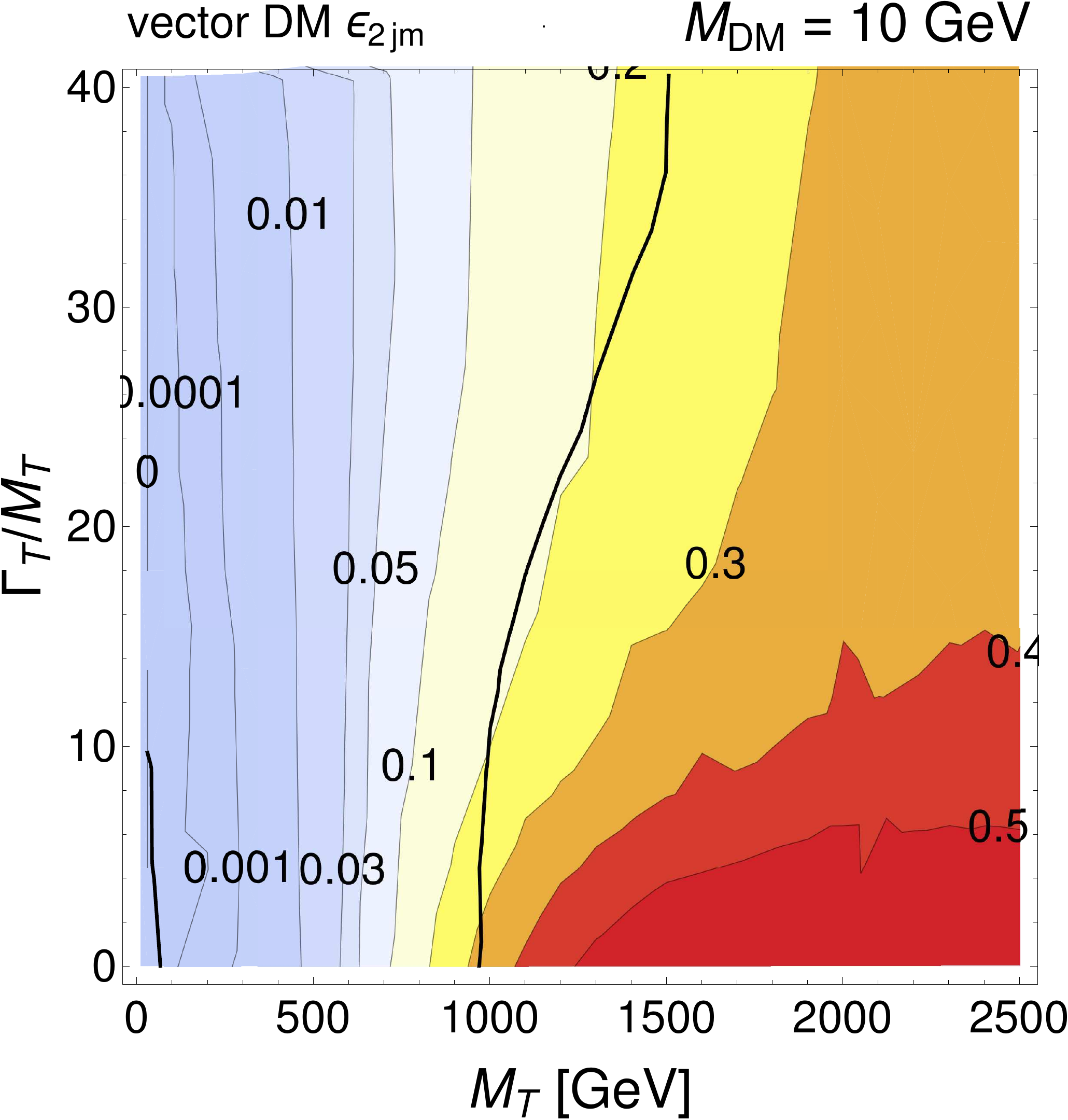} 
\includegraphics[width=.32\textwidth]{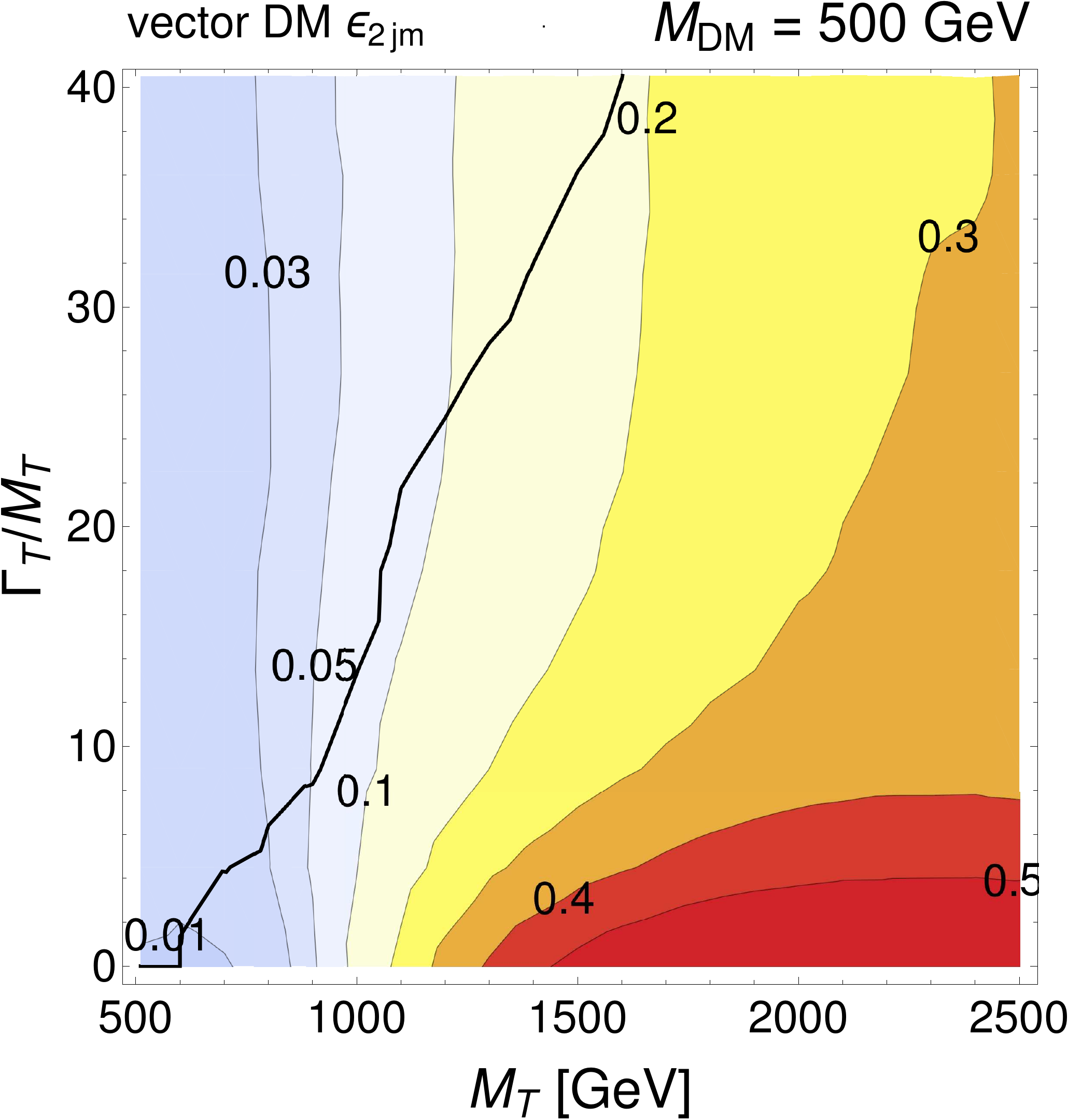} 
\includegraphics[width=.32\textwidth]{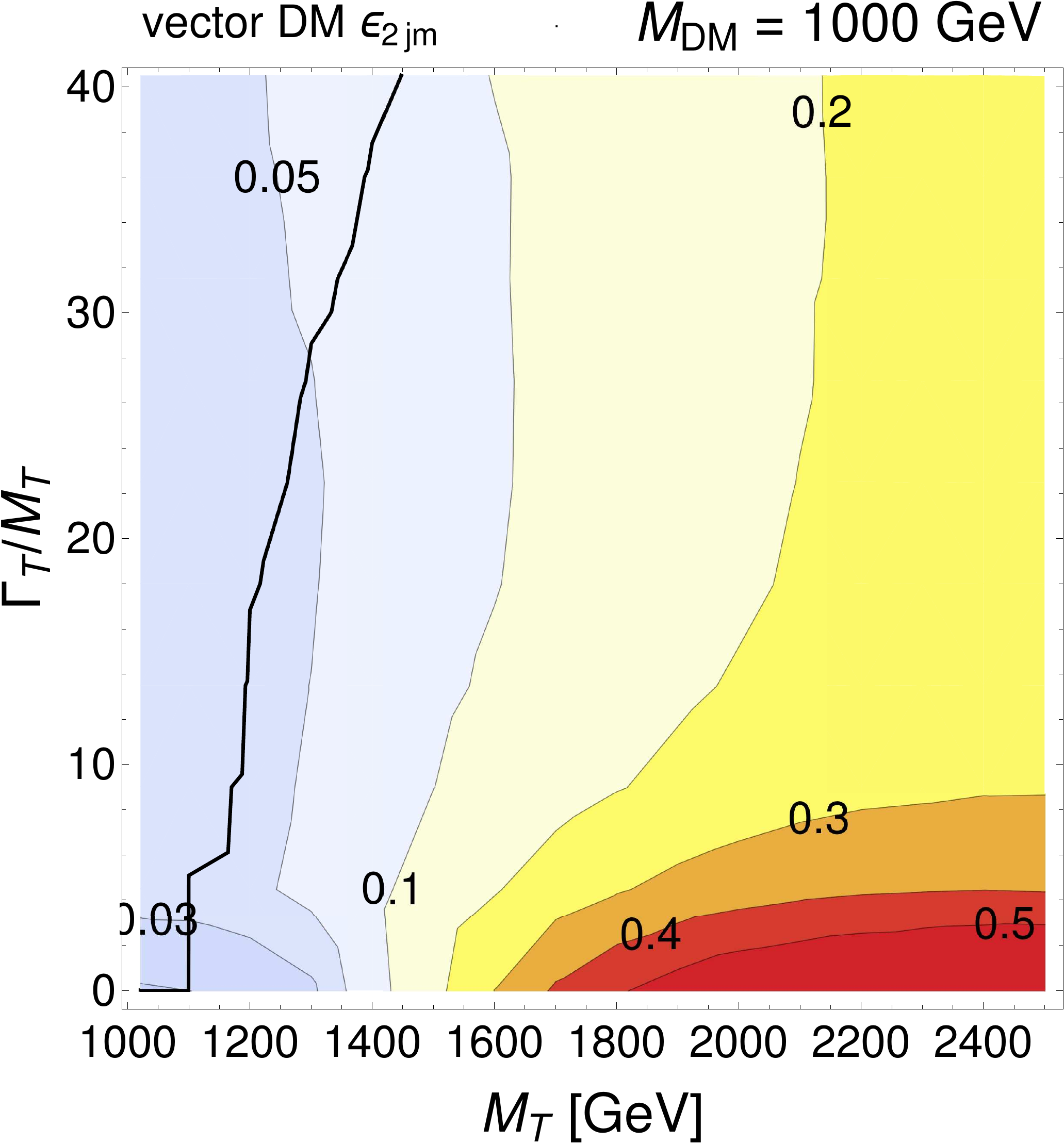}
\caption{\label{fig:sigmaEffv1} Top row: full signal cross sections for the vector DM case. Bottom row: efficiencies of the SR 2jm from the ATLAS search~\cite{Aaboud:2016zdn} for different scalar DM masses.}
\end{figure}

In Fig~\ref{fig:Exclusion1} the exclusion bound and the best SR are shown in the $(M_T, \Gamma_T / M_T)$ plane for both scalar and vector DM scenarios and for the same value of the DM mass considered in Fig.\ref{fig:SXfirst}. In Figs.~\ref{fig:sigmaEffs1} and \ref{fig:sigmaEffv1} the exclusion bounds for scalar and vector DM respectively are shown together with the full signal cross sections and with the efficiencies of the most relevant signal regions for the two DM spin hypotheses.

The main results for the case of $T$ coupling to first generation quarks are the following.
\begin{itemize}
\item For DM masses below to the TeV the bounds have a qualiltatively similar behaviour, the width dependence is always sizable, the bounds for small width are similar between scalar and vector DM and as the width increases the different DM spins exhibit different behavours, where scalar DM scenarios show a stronger dependence on the $T$ width. 
\item The most sensitive SRs for the determination of the bounds are almost always 2jl, 2jm or 2jt of the ATLAS search \cite{Aaboud:2016zdn}, which are optimised for signals with two jets and $\MET$ in the final state. 
\item For DM masses around the TeV or higher the width dependence of the bound is still present but the difference between the scalar and the vector DM scenarios becomes weaker. Furthermore, the NWA region is never excluded. Analogously to the case of coupling with third generation, this is a consequence of a combination between larger phase space and width dependence of the experimental acceptances.
\end{itemize}

% \clearpage

%\vspace{\baselineskip}
\subsubsection{Dependence on the chirality of the couplings}

% \ \vspace{\baselineskip}

\begin{figure}[ht!]
\centering
\includegraphics[width=.45\textwidth]{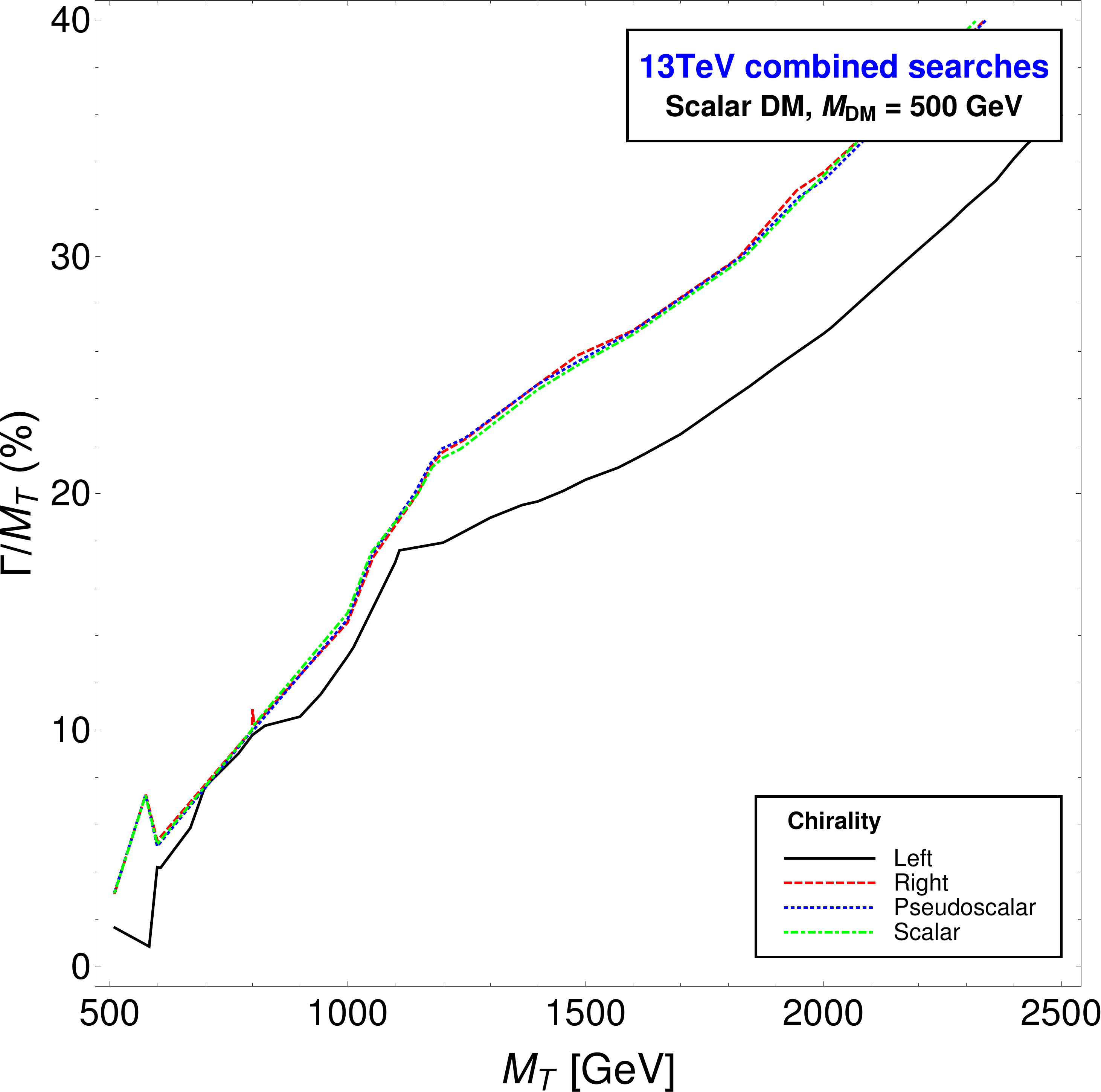} 
\includegraphics[width=.45\textwidth]{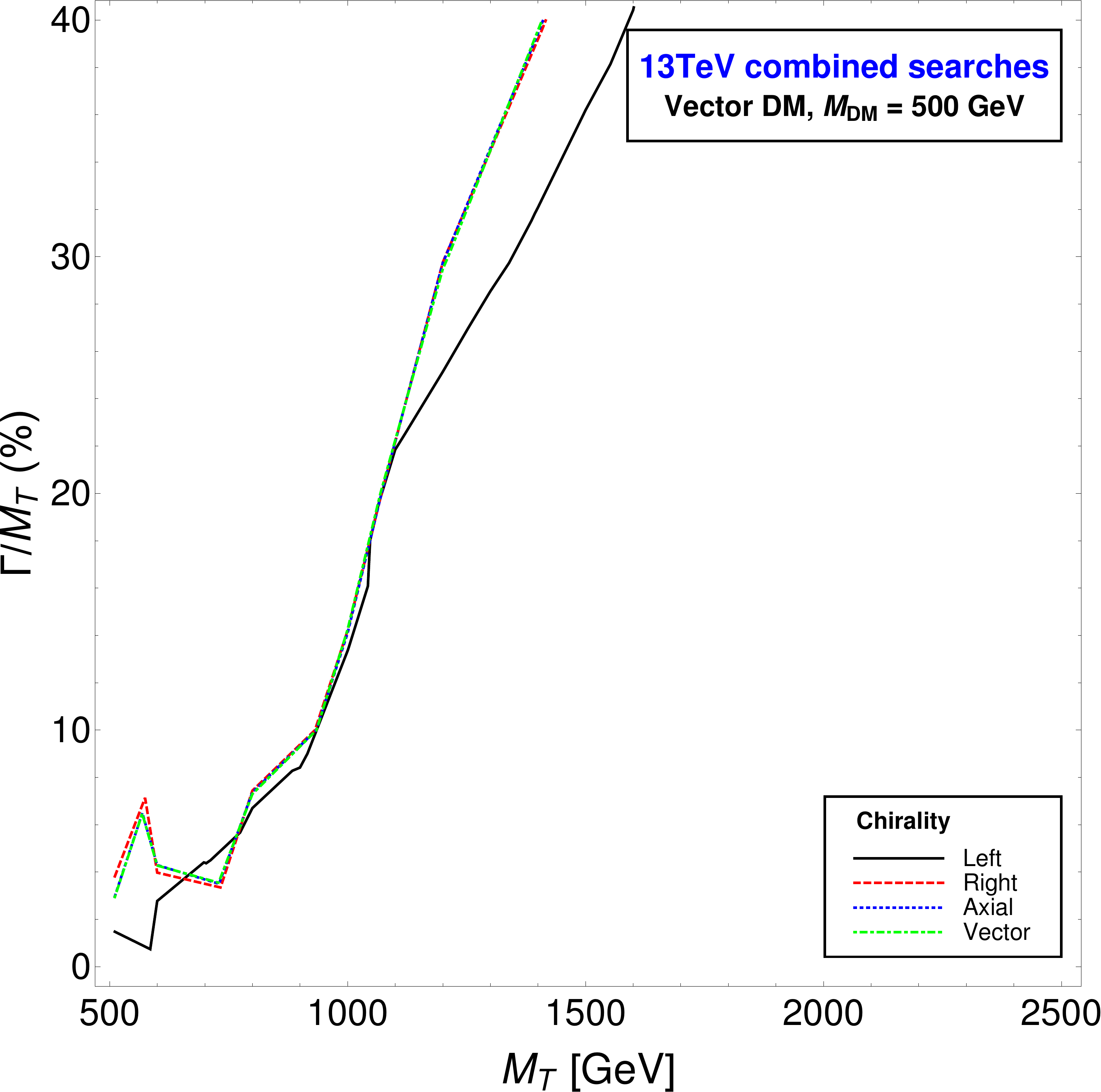} 
\caption{\label{fig:1Gchirality}Exclusion bounds for a $T$ interacting with the SM up quark and DM for different hypotheses on the chirality of the couplings: for a VLQ $T$ pure left-handed and pure right-handed couplings, and for a ChQ $T$ pure scalar (vector) or pseudoscalar (axial-vector) couplings if $T$ interacts with scalar (vector) DM.}
\end{figure}

Analogously to the case of $T$ coupling with third generation quarks, the analysis of the dependence of the limits on the chirality of the couplings (and therefore on the hypotheses about the properties and representations of $T$) is presented. In Fig.~\ref{fig:1Gchirality} the exclusion bounds for different couplings are shown. Once again even if the uncertainty due to the use of a recasting tool is quite large, we observe that the scenario with pure left-handed coupling exhibits a slightly stronger width dependence than the rest of the scenarios in the large width regime, especially for the scalar DM scenario. This can be due again to the milder dependence of the transverse components of vector DM on the chirality of the couplings, which is however not completely diluted, as scenarios with left-handed couplings and vector DM still exhibit a slight stronger width dependence. Even if the bounds are in the same regions, the most sensitive SRs of (the subset of) current searches could be in principle used to distinguish the scenario where the $T$ is a VLQ doublet from the others, in case of discovery. We are not going, however, to explore this potentiality in the present study, as it goes beyond the scope of our analysis, and we defer it to a subsequent one.

%%%%%%%%%%%%%%%%%%%%%%%%%%%%%%%%%%%%%%%%%%%%%%%%%%%%%%%%%%%%%%%%%%%%%%%%%
%%%%%%%%%%%%%%%%%%%%%%%%%%%%%%%%%%%%%%%%%%%%%%%%%%%%%%%%%%%%%%%%%%%%%%%%%

\clearpage

\section{Exclusion limits in the $M_T-M_{DM}$ plane}
\label{sec:BellPlots}

\begin{figure}[ht!]
\centering
\includegraphics[width=.32\textwidth]{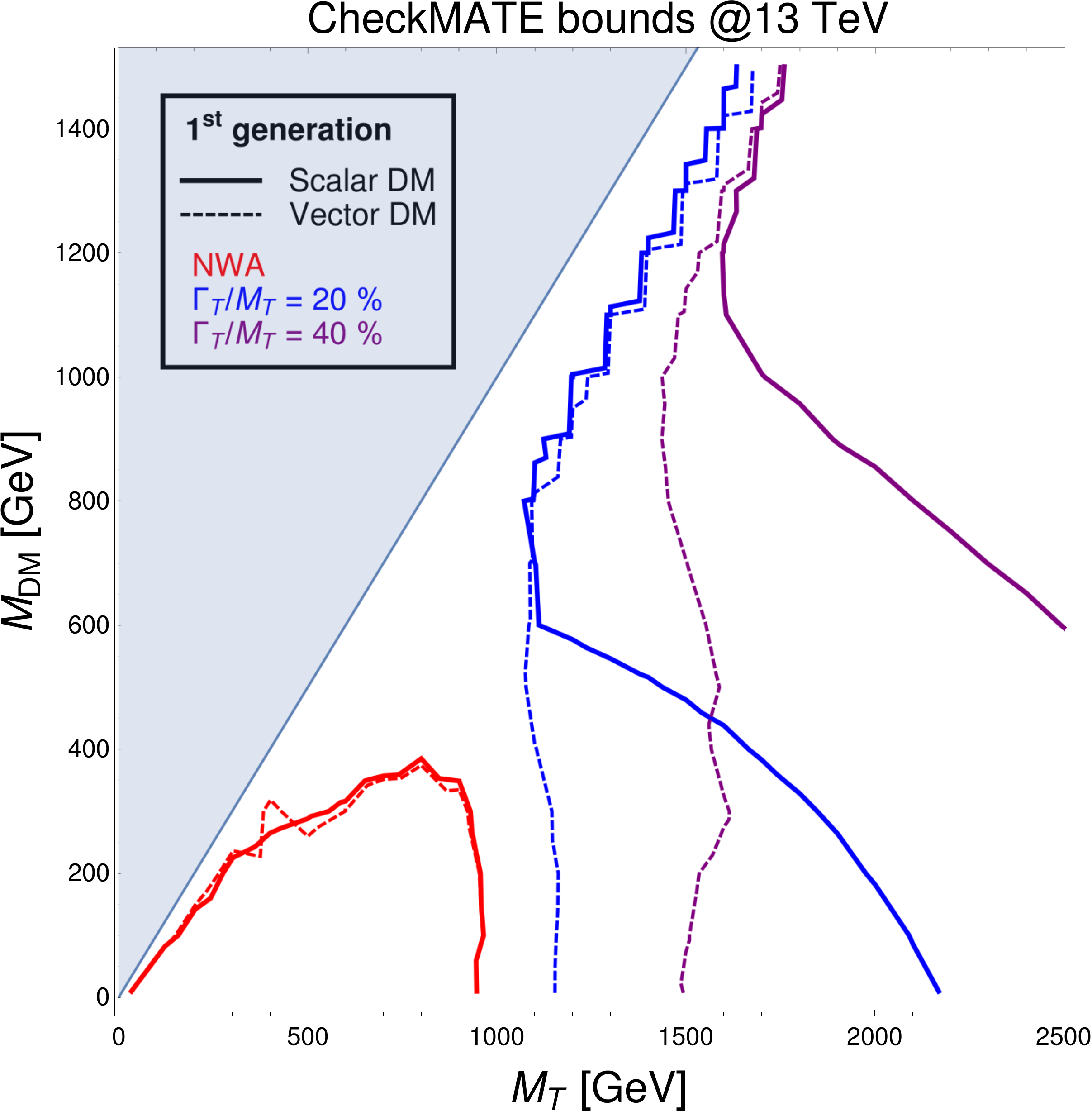} 
\includegraphics[width=.32\textwidth]{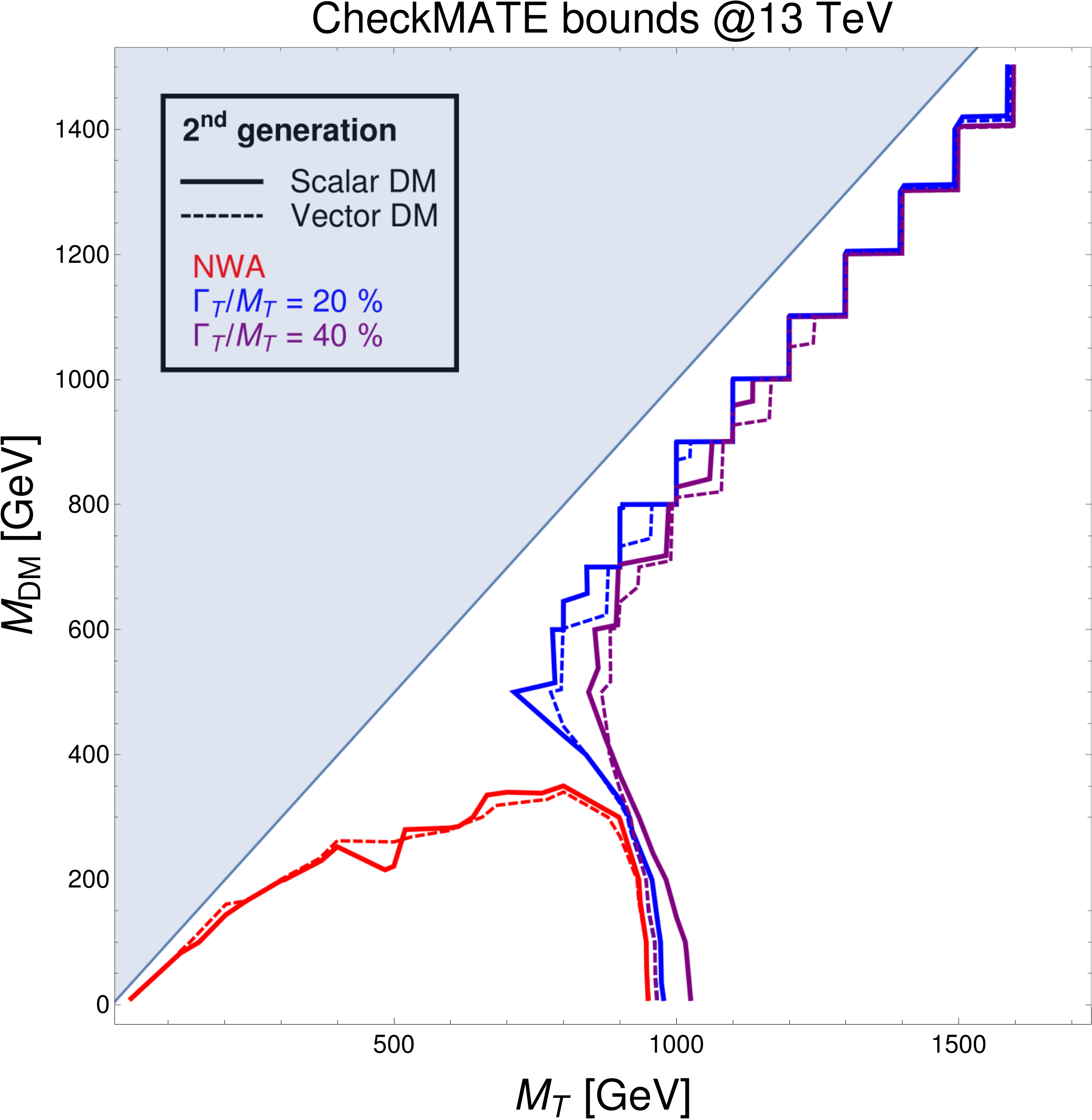} 
\includegraphics[width=.32\textwidth]{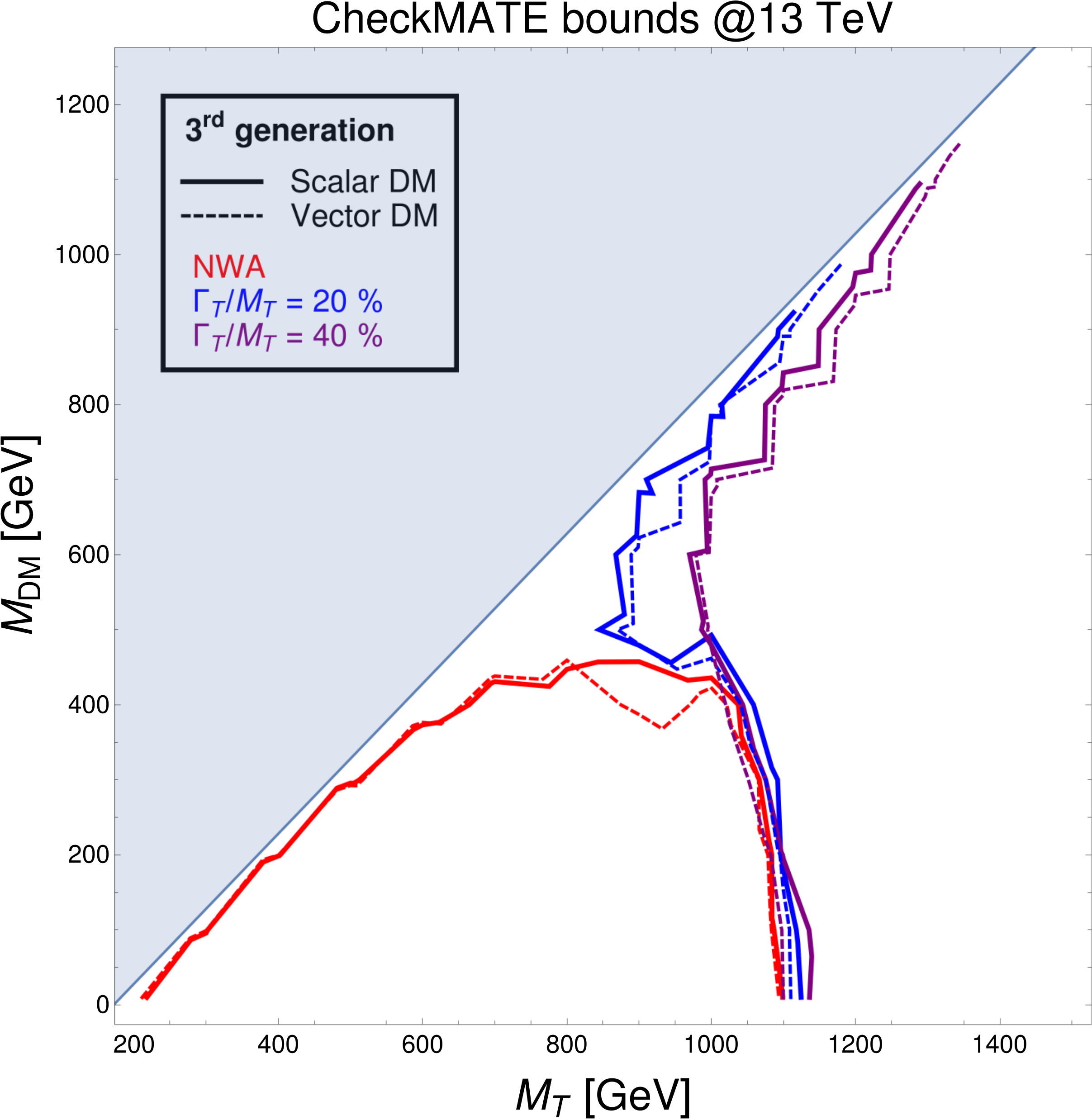} 
\caption{\label{fig:BellCombined}Bounds in the $(M_T, M_{\rm DM})$ plane for $T$ quark coupling DM particle and first (left panel), second (center panel) and third (right panel) generations of SM quarks for different values of $\Gamma_T / M_T$.}
\end{figure}

The scenarios we are considering have three parameters: the mass of the $T$, the width of the $T$ and the mass of the DM, with the only constraints given by the kinematical limit between the masses ($M_T > M_{\rm DM} + m_q$) and by the fact that the width should not really exceed 50\% of the mass, otherwise the concept of resonant state is essentially lost. The exclusion bound at 2$\sigma$ will therefore identify a 3D surface in the space defined by the three parameters (where the width is substituted by the $\Gamma_T/M_T$ ratio) and therefore it is instructive to analyse the projections of this surface on the plane identified by the masses of $T$ and DM for different values of the $\Gamma_T/M_T$ ratio. Such representation is also useful to directly compare bounds on $T$ and bosonic DM with analogous results in other models, such as SUSY. Indeed the exclusion limits of SUSY searches are often presented in the $(M_{\tilde{t}}, M_{\chi_0})$ plane. We show in Fig.~\ref{fig:BellCombined} the bounds in the $(M_T, M_{\rm DM})$ plane for specific values of $\Gamma_T / M_T$: the NWA case, 20\% and 40\%. We included in this figure the results for a $T$ quark coupling to DM and the charm quark. 

The qualitative behaviours of the exclusion limits strongly depend on the assumption about which SM quark generation the $T$ couples to.

\begin{itemize}

\item {\bf $T$ coupling to DM and up quark}: in the NWA the exclusion limits for scalar and vector DM are not distinguishable in practice (barring numerical fluctuations). When the width of the $T$ increases, however, the bounds for scalar and vector exhibit a sizably different dependence on the $T$ and DM masses. If the DM mass is below a width-dependent threshold, the scalar DM case excludes a much wider region of the parameter space. This behaviour can be understood by looking again at Figs.~\ref{fig:sigmaEffs1} and \ref{fig:sigmaEffv1}, which show that the full signal cross section has a largely different trend with changing width depending on the scalar or vector nature of the DM. For high enough DM masses, the dependence on the width is less pronounced and this erases the differences between the bounds above a certain value of the DM mass. A further peculiarity of the large width regime, with respect to the NWA, is that the region where the mass gap between $T$ and DM is small is always excluded.

\item {\bf $T$ coupling to DM and charm quark}: in both NWA and large width regime it is not possible to distinguish scalar from vector scenarios. As the width increases, the region close to the kinematics limit ($M_T=M_{\rm DM}+m_c$) becomes excluded, while it would be allowed in the NWA. If the DM mass is below 300 GeV and far from the kinematics limit, the bound depends very weakly on the width.

\item {\bf $T$ coupling to DM and top quark}: the mass bounds for scalar and vector DM are very similar in both the NWA and large width regime. The increase of the width modifies the bound (with respect to the NWA) if the mass of $T$ is close enough to the kinematics limit ($M_{\rm DM}+m_t$): unlike in the NWA case, as the values of the $T$ mass approaches the kinematics limit, they become more and more excluded by experimental data as the $T$ width increases. Moreover, if the DM mass is below $\sim 400$ GeV and far from the kinematics limit, the bound on the $T$ mass does not depend on the width. Designing new specific cuts could allow a more optimised exploration of the large width regime of XQs decaying into DM and third generation SM quarks, especially considering the fact that efficiencies for the most sensitive SRs exhibit a general decrease along the bound region as the width increases (as shown in Fig.~\ref{fig:sigmaEffs3}). \\

\end{itemize}

To conclude this section, the bounds obtained under the NWA are less stringent than the bounds obtained when the NWA is relaxed and the width is allowed to have large values, relative to the $T$ mass. This results can be intuitively expected when considering that larger widths correspond to larger cross sections and, unless the selection and cut efficiencies compensate the cross section enhancement, the number of signal events increases with respect to the NWA scenario. It is remarkable, though, that different assumptions about the couplings of $T$ with different SM quark generations produce either negligible or sizably different bounds if the DM is scalar or vector. This result could be exploited for the design of new experimental searches which are not only meant to discover new signals in channels with \MET\ but also to characterise the signal.

%%%%%%%%%%%%%%%%%%%%%%%%%%%%%%%%%%%%%%%%%%%%%%%%%%%%%%%%%%%%%%%%%%%%%%%%%
%%%%%%%%%%%%%%%%%%%%%%%%%%%%%%%%%%%%%%%%%%%%%%%%%%%%%%%%%%%%%%%%%%%%%%%%%

%\newpage

\section{Conclusions}
\label{sec:conclusions}

Experimental analyses aimed at assessing current LHC limits on the possible existence of XQs, or else at establishing the 
machine potential in unravelling them, rely upon a NWA emulation of the signal yield, the latter being limited to the contribution of QCD induced pair production of such new states of matter. While this approach is essentially model-independent, as the XQ BRs are the only physics observables carrying the model dependence, and further offers  one  the possibility of eventually enabling the interpretation of limits (or indeed evidences) in a variety of BSM scenarios, it suffers from the drawback that the accuracy of the ensuing results worsen considerably as the width of the XQ grows larger and/or all (gauge invariant) topologies relevant to the final state searched for (in addition to those pertaining to XQ pair production), all containing at least one XQ state (hence rightfully classifiable as signal), are accounted for. Conversely, the accurate emulation of such effects is necessarily model dependent, since the standard factorisation possible in NWA between QCD pair production topologies and decay BRs is no longer possible (i.e., model independent couplings now enter both the production and decay dynamics). 

Herein, we have estimated such effects in a rather simple model with only one XQ decaying into DM and a SM quark. The XQ was taken to have electric charge $+2/3$, with either a VLQ or ChQ nature. The DM candidate was assumed to be either a scalar or a vector state. Hence, necessarily, the SM quark is a $t$-quark in the case the XQ only couples to a third generation SM quark or a $u(c)$-quark in case it does with a first(second) generation one instead. In all such cases, we looked at the differences, initially at the parton level and eventually at the detector one as well (including cuts in this case), between the complete XQ process and the one only including $T\bar T$ hadron-production at the LHC Run 1 and 2, in the final states $t\bar t+E_T^{\rm miss}$ and $jj+E_T^{\rm miss}$, respectively.  

Upon choosing discrete values for the DM mass, of 10 GeV, 500 GeV, 1000 GeV and 1500 GeV, and an XQ of mass $M_T > M_{\rm DM} + m_q$, with $q \in \{u(c),t\}$ (such that its on-shell decay is kinematically allowed) up to $M_T^{\rm max}$ = 2500 GeV (essentially the ideal kinematics reach of the 13 TeV LHC for quark pair production), we have ascertained the following, as a function of
$\Gamma_T/M_T$ taken between 0 and 40\%. As a general result, we have concluded that the XQ nature, whether it be VLQ or ChQ, does not play a significant role in the phenomenology we have studied, primarily because one can be turned into the other by simply changing the left and right fermion couplings suitably and the observables normally adopted in experimental analyses do not resolve their relative size and/or sign. Furthermore, we have established that, for the same choice of $M_{\rm DM}$, there occur sizable differences between the two aforementioned approaches (NWA versus full result) depending on whether one adopts the scalar or vector nature of the DM candidate, the more so the larger the value of $\Gamma_T /M_T$. However, are the coupling properties of the $T$ state that are most responsible for the largest differences seen between the simplisitic (model-independent) and realistic (model-dependent) approaches outlined. 
On the one hand, when coupling is allowed to the third generation only, the exclusion limits depend only slightly upon $\Gamma_T/M_T$, with a general trend pointing towards the  
cross section becoming larger when the width increases, yet with the additional contributions with respect to the NWA being generally suppressed by the cuts on missing transverse energy normally adopted in experimental searches. On the other hand, when coupling is allowed to the
first(second) generation only, exclusion limits massively depend upon the width because the aforementioned additional topologies are not suppressed by such cuts in missing transverse energy, the more so the larger both $M_Q$ and $\Gamma_Q/M_Q$ are.
(In fact, differences between the DM nature are significantly more prominent in the case of coupling to first(second) generation than in the third generation one.) Clearly, a fully-fledged model incorporating coupling to any generation will fall in between these two extreme conditions, with further subtleties induced by the PDF behaviour, as one can already see by comparing our results for the first and second generation cases.

In conclusion then, results from LHC searches for any XQs, when decaying to DM (whether spin 0 or 1) and either a heavy or light SM quark, should be taken with caution, as they do not account for effects induced by either the large XQ width, the additional (to the pair production ones) topologies or both, which can be very large even in a simplified model with only one XQ. Hence, one should rescale the observed limits from established experimental analyses to the actual ones upon accounting for such effects (as we have done here) or else attempt  deploying new ones adopting different selection strategies which minimise (in the case of exclusion) or indeed exalt (in the case of discovery) such effects  (which will be the subject of a future publication). At any rate, the time-honoured assumption that the NWA  is a reliable investigative approach applicable over most of the parameter space of the BSM scenarios dealt with here should be dismissed. In fact, we also have cautioned that, despite cancellations may exist between the various effects described here, which in the end might not change sizably the inclusive cross section for certain values of $\Gamma_Q/M_Q$, these are only accidental and do not apply to the exclusive observables used in experimental searches, so that, again, limits obtained in the NWA would be inaccurate, owing to mis-estimated efficiencies.

\newpage

%%%%%%%%%%%%%%%%%%%%%%%%%%%%%%%%%%%%%%%%%%%%%%%%%%%%%%%%%%%%%%%%%%%%%%%%%
%%%%%%%%%%%%%%%%%%%%%%%%%%%%%%%%%%%%%%%%%%%%%%%%%%%%%%%%%%%%%%%%%%%%%%%%%
\section*{Acknowledgements}
SM and LP have been funded in part through the NExT Institute.

\bibliographystyle{JHEP}
\bibliography{XQCAT}

\noindent

\end{document}